\documentclass[12pt]{article}

\usepackage{fullpage}
\usepackage{indentfirst}

\usepackage{mathptmx}

\usepackage{authblk}

\usepackage{amsmath}
\usepackage{amssymb}
\usepackage{mathtools}

\usepackage{graphicx}
\usepackage{xcolor}

\usepackage{bm}
\usepackage{ulem}
\usepackage{float}
\usepackage{cite}
\usepackage{soul}
\usepackage{caption}
\usepackage{subcaption}

\usepackage{hyperref}

\newcommand\me{\mathrm{e}}

\newcommand{\dif}{\mathrm{d}}
\DeclareMathOperator{\diag}{diag}
\newcommand{\Tr}{\mathrm {Tr}}

\begin{document}

\title{Acoustic regular black hole in fluid and its similarity and diversity to a conformally related black hole}
	
\author{Chen Lan\thanks{lanchen@nankai.edu.cn}}
\author{Yan-Gang Miao\thanks{Corresponding author, miaoyg@nankai.edu.cn}}
\author{Yi-Xiong Zang\thanks{zangyx@mail.nankai.edu.cn}}
	
\affil{School of Physics, Nankai University, Tianjin 300071, China}

\maketitle

\begin{abstract}
We address an interesting question in the present paper that whether the acoustic gravity can be applied as a 
tool to the study of regular black holes. For this purpose, 
we construct a general acoustic regular black hole in the spherically symmetric fluid, where its  regularity is
verified from the perspective of  finiteness of curvature invariants and completeness of geodesics. In particular,
we find that the acoustic interval not only looks like a line element of a conformally
related black hole in which the fluid density can be regarded as a conformal factor, but also
gives rise to a non-vanishing partition function which coincides with that of a conformally related black hole. As an application, we provide a specific acoustic regular black hole model, investigate its energy conditions
and compute its quasinormal modes. We note that the strong energy condition of our model is violated completely outside the horizon of the model but remains valid in some regions inside the horizon, which may give a new insight into the relation between the regularity and strong energy condition. Moreover, we analyze the oscillating and damping features of our model when it is perturbed.
\end{abstract}
	
\maketitle

\newpage
	
\section{Introduction}
	
Since the Hawking radiation from black holes (BHs) was discovered \cite{hawking1974black},
it has become one of the central
subjects to study the quantum behaviors of BHs. However, this thermal radiation is too small to be directly detected by
any conceivable experiments. When the Schwarzschild BH with one solar mass is taken as an example, its radiation
temperature is
approximately $6\times 10^{-8} \rm{K}$, while the cosmic background is of $3  \rm{K}$ microwave radiation. Therefore, the former is completely covered up by the latter.
In other words, even if the thermal radiation is emitted,
it will be drowned out by the background noise.
This situation motivates the researches to shift from astrophysical phenomena to their analogues in laboratories on
Earth,
which was pioneered by Unruh \cite{unruh1981experimental} who proposed an acoustic analogy.
	
An acoustic black hole (ABH), being one of the realizations of analogue BHs, can be formed in laboratories on Earth
when the velocity of moving fluid exceeds the local velocity of sound, where the horizon is located~\cite{visser1998acoustic} at the junction of the supersonic and subsonic regions. Several attempts have been done in recent decades, including surface waves in Bose-Einstein
condensates \cite{garay2000sonic},
water flows 
\cite{weinfurtner2011measurement}, optical systems \cite{drori2019observation}, quantum many-body 
systems~\cite{yang2020simulating}, and so on.
For the early progress in analogue BHs, see, for instance, the review article \cite{barcelo2011analogue} and the references therein. Recently, there have been many theoretical and experimental advances in various aspects of analogue gravity, such as in the Hawking 
	radiation~\cite{Euve:2015vml,Steinhauer:2015saa,MunozdeNova:2018fxv,Kolobov:2019qfs,Euve:2018uyo}, the  superradiation \cite{ge2012acoustic,Torres:2016iee,Braidotti:2021nhw}, the quasinormal modes 
	(QNMs)~\cite{Torres:2020tzs}, and the
Lyapunov exponent \cite{wang2020geometrya}, etc. Moreover, ABHs have been  generalized~\cite{ge2019acoustic} to curved spacetimes. 
In particular, the experimental advances reflect the applicability of analogue gravity.

Although the analogue gravity has been developed and regarded as a tool of gaining insight into 
general relativity, the first simulation of Schwarzschild and Reissner-Nordstr\"om BHs was not realized 
until 2021 \cite{deOliveira:2021edr}.
Prior to this work, some analogue BH models, such as the draining bathtub model~\cite{visser1998acoustic,berti2004quasinormala}, may contain the necessary features that give rise to the astronomical phenomena, but can hardly have the direct counterparts in the universe.
And the differences between astronomical black holes and their acoustic counterparts may appear distinctly in the desired phenomena in the earth laboratory. 
For instance, in the acoustic simulation of the Painlev\'e-Gullstrand spacetime \cite{barcelo2011analogue},
the astronomical metric differs from the acoustic one by a conformal factor.
Thus, the study of the quasi-normal modes from the acoustic counterpart may not provide the full information of spectra
for the Painlev\'e-Gullstrand geometry, 
because the conformal factor affects the quasi-normal modes except in the eikonal limit \cite{Chen:2019iuo}.


Moreover, it is widely known that singular black holes (SBHs) suffer \cite{Frolov:2016pav} from the UV incompleteness at both classical and 
quantum levels
because  of the spacetime singularity. Many phenomenological models have been proposed for avoiding the
singularity at the center of BHs, see, for instance, the review~\cite{ansoldi2008spherical}. These nonsingular
solutions of general relativity are called regular black holes (RBHs), being of finite curvature invariants on the
entire manifold of spacetime. In fact, Bardeen proposed \cite{bardeen1968non} the first RBH which was recognized
\cite{ayon-beato2000bardeen} later on
as a product created by nonlinear electrodynamics (NED).
This model is currently dubbed as Bardeen black hole (BBH).
The further developments of the BBH have been presented, see, e.g.,
Refs.~\cite{Bronnikov:1979ex,ayon-beato1998regular,Fan:2016hvf}.
Besides the BBH, the other RBHs have also been proposed~\cite{Hayward:2005gi,Dymnikova:1992ux,Frolov:2017rjz}.

Our aim in the current work is to construct RBHs in acoustic
gravity named as acoustic RBHs (ARBHs) and to investigate their energy conditions and dynamic properties, 
such as QNMs~\cite{konoplya2011quasinormal}. Here we note that the energy conditions refer to the 
constrains
on the matters generating the RHBs in the Universe, not on the fluid for simulation in a laboratory. Since we
dedicate to study the RBHs with the aid of analogue gravity, we investigate whether the
RBHs we construct in fluid are reasonable or not, that is, if their counterparts in the Universe have the possibility of existence. As a by-product, we find that ARBHs have different characteristics from those of
RHBs generated by NED, in particular, they should be classified into conformal
gravity~\cite{Bambi:2016wdn,Toshmatov:2017kmw,Li:2021qim}. The seeming reason is that the acoustic
interval looks like a line element of a conformally related black hole, where the fluid density can be regarded as
a conformal factor, but the virtual reason is that the acoustic interval leads to a non-vanishing partition function
if it is interpreted in the context of conformally invariant theory.

In general relativity, the energy conditions give~\cite{hawking1973large,poisson2004relativist} constraints 
upon the energy-momentum tensor of matter fields, such as positivity of energy density and validity of causality.
For instance, one can determine whether the matter field of RBHs created by NED is physically 
reasonable in terms of the dominant energy condition and whether the superradiance occurs by checking the weak energy 
condition which is also associated with the second law of BH mechanics \cite{Unruh:1973bda}.
In the context of ARBHs, we define the analogue energy-momentum $T_{\mu\nu}$ and thus explore the 
corresponding energy conditions by supposing the linear relation between the analogue Einstein tensor and  
energy-momentum tensor. We find that the energy conditions of ARBHs have novel properties when ARBHs are 
dealt with in the framework of conformal gravity.
	
As QNMs play an important role in the stability analysis of analogue BHs, see, for instance, an example of optical BHs
\cite{Guo:2019tmr},  we focus on the QNMs of ARBHs by studying the propagation of scalar fields in the effective curved
spacetime manifested as the
acoustic disturbance.  As shown in Ref.~\cite{unruh1981experimental}, the equation of motion for the
acoustic disturbance is identical to the d'Alembertian equation of a massless scalar field propagating in a curved
spacetime. We can thus compute the QNM frequencies of ARBHs by using the WKB method~\cite{schutz1985black,iyer1987blackhole,iyer1987blackholea,berti2009quasinormal,konoplya2003quasinormal} as usual.
	
This paper is organized as follows. We propose a general method to construct ARBHs  in Sec.~\ref{sec2}, where
the regularity is verified  in the perspective of finiteness of curvature invariants and completeness of geodesics.
We then give one specific ARBH model in Sec.~\ref{sec:bardeen}. 
In Sec.\ \ref{sec:potential}, based on the complete form of Euler's equation we analyze the importance of an external-force term in the realization of acoustic analogy.
The energy conditions of the model are
discussed and compared with those of the conformally related Schwarzschild black holes
(CRSBHs)~\cite{Toshmatov:2017kmw}
in Sec.~\ref{sec:energy-condition}. In Sec.~\ref{sec3}, we analyze the effective potential and calculate the
QNMs for the ARBH model. Finally, we give our summary in Sec.~\ref{sec4}.
The Apps.~\ref{appendix:A} and \ref{appendix:B} include the detailed analyses of energy
conditions of CRSBHs and the repulsive interaction of the specific ARBH model outside the model's event horizon.
Throughout this paper, we adopt the units with the speed of sound $c = 1$ and the sign convention $(-,+,+,+)$.
	
\section{Acoustic regular black hole in fluid}
\label{sec2}

In this section, we construct a general ARBH in the spherically symmetric fluid.
The fluid is assumed to be locally irrotational, barotropic, inviscid, and compressible. The acoustic
interval then takes the form~\cite{barcelo2011analogue},
\begin{equation}
	ds^2 = \frac{\rho}{c}  \left[- c^2 dt^2 + (d\bm{x}-\bm{v}dt)^2\right],
	\label{eq3}
\end{equation}
which can be obtained by combining the equation of continuity,
\begin{equation}
	\partial_t\rho + \nabla \cdot (\rho \bm{v})=0, \label{eq1}
\end{equation}
and Euler's equation,
\begin{equation}
	\rho \left[ \partial_t \mathbf{} \bm{v}+ (\bm{v} \cdot
	\bm{\nabla})\bm{v} \right] = - \bm{\nabla} p
	-\rho\bm{\nabla} \psi, \label{eq2}
\end{equation}
where $\rho$, $\bm{v}$, and  $p$ are density, velocity, and pressure
 of the fluid, respectively, and $c\equiv \sqrt{\left|{\partial p}/{\partial \rho}\right|}$ is local speed of sound. In the following discussions $c$ is
normalized to unity,\footnote{In general, the local speed of sound depends mainly on the temperature of fluid. Here the temperature of fluid is constant, so it is usual to set $c=1$.} and the density $\rho$ and velocity $\bm{v}$ are supposed to be functions of radial coordinate $r$ only. 
In addition, the last term of Eq.\ \eqref{eq2} represents~\cite{visser1998acoustic} an external driving force and $\psi$ is the corresponding potential.
This term does not affect~\cite{visser1998acoustic} the wave equation of sound and the acoustic metric, but it is indispensable in the acoustic analogue of an astronomical black hole because $\psi$ provides an external field for realizing the specific fluid, which  
will be explained in detail in Sec.\ \ref{sec:potential}.

If we consider the spherically symmetric fluid with only non-vanishing radial velocity, $v_r\ne 0$,
and perform the following transformation,
\begin{equation}
	dt \rightarrow d \tilde{t} - \frac{v_r}{1 - v_r^2} dr,
\end{equation}
we rewrite Eq.~\eqref{eq3} as follows,
\begin{equation}
ds^2 = \rho\left(-fd {\tilde{t}}^2+f^{-1} dr^2+r^2d{\theta}^2+r^2 \sin^{2 } \theta d{\phi}^2\right),\label{lineele}
\end{equation}
or write the metric explicitly,
\begin{equation}
g_{{\mu}{\nu}}=\rho{\tilde g}_{{\mu}{\nu}}, \qquad {\tilde g}_{{\mu}{\nu}}\equiv {\rm diag}\left\{-f, f^{-1}, r^2, r^2 \sin^{2 } \theta\right\}, \qquad f\equiv 1 - v_r^2. \label{eq14}
\end{equation}
The density $\rho$ plays the role of a conformal factor if ${\tilde g}_{{\mu}{\nu}}$ describes a static spherically symmetric black hole.
In the above specific setting, $\rho$ and $v_r$ are constrained by the relation,\footnote{This represents the peculiarity of acoustic intervals which will be utilized to pick ARBHs out.}
\begin{equation}
	\label{eq:bound}	\rho v_r=\frac{A}{r^2},
\end{equation}
which can be derived by integrating Eq.~\eqref{eq1} with respect to the radial coordinate, where  $A$ is integration constant. Note that $\rho v_r$ is divergent at $r= 0$ in the manner of $r^{-2}$. This divergence appears at $r= 0$ in the following three cases:
\begin{itemize}
\item (i)  $\rho$ is divergent, while $v_r$ is finite;\footnote{Here ``finite" includes zero and nonzero constants.}
\item (ii) $\rho$ is finite, while $v_r$ is divergent;
\item (iii) Both $\rho$ and $v_r$ are divergent.
\end{itemize}
Such a classification will help us construct ARBHs.

In order to check whether $g_{{\mu}{\nu}}$, see Eq.~(\ref{eq14}),  together with Eq.~(\ref{eq:bound}) describes an ARBH or not, we have to investigate the finiteness of curvature invariants and completeness of
geodesics at the center of this ARBH. Next, we discuss the two issues in two separate subsections.

\subsection{Finiteness of curvature invariants}
Using Eq.~(\ref{eq14}) and the definitions of the three curvature invariants, the Ricci scalar $R\equiv g^{\mu\nu}R_{\mu\nu}$, the contraction of two Ricci tensors $R_2\equiv R_{\mu \nu} R^{\mu \nu}$, and the Kretschmann scalar $K\equiv R_{\mu \nu \rho \sigma} R^{\mu \nu \rho \sigma}$, we obtain
\begin{eqnarray}
	R &=& \frac{3 f \rho^{\prime 2}}{2 \rho^3} - \frac{3 \left(rf' \rho' + 2 f \rho' +
		rf \rho''\right)}{r \rho^2} - \frac{r^2 f'' + 4 r f' + 2 f - 2}{r^2 \rho},\label{rb}\\
		R_2 &= & \frac{\left(2 r \rho \rho' f' + 2 f \rho \rho' + 3 r f \rho \rho'' + r
			\rho^2 f'' + 2 \rho^2 f' - 3 r f \rho^{\prime 2}\right)^2}{4 r^2 \rho^6} \nonumber\\
		& &+  \frac{\left(r^2 f' \rho' + 4 r f \rho' + r^2 f \rho'' + 2 r \rho f' + 2 f
			\rho - 2 \rho\right)^2}{2 r^4 \rho^4} \nonumber\\
		& & +\frac{\left(2 rf' \rho' + 2 f \rho' + r \rho f'' + 2 \rho f' + rf
			\rho''\right)^2}{4 r^2 \rho^4},\label{rcont}\\
		K &= &  \frac{15 f^2 \rho^{\prime 4}}{4 \rho^6} - \frac{3 f \rho^{\prime
				2}  \left(f' \rho' + 2 f \rho''\right)}{\rho^5} \nonumber\\
		& & +\frac{4 rf \rho' \rho''  \left(rf' + f\right) + 2 r^2 f^{\prime 2}
			\rho^{\prime 2} + 2 f \rho^{\prime 2} \left (rf' - r^2 f'' + 5 f - 1\right) + 3
			r^2 f^2 \rho^{\prime \prime 2}}{r^2 \rho^4} \nonumber\\
		 &  & +\frac{2 \left[r^3 ff'' \rho'' + 2 r^2 ff' \rho'' + \rho' \left (4 rff' - 4 f
			+ r^3 f' f'' + 2 r^2 f^{\prime 2} + 4 f^2\right)\right]}{r^3 \rho^3} \nonumber\\
		& & +\frac{r^4 f^{\prime \prime 2} + 4 r^2 f^{\prime 2} + 4 f^2 - 8 f +
			4}{r^4 \rho^2},\label{kb}
\end{eqnarray}
where the prime denotes the derivative with respect to the radial coordinate.

Now let us analyze whether the three curvature invariants are finite or not when $r\to 0$ in the first case mentioned above.
Substituting Eq.~(\ref{eq:bound}), i.e., $\rho=A/(r^2v_r)$, into Eqs.~(\ref{rb}), (\ref{rcont}), and (\ref{kb}),
we express explicitly the leading orders of  the three curvature invariants,
\begin{eqnarray}
	R  &=& \frac{2 v_{0}^3 }{A} + O (r), \label{eq51}\\
	R_2  &=& \frac{2 v_{0}^2 }{A^2}  \left(v_{0}^4  - 2 v_{0}^2  + 2\right)+ O (r^2),\label{eq52}\\
	K &=& \frac{4v_{0}^2}{A^2}  \left(v_{0}^4 - 2v_{0}^2  + 2  \right) + O (r^2),\label{eq53}
\end{eqnarray}
where $v_{0}\equiv \lim_{r\to 0}v_r$. They are obviously finite as $r$ goes to zero.
As to the asymptotic behaviors of $\rho$ at $r\to 0$, we know from Eq.~(\ref{eq:bound}), $\rho(r)\sim 1/r^{2+a}$  with $a\ge 0$, where $a> 0$ corresponds to that $v_r$ goes to zero in the manner of $r^a$ and $a= 0$ corresponds to that $v_0$ is a nonzero constant.
Moreover,  we have to require the asymptotic flatness of the metric (Eq.~(\ref{eq14}) associated with Eq.~(\ref{eq:bound})) in the first case. Let us analyze the leading orders of $v_r (r)$ and $\rho(r)$. If $v_r (r)\to A/r^2$ and $\rho(r)\to 1$ when $r\to \infty$, the asymptotic flatness is ensured.
As a result, the models constructed in the first case can be regarded as a candidate of ARBHs.\footnote{We note that the density $\rho$ can indeed be regarded as a conformal factor due to its asymptotic behaviors: $\rho(r)\sim 1/r^{2+a}$ at zero and $\rho(r)\sim 1$ at infinity. Based on such asymptotic behaviors, one of the possible forms reads, $\rho(r)=\left(1+\frac{L^2}{r^2}\right)^{2b}$, where $b\equiv \frac{2+a}{4}$ and $L\equiv A^{1/{(4b)}}$, see, for instance, the conformal factors chosen in Refs.~\cite{Chen:2019iuo,Li:2021qim}.}

For the second case in which $\rho$ is finite, while $v_r$ is divergent at $r=0$, we can judge by following the way for the first case that the three curvature invariants are  divergent as $r$ goes to zero. In fact, we have a shortcut to reach the goal. If we choose the asymptotic behaviors of $\rho$ and $v_r$, for instance, to be $\rho(r)\sim 1$ and $v_r(r)\sim A/r^2$ as $r\to 0$, respectively, the shape function of Eq.~(\ref{eq14}) tends to $1-A^2/r^4$, which definitely describes a singular spacetime. Thus, no ARBHs can be given in the second case.

As to the third case where both $\rho$ and $v_r$ are divergent as $r\to 0$, we can easily determine from Eqs.~(\ref{rb})-(\ref{kb}) that no ARBHs can be constructed in this case, either.

In summary,  Eq.~(\ref{eq14}) associated with Eq.~(\ref{eq:bound}) indeed describes an ARBH when the fluid density is divergent while the radial velocity is finite at $r=0$, where the fluid density plays the role of a conformal factor, see footnote 4 for a detailed explanation.

\subsection{Completeness of geodesics}
\label{sec:geodesic}

To check the geodesic completeness of the metric Eq.~\eqref{eq14}, we start with the Lagrangian~\cite{Bambi:2016wdn} of a test particle constrained in the equatorial orbit $\theta=\pi/2$,
\begin{equation}
	\label{eq:lagrange}
	2\mathcal{L} =  \rho
	\left( f  \dot{\tilde{t}}^2 -
	\frac{\dot{r}^2}{f } - r^2  \dot{\phi}^2 \right),
\end{equation}
where the dot stands for the derivative with respect to affine parameter $\tau$. Since $t$ and $\phi$ are cyclic
coordinates, one has two integrations of motion,
\begin{equation}
	P_t = f  \rho   \dot{\tilde{t}} \equiv {\mathbb E},\qquad
	P_{\phi} = - r^2 \rho   \dot{\phi} \equiv - {\mathbb L},\label{eq43}
\end{equation}
where the energy ${\mathbb E}$  and angular momentum ${\mathbb L}$  are conserved quantities for a free radially infalling particle in static spacetimes.
Then replacing the velocities in Eq.~\eqref{eq:lagrange} by Eq.~\eqref{eq43} we obtain
\begin{equation}
	\dot{r}^2=V_{\rm eff},\qquad
	V_{\rm eff}=\frac{{\mathbb E}^2}{ \rho^2}
	-\frac{{\mathbb L}^2f}{r^2 \rho^2 }
	-\delta  \frac{f}{\rho },
	\label{eq54}
\end{equation}
where $\delta=0$ corresponds to null and $\delta=1$ to timelike geodesics, respectively. For simplicity, we consider the radial geodesic
motion, which implies that the angular momentum vanishes, ${\mathbb L}= 0$.
Now we can write down the affine parameter by the following integral,
\begin{equation}
	\label{eq:affine-integral}
	\tau =  \int^{r_i}_{r_f} \frac{\dif r}{\sqrt{V_{\rm eff}}},
\end{equation}
where $r_i$ and $r_f$ represent the initial and final positions, respectively.

For a null geodesic, $\delta=0$, the integrand of Eq.~\eqref{eq:affine-integral} can be written as follows:
\begin{equation}
	\label{eq:integrand delta=0}
	\frac{1}{\sqrt{V_{\mathrm{eff}}}} = \frac{\rho}{{\mathbb E}}.
\end{equation}
Since $\rho$ diverges at $r=0$, Eq.~\eqref{eq:integrand delta=0} implies that the proper time is also divergent.

For a timelike geodesic, $\delta=1$, the integrand can be written as
\begin{equation}
		\label{eq:integrand delta=1;transform }
	\frac{1}{\sqrt{V_{\mathrm{eff}}}} = \frac{\rho}{\sqrt{\mathbb{E}^2 - f
			\rho}}.
\end{equation}
From Eq.~(\ref{eq54}), we deduce that ${\mathbb{E}^2 - f	\rho}\geqslant0$, which means that $\mathbb{E}$ goes to infinity if $f>0$  inside
the innermost horizon. That is to say, the test particle needs infinite energy to reach the center of ARBHs, so
there are no particles that can reach the center. Alternatively, considering that $f<0$ inside horizons and $\mathbb{E}$ is
finite but ${\rho}$ goes to infinity when $r\to0$, we have $ {1}/{\sqrt{V_{\mathrm{eff}}}}
\to\sqrt{\rho}/\sqrt{-f} $. Thus, the integrand is also divergent, i.e., the timelike geodesic is complete as well.

As a matter of fact, Eq.~\eqref{eq54} describes a particle that is moving in a negative
potential well but has vanishing total energy.
Intuitively, this test particle cannot reach the center of ARBHs within finite ``time" because $V_{\rm eff}$ vanishes at $r=0$.

In this section, we have proven that the Ricci scalar $R$, the contraction of two Ricci tensors $R_{2}$, and the Kretschmann scalar $K$ are
finite at $r=0$, and both the null and timelike geodesics are complete in the ARBH spacetimes, which
means that the ARBHs we constructed have no spacetime singularity.

\section{A specific model}
\label{sec:bardeen}

A direct way to construct a RBH is to substitute a shape function into Eq.~\eqref{eq14},
which is similar to the case of Schwarzschild BHs, then one can determine
$\rho$ and the metric $g_{\mu\nu}$ with the help of
Eq.~\eqref{eq:bound}. Nevertheless, such a RBH is the lack of  asymptotic flatness. 
Therefore, considering the
asymptotic behaviors of the fluid density at zero and at infinity together with the constraint
between the density and the radial velocity, we give such an ARBH model,
\begin{eqnarray}
	{\rho} = \rho_{\ast} \left( 1 + \frac{L^2}{r^2} \right)^{2 N},\qquad
	{v}_r = \frac{A}{\rho_{\ast}r^2  \left( 1 + \frac{L^2}{r^2} \right)^{2 N}},\label{eq:v}
\end{eqnarray}
where $\rho_{\ast}$ is a constant with the dimension of density and the integration constant $A$ has been  introduced in
Eq.~\eqref{eq:bound}. As explained in
Refs.~\cite{Toshmatov:2017kmw,Li:2021qim,1605}, $L$ is a typical length scale of this model, such as the horizon radius or the
Planck length, and $N$, a dimensionless constant, determines whether the scalar curvatures are regular at the center of this model. Further, we perform such a
transformation,
\begin{eqnarray}
r\rightarrow \sqrt{\frac{A}{\rho_{\ast}}} r, \qquad  L \rightarrow \sqrt{\frac{A}{\rho_{\ast}}}L,
\end{eqnarray}
in Eq.~\eqref{eq:v}, and substitute the transformed Eq.~\eqref{eq:v} into the line element, Eq.~(\ref{lineele}), and then let the line element absorb $\rho_{\ast}$. In this way,  we make the new line element look like Eq.~(\ref{lineele}) but associate with the dimensionless density and radial velocity\footnote{Due to the setting, $c = 1$, the radial velocity is dimensionless, which gives rise to the dimensionless length and new line element in our units.}
as follows:
\begin{eqnarray}
	\rho =  \left( 1 + \frac{L^2}{r^2} \right)^{2 N},\qquad
	v_r =  \frac{1}{ r^2\left( 1 + \frac{L^2}{r^2} \right)^{2 N}}.\label{eq:vre}
\end{eqnarray}
We emphasize that the new line element is independent of the constant density $\rho_{\ast}$ and the integration constant $A$ but dependent only on the parameters $L$ and $N$.

Now we substitute Eqs.~(\ref{lineele}), (\ref{eq14}), and (\ref{eq:vre}) into Eqs.~\eqref{rb}-\eqref{kb} and thus derive the leading
orders of curvature invariants near $r=0$. We notice that the leading orders depend on $N$.  When $N\le 1/2$,
the leading orders near $r=0$ are
\begin{eqnarray}
	\label{eq:invariantsR}
	R &=& \frac{6 - 4 N (2 N + 1)}{L^{12 N}}\, r^{6 (2 N - 1)},\\
	\label{eq:invariantsR2}
	R_2 &=& \frac{8 N\left [\left(4 N^2 (6 N - 17) + 86 N\right) - 51\right] + 90}{L^{24 N}}
	\,r^{12 (2 N - 1)},\\
	\label{eq:invariantsK}
	K &=& \frac{16 N \left[\left(4 N^2 (27 N- 76) + 329 N\right) - 160\right] + 468}{L^{24
			N}} \,r^{12 (2 N - 1)};
\end{eqnarray}
when $N>1/2$, they have the following forms,
 \begin{eqnarray}
 	R &=& \frac{12 (1 - 2 N) N}{L^{4 N}} r^{2 (2 N - 1)},\\
 	R_2 &=& \frac{16 N^2  [2 N (6 N - 7) + 5]}{L^{8 N}} r^{4 (2 N - 1)},\\
 	K &=& \frac{16 N^2  [4 N (3 N - 4) + 7]}{L^{8 N}} r^{4 (2 N - 1)}.\label{eq:K}
 \end{eqnarray}
From Eqs.~\eqref{eq:invariantsR}-\eqref{eq:K}, we can confirm that the curvature invariants are finite when $N\ge 1/2$.

To illustrate the finiteness of curvature invariants and completeness of geodesics for the specific model, we take
two different cases, $N=1/2$ and $N=1$, where the former is critical while the latter is a sample of $N> 1/2$.
\begin{itemize}
\item (i) $N=1/2$

In this case, there exists only one horizon whose radius equals $r_+= \sqrt{1 - L^2}$, where the existence of horizons requires $L^2<1$.
Eq.~(\ref{eq:vre}) reduces to
\begin{equation}
	\label{eq:metric s=1/2}
	\rho = 1 + \frac{L^2}{r^2},\qquad  v_r = \frac{1}{r^2+L^2}.
\end{equation}
Correspondingly, the leading orders of the three curvature invariants near $r=0$ read
\begin{equation}
	R = \frac{2}{L^6} + O (r^2), \qquad R_2 = \frac{4 L^8 - 4 L^4 + 2}{L^{12}} +
	O (r^2), \qquad K = \frac{8 L^8 - 8 L^4 + 4}{L^{12}} + O (r^2).\label{eq21}
\end{equation}
They are obviously finite. As to the  completeness of geodesics, for the null geodesics with $\delta=0$, substituting Eq.~\eqref{eq:metric s=1/2} into Eqs.~\eqref{eq:affine-integral} and \eqref{eq:integrand delta=0}, we obtain the affine parameter,
\begin{equation}
\tau = \frac{1}{{\mathbb E}} \left( r_i - r_f - \frac{L^2}{r_i} + \frac{L^2}{r_f}
\right),\label{taun05}
\end{equation}
which goes to infinity when the initial position is fixed and the final position goes to zero. Moreover, for the timelike
geodesics with $\delta=1$, Eq.~\eqref{eq:affine-integral} cannot be expressed analytically because of the
complicated integrand, but the expansion of the integrand near $r=0$ can be written as
\begin{equation}
	\label{eq:leading oeder s=1/2}
	 \frac{1}{\sqrt{V_{\mathrm{eff}}}}=\frac{L^3}{\sqrt{(1 - L^4) }} \frac{1}{r}+O(r),
\end{equation}
which implies that the affine parameter diverges when the final position goes to zero.
\item (ii) $N=1$

For this case, the horizon radii are $r_{\pm}=\sqrt{ (1 - 2 L^2) / 2 \pm \sqrt{1 - 4 L^2} / 2}$, where $``+"$ means the
outer horizon and $``-"$ the inner horizon, and the existence of horizons gives the condition, $L^2\le 1/4$. Eq.~(\ref{eq:vre}) gives the density and radial velocity as follows:
\begin{equation}
		\label{eq:metric s=1}
	\rho = \left(1+ \frac{L^2}{r^2} \right)^2, \qquad v_r = \frac{1}{r^2
		\left(1+ \frac{L^2}{r^2} \right)^2},
\end{equation}
and thus the expansions of curvature invariants near $r=0$ read
\begin{equation}
	R = - \frac{12 }{L^4}r^2 + O (r^4), \qquad R_2 = \frac{48 }{L^8}r^4 + O(r^6), \qquad K = \frac{48 }{L^8}r^4 + O (r^6).
\end{equation}
It is obvious that the curvature invariants converge  at $r=0$. For the completeness of the null geodesics with $\delta=0$,
we derive the affine parameter,
\begin{equation}
	\tau = \frac{1}{{\mathbb E}} \left( \frac{L^4 + 6 L^2 r_f^2 - 3 r_f^4}{3 r_f^3} -
	\frac{L^4 + 6 L^2 r_i^2 - 3 r_i^4}{3 r_i^3} \right),\label{taun1}
\end{equation}
which is divergent when $r_f\to 0$, i.e.,  the particles moving along the radial geodesic can never reach the center within a
finite proper time. For the completeness of the timelike geodesics with $\delta=1$, we give the expansion of the
integrand of Eq.~\eqref{eq:affine-integral} near $r= 0$,
\begin{equation}
	\label{eq:leading oeder s=1}
	 \frac{1}{\sqrt{V_{\mathrm{eff}}}}=\frac{L^2}{r^2} + O(r^2),
\end{equation}
which diverges at $r= 0$ as expected.
\end{itemize}

\begin{figure}[!ht]
	\centering
	\begin{subfigure}[b]{0.45\textwidth}
		\centering
		\includegraphics[width=\textwidth]{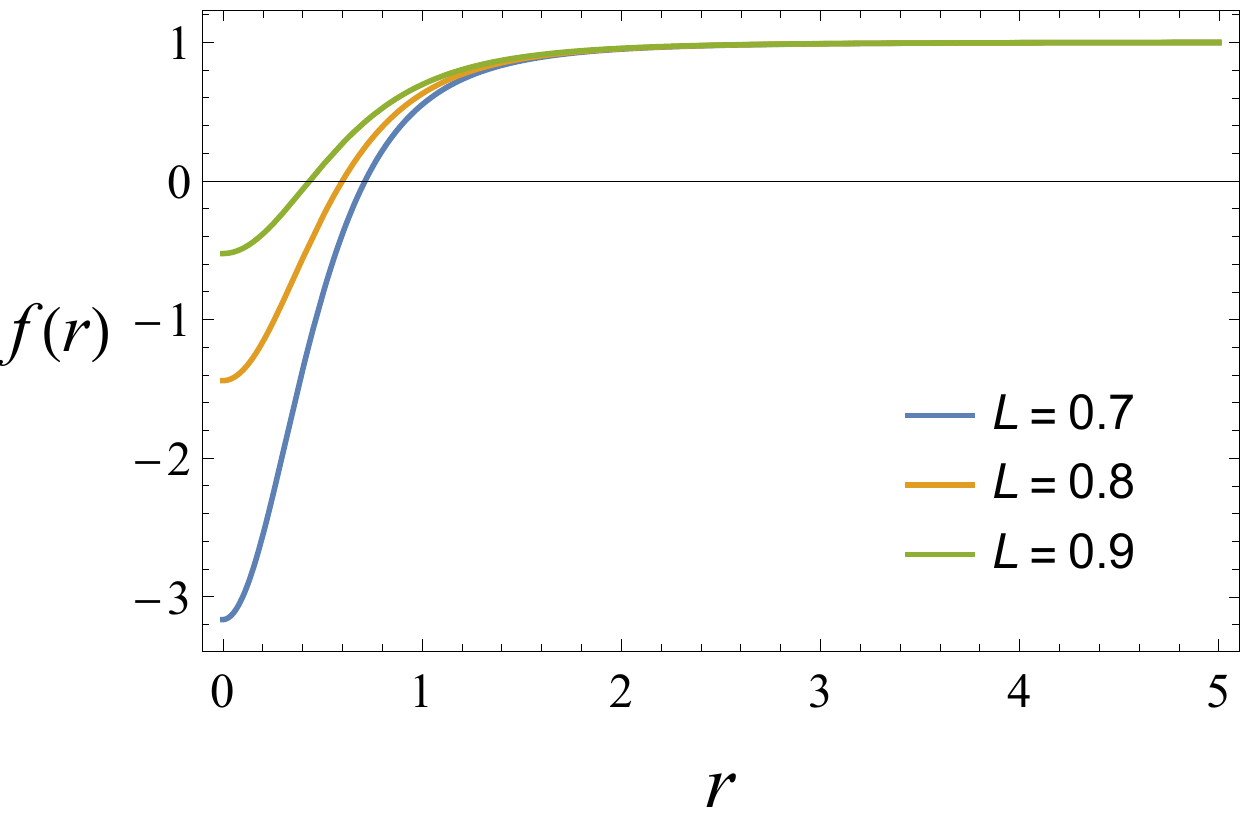}
	\end{subfigure}
	\begin{subfigure}[b]{0.45\textwidth}
		\centering
		\includegraphics[width=\textwidth]{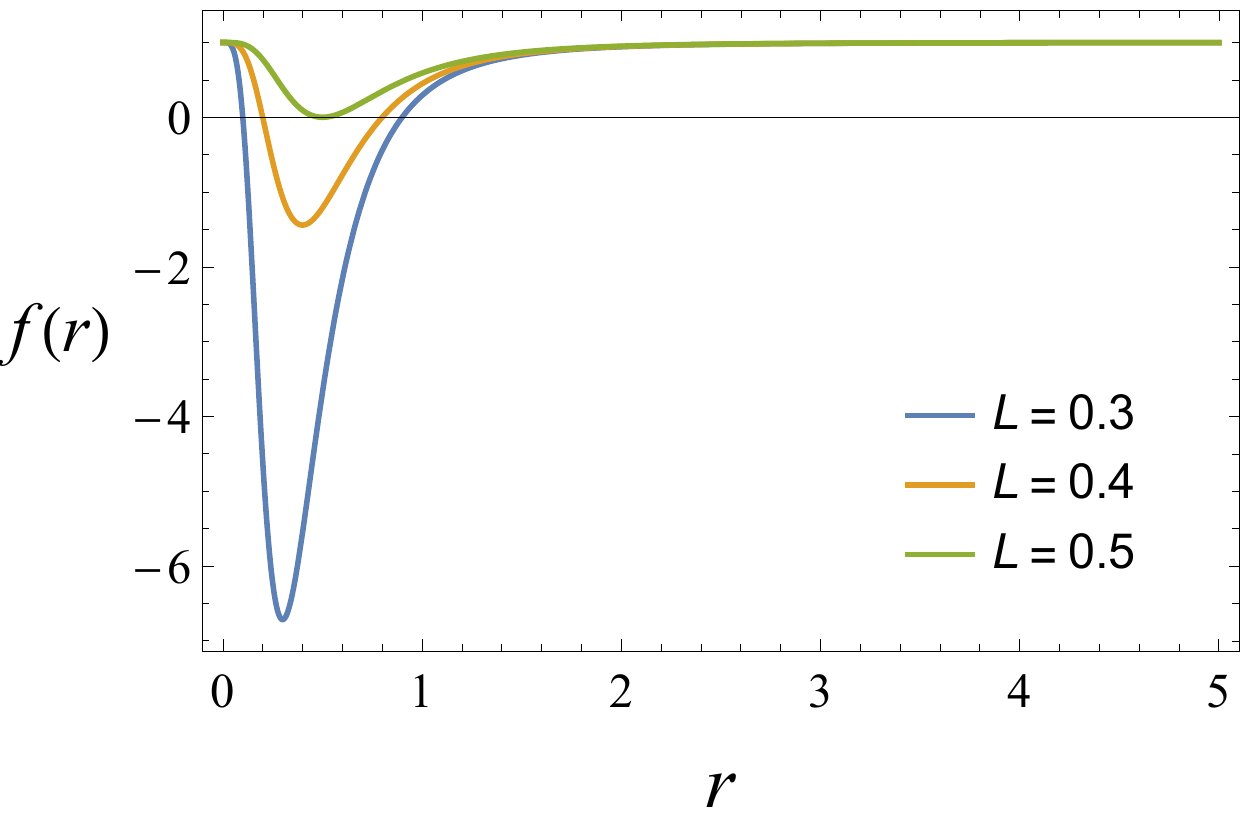}
	\end{subfigure}
	\captionsetup{width=.9\linewidth}
	\caption{$f(r)$ with respect to $r$ for the cases of $N=1/2$ (left) and $N=1$ (right), only one horizon in the former case but normally two horizons in the latter. Note that the values of $L$ satisfy $L^2<1$ in the left graph and $L^2\le1/4$ in the right graph, respectively.}
	\label{fig:f-0.5and1}
\end{figure}

\begin{figure}[!ht]
	\centering
	\begin{subfigure}[b]{0.45\textwidth}
		\centering
		\includegraphics[width=\textwidth]{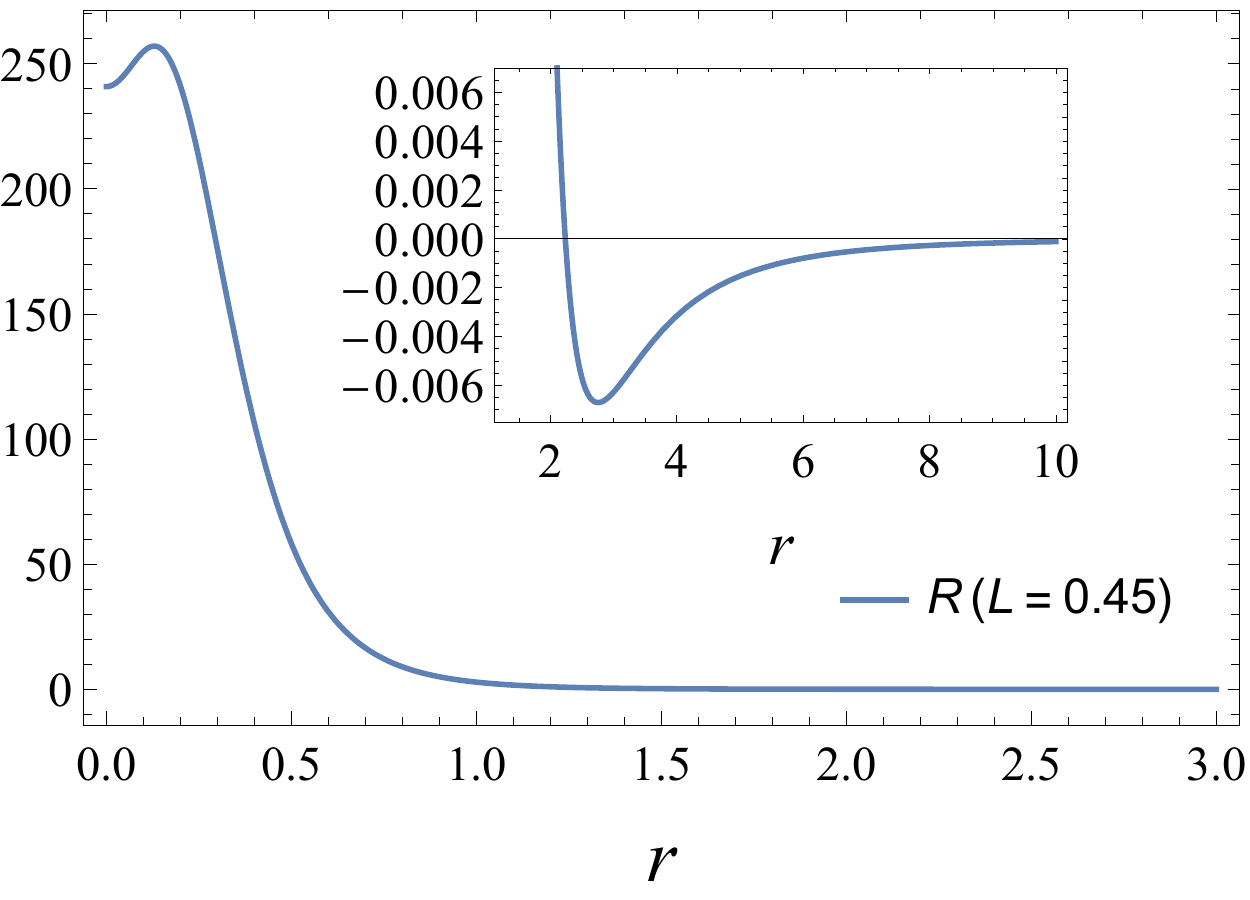}
		\caption{$R$ with respect to $r$.}
		\label{fig:s=0.5}
	\end{subfigure}
	\begin{subfigure}[b]{0.46\textwidth}
		\centering
		\includegraphics[width=\textwidth]{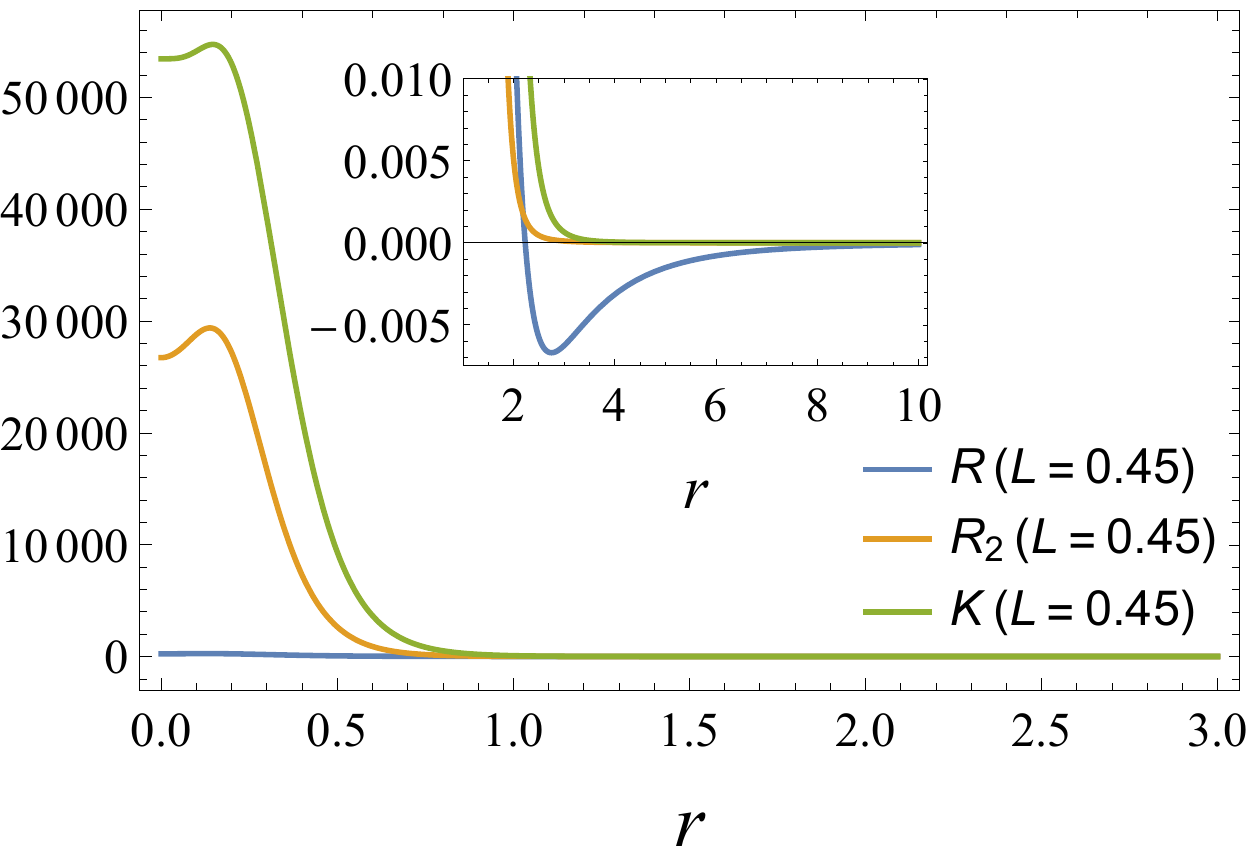}
		\caption{$R$, $R_2$, and $K$ with respect to $r$.}
		\label{fig:s=1}
	\end{subfigure}
	\captionsetup{width=.9\linewidth}
	\caption{$R$, $R_2$,
		and $K$ with respect to $r$ for the case of $N=1/2$, where $L=0.45$ which satisfies $L^2<1$. Note that $R$ is separated from the right graph and presented in the left graph in order to show its detail features.}
	\label{fig:scurvatures}
\end{figure}

\begin{figure}[!ht]
	\centering
	\includegraphics[width=0.45\textwidth]{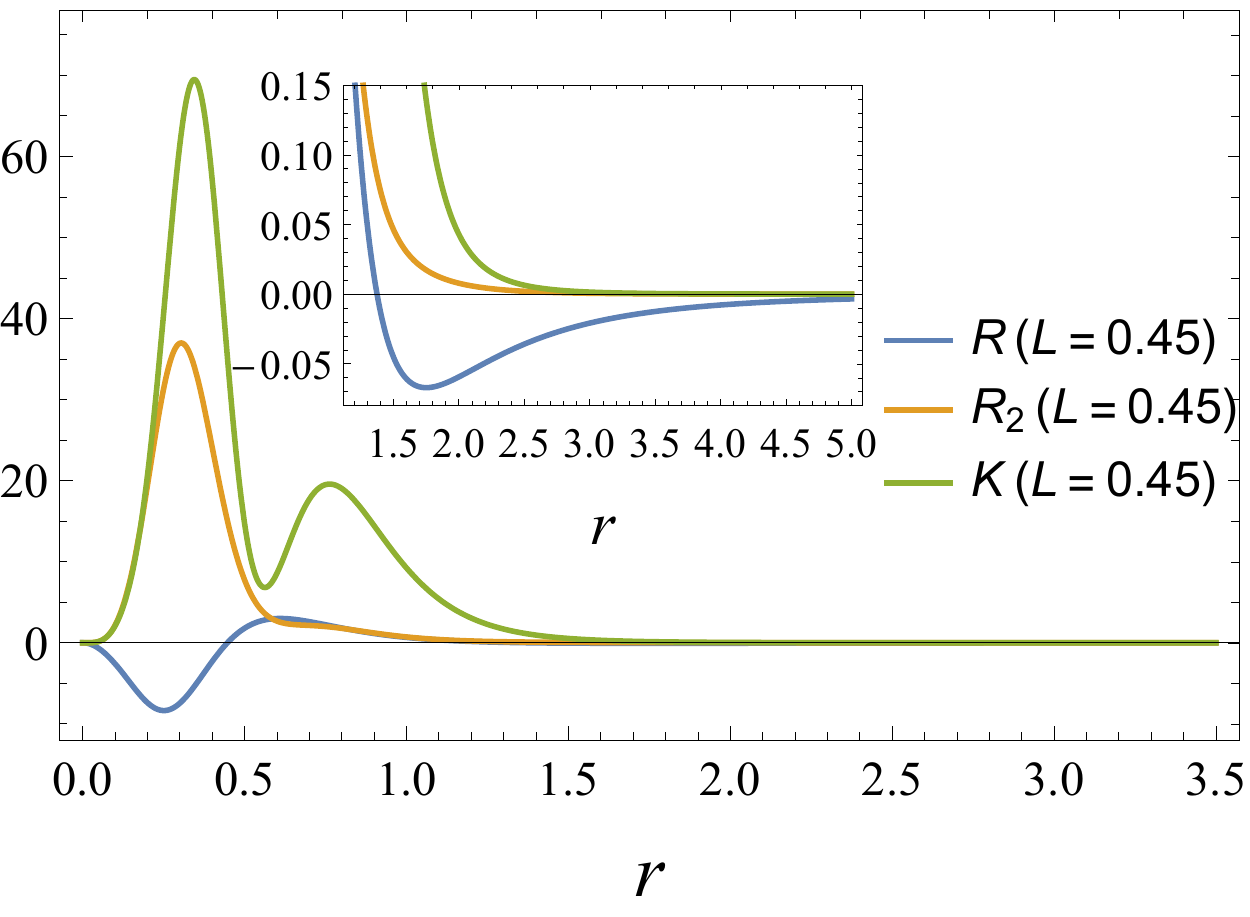}
	\captionsetup{width=.9\linewidth}
	\caption{$R$, $R_2$,
		and $K$ with respect to $r$ for the case of $N=1$, where $L=0.45$ which satisfies $L^2\le 1/4$.}
	\label{fig:newpicture-1}
\end{figure}

\begin{figure}[!ht]
	\centering
	\includegraphics[width=0.47\textwidth]{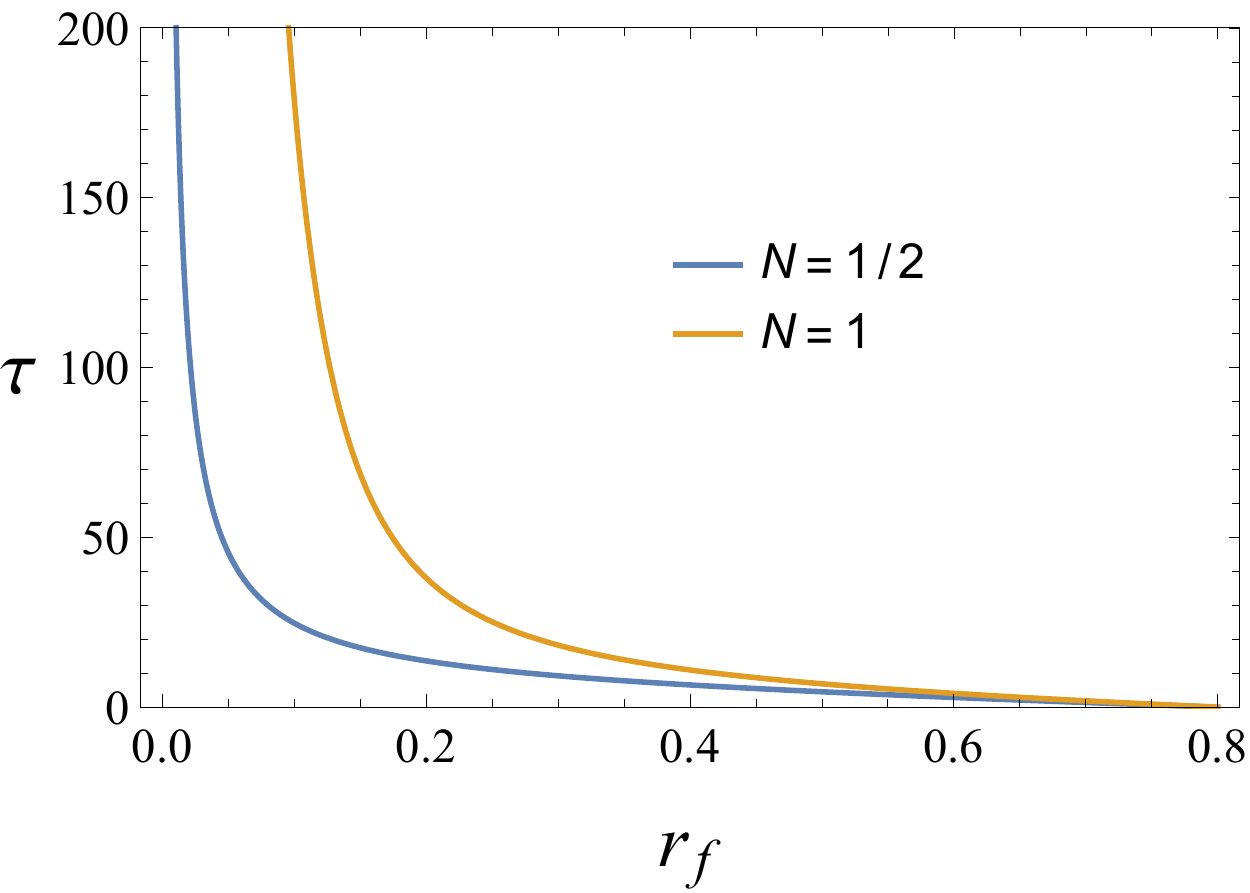}
	\captionsetup{width=.9\linewidth}
	\caption{$\tau$ with respect to $r_f$ for the two cases of $N=1/2$ and $N=1$, where $L=0.45$, the initial
		position $r_i=0.8$, and energy ${\mathbb E}=0.1$.}
	\label{fig:nullgeodesic0.5}
\end{figure}

Now we illustrate the regularity of this specific ARBH model in four figures. We plot the graphs of shape function $f(r)$ in Fig.~\ref{fig:f-0.5and1} for the cases of $N=1/2$ and $N=1$. The three curvature invariants as a function of the radial coordinate are plotted in Fig.~\ref{fig:scurvatures} for
the case of $N=1/2$ and in Fig.~\ref{fig:newpicture-1} for the case of $N=1$ according to Eqs.~(\ref{rb})-(\ref{kb}) and Eqs.~(\ref{eq:metric s=1/2}) and (\ref{eq:metric s=1}).
Moreover, we plot the graph of the affine parameter of null geodesics as a function of the final position in
Fig.~\ref{fig:nullgeodesic0.5} according to Eqs.~(\ref{taun05}) and (\ref{taun1}).

\newpage
\section{Potentials of external driving force}
\label{sec:potential}

Our approach to construct the acoustic metric in Sec.\ \ref{sec:bardeen} is based on the following assumptions:
\begin{itemize}
    \item The speed of sound is a position-independent constant and can be normalized to unity, $c=1$;
    \item The fluid is irrotational, i.e., its vorticity $w$ vanishes, $w\equiv\nabla\times \bm{v}=0$;
    \item The fluid is spherically symmetric, i.e., the velocity $\bm{v}$ has only radial component $v_r$ and all physical quantities, such as $\rho$, $v_r$, etc., depend only on radial coordinate $r$.
\end{itemize}
Therefore, if Euler's equation Eq.\ \eqref{eq2} did not involve an external-force term, the above items would lead to a problem on consistency when we are going to establish the acoustic counterpart of a gravitational metric. 
On the premise of the above three assumptions, 
the continuity equation and Euler equation are reduced to
\begin{equation}
\partial_r\left(r^2 \rho v_r\right) =0,
\end{equation}
and
\begin{equation}
\label{eq:reduced-sys}
\partial_r\left(\frac{v_r^2}{2}\right) + \partial_r \ln(\rho)=-\partial_r \psi,
\end{equation}
respectively. Thus, if there were no the external-force term,  $-\partial_r \psi$, 
one would fix $v_r$ (or $\rho$) via the continuity equation when $\rho$ (or $v_r$) is given to mimic a gravitational metric, but such a treatment would probably contradict to the Euler equation.
In other words, we have actually only one unknown variable $v_r$ (or $\rho$) but two dynamical equations, i.e., one redundant condition appears. Nonetheless, this case will never happen when the external-force potential exists.


Now
we calculate the external potential for our ARBH model established in Sec.\ \ref{sec:bardeen}.
The first integral of Euler's equation in Eq.\ \eqref{eq:reduced-sys} provides
\begin{equation}
\psi=\psi_0-\ln(\rho)-\frac{v_r^{2}}{2},\label{solpsi}
\end{equation}
where $\psi_0$ is an integration constant. 
Then, substituting Eq.\ \eqref{eq:vre} into Eq.~(\ref{solpsi}), we arrive at
\begin{equation}
    \psi=\psi_0-2N\ln \left( 1 + \frac{L^2}{r^2} \right)-\frac{1}{2 r^4\left( 1 + \frac{L^2}{r^2} \right)^{4N}},
\end{equation}
whose asymptotic behaviors at $r\to 0$ and $r\to \infty$ take the forms,  
\begin{equation}
   \psi \xrightarrow{r\to 0} 4 N \ln (r),\qquad
   \psi \xrightarrow{r\to \infty} \psi_0 -2N\frac{L^2}{ r^2},
\end{equation}
respectively.
In other words, the external force is asymptotic to $-4N/r$ around the center and vanishes at infinity. It is obvious that the Euler equation of our ARBH model,  Eq.~(\ref{solpsi}), has a consistent asymptotic behavior when $v_r$ is finite and $\rho$ divergent as $r\to 0$. 

Because the external-force term in Euler's equation does not affect~\cite{visser1998acoustic} acoustic metrics, so it has rarely been drawn much attention~\cite{barcelo2011analogue}. As we have discussed above, this term suggests a way to realize the specific fluid when we study the acoustic analogue of an astronomical black hole, so it is critical.
\section{Energy conditions}
\label{sec:energy-condition}

As is known, the energy conditions can examine cosmological models and strong gravitational fields, and give restrictions on the forms of energy-momentum tensors of matter fields. In general, the energy conditions are classified~\cite{hawking1973large} into four categories: Null energy condition (NEC), weak energy condition (WEC), strong energy condition (SEC), and dominant energy condition (DEC).

Based on Refs.~\cite{poisson2004relativist, Toshmatov:2017kmw}, we briefly explain the meanings of the four
energy conditions. The NEC requires that both energy density and pressure cannot be negative when measured
by an observer traversing a null curve, or if one of them is negative, the other must be positive and its
magnitude must be larger than the absolute value of the negative quantity. The WEC states that the energy density of any matter distribution
measured by any observer traversing a timelike curve must be nonnegative. The SEC requires
\begin{equation}
	\left( T_{\mu \nu} - \frac{1}{2} Tg_{\mu \nu} \right) v^{\mu} v^{\nu} \geq 0,
\end{equation}
where $v^{\mu}$ is future-directed, normalized, and timelike vector,  $T_{\mu \nu} $ is energy-momentum
tensor, and $T=g^{\mu \nu} T_{\mu \nu} $. The DEC states that the energy flow cannot be
faster than the speed of light, i.e., it ensures the causality. 
The energy-momentum tensor can be written as
${T^{\mu}}_{ \nu}\equiv g^{\mu \alpha} T_{\alpha
\nu}=\diag\{-\rho_0, P_1,P_2, P_3\}$, see App.~\ref{appendix:A} for the derivation and discussion. Thus, the four energy conditions can be expressed in terms of the components of the energy-momentum tensor as follows:
\begin{equation}
	\begin{array}{ll}
		\text{NEC: } & \rho_0 + P_i \geq 0, \qquad i = 1, 2, 3,\\
		& \\
		\text{WEC: } & \rho_0 \geq 0,\qquad \rho_0 + P_i \geq 0, \qquad i = 1, 2, 3,\\
		& \\
		\text{SEC: } & \rho_0 + \sum_{i = 1}^3 P_i \geq 0,\qquad \rho_0 + P_i \geq 0, \qquad i = 1, 2, 3,\\
		& \\
		\text{DEC} :&\rho_0 \geq 0,\qquad \rho_0 -| P_i |\geq 0, \qquad i = 1, 2, 3.\label{encon}
	\end{array}
\end{equation}

\subsection{Energy conditions of our ARBH model}

Let us investigate various energy conditions for the ARBH model we just constructed. 
We suppose the energy-momentum tensor is proportional to the Einstein tensor of the acoustic gravity 
because	our strategy is to investigate the physicality of a gravitational BH equivalent to our 
ARBH, and therefore derive the four components of  ${T^{\mu}}_{ \nu}$. Using Eqs.~\eqref{eq14}
and \eqref{eq:bound} together with Eq.~\eqref{eq:metric s=1/2} for the case of $N=1/2$ or
Eq.~\eqref{eq:metric
s=1} for the case of $N=1$, we can verify the relation,\footnote{In fact, this condition is valid for a general static
and spherically symmetric BH.} $P_2=P_3$, so there are only six independent inequalities in Eq.~(\ref{encon})
that are listed below.

\begin{figure}[!ht]
	\centering
	\begin{subfigure}[b]{0.3\textwidth}
		\centering
		\includegraphics[width=\textwidth]{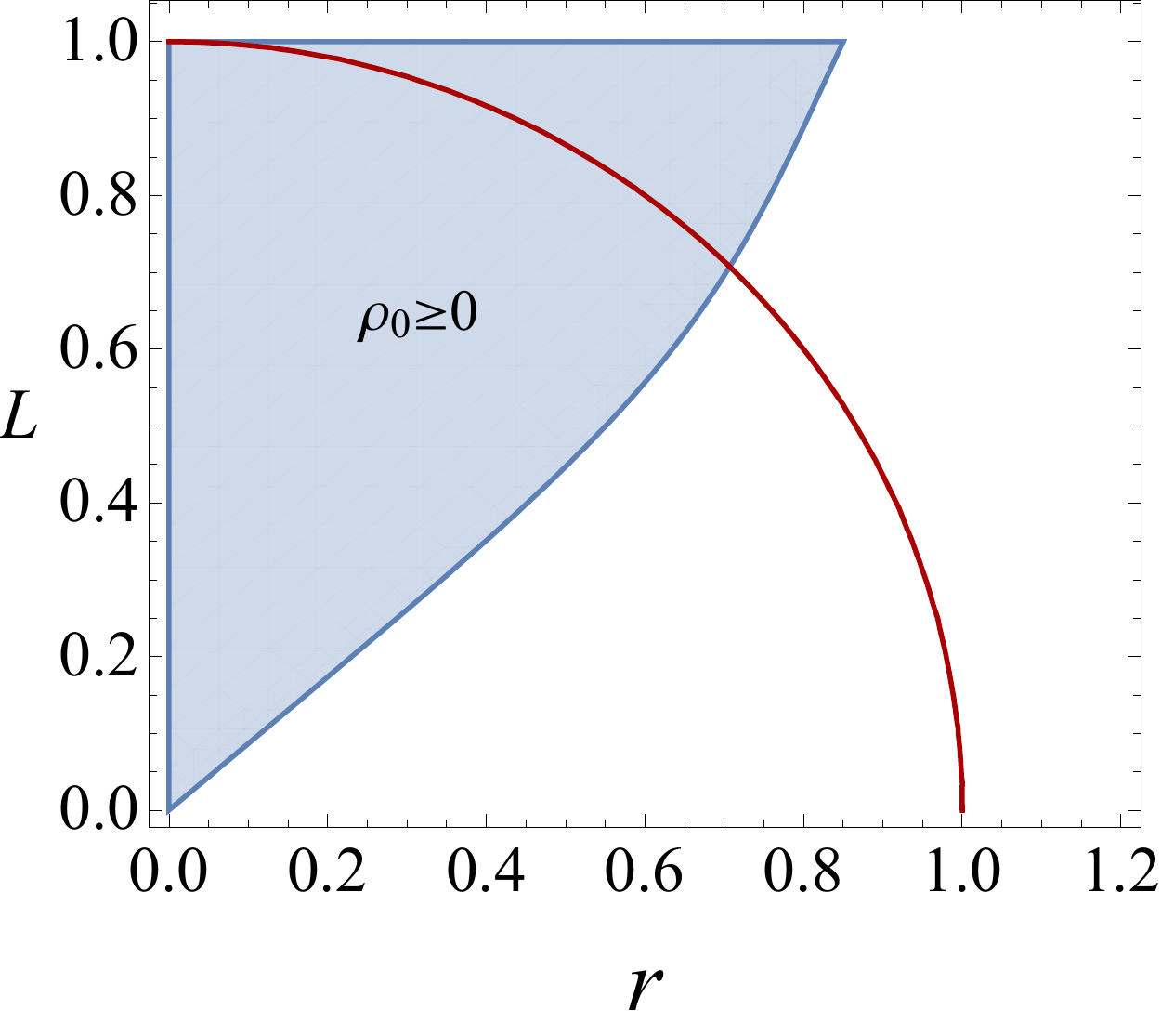}
		\caption{$\rho_0\ge0$}
		\label{fig:ec-s0-w1}
	\end{subfigure}
	\begin{subfigure}[b]{0.3\textwidth}
		\centering
		\includegraphics[width=\textwidth]{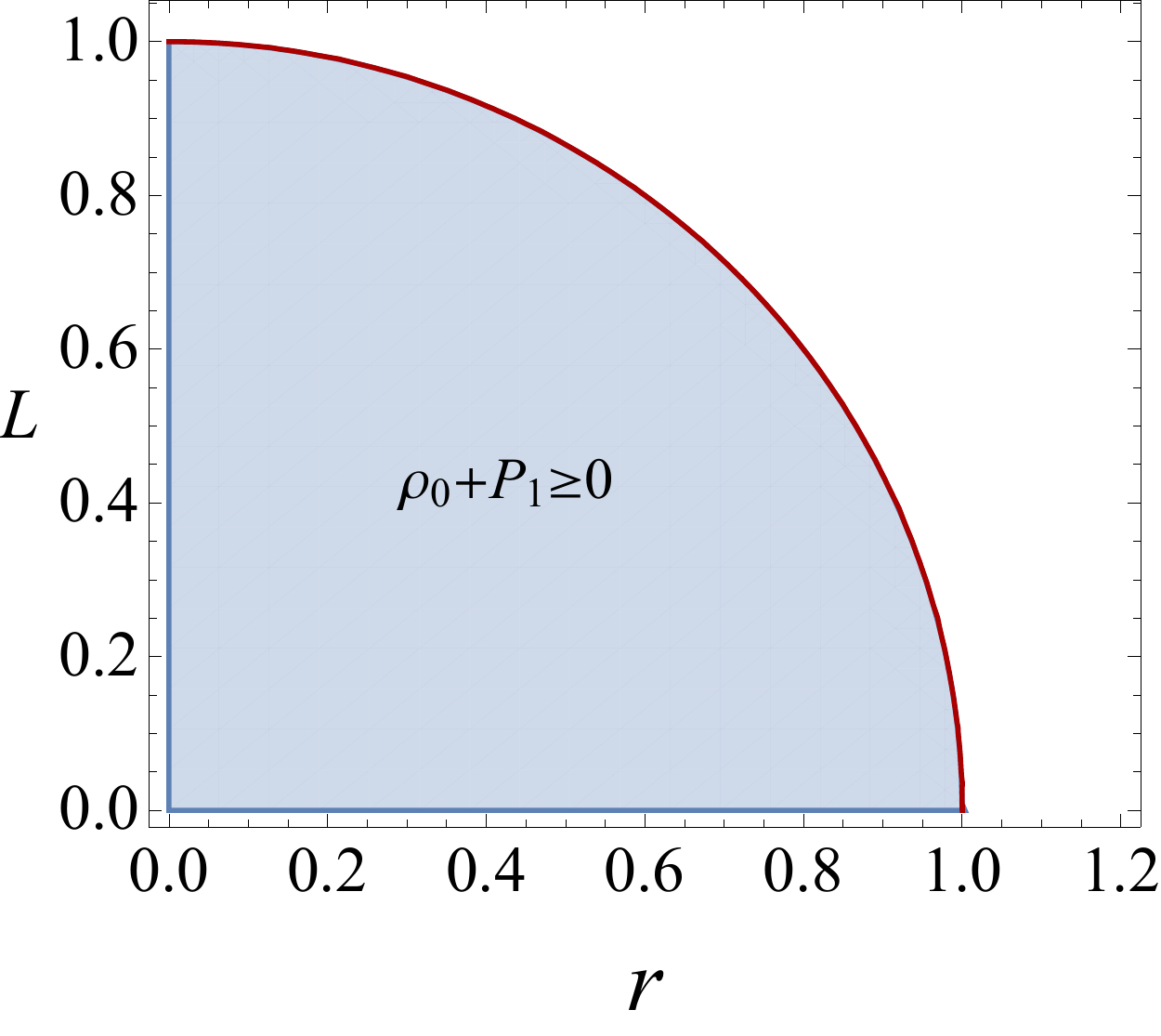}
		\caption{$\rho_0+P_1\ge0$}
		\label{fig:ec-s0-w2}
	\end{subfigure}
	\begin{subfigure}[b]{0.3\textwidth}
		\centering
		\includegraphics[width=\textwidth]{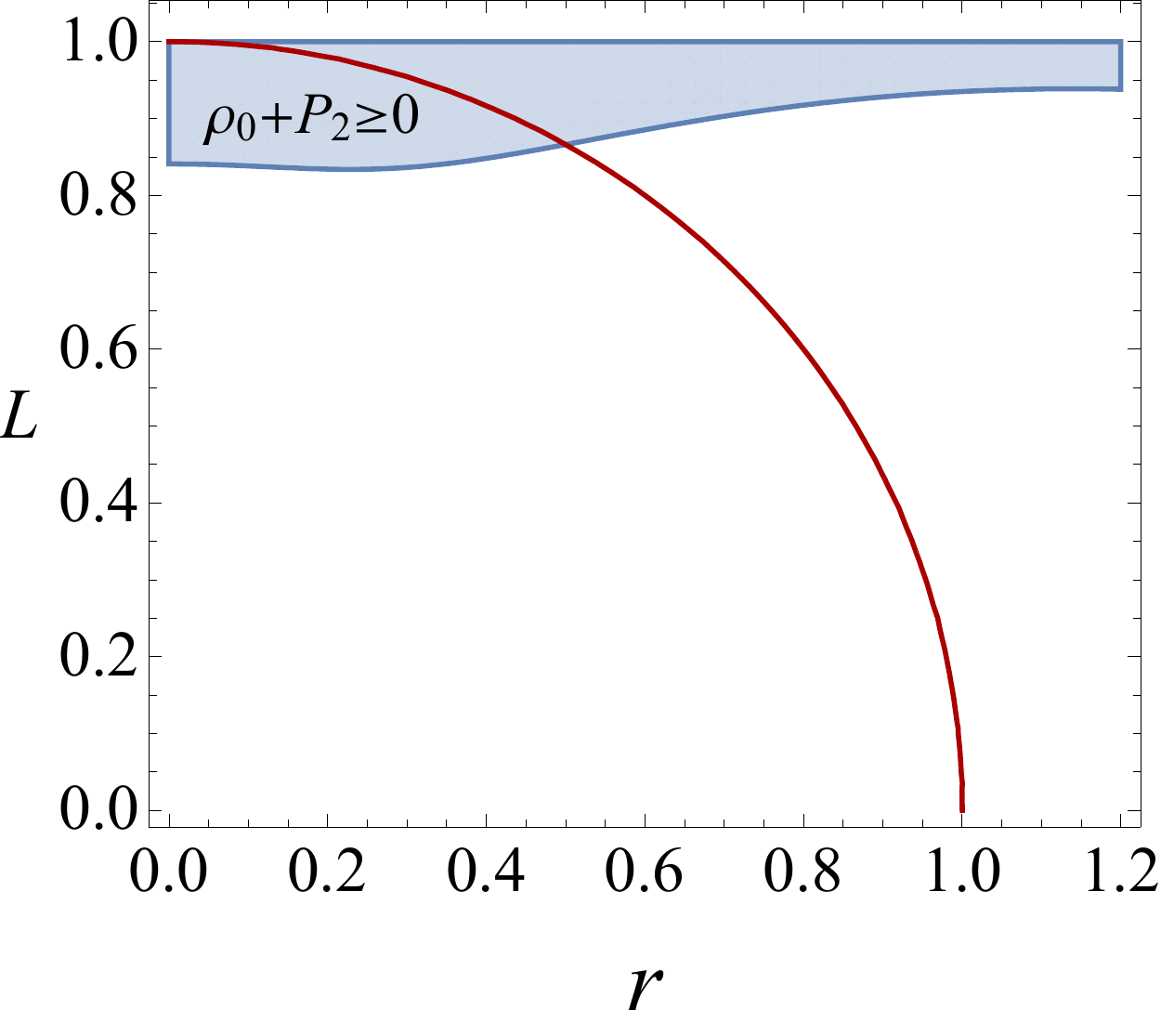}
		\caption{$\rho_0+P_2\ge0$}
		\label{fig:ec-s0-w3}
	\end{subfigure}
\begin{subfigure}[b]{0.68\textwidth}
	\centering
	\includegraphics[width=\textwidth]{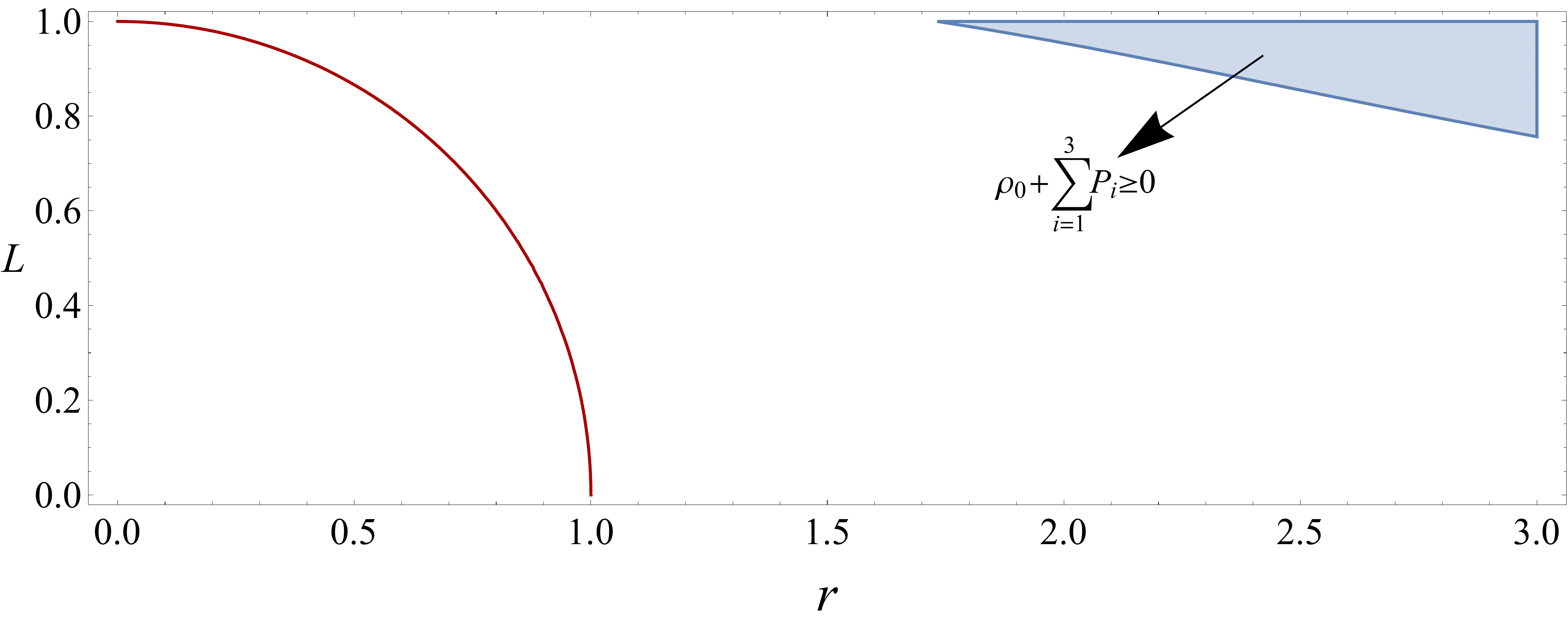}
	\caption{$\rho_0+\sum_{i=1}^3 P_i\ge0$}
	\label{fig:ec-s0-s1-Inc}
\end{subfigure}

\vskip 3mm	
\begin{subfigure}[b]{0.3\textwidth}
		\centering
		\includegraphics[width=\textwidth]{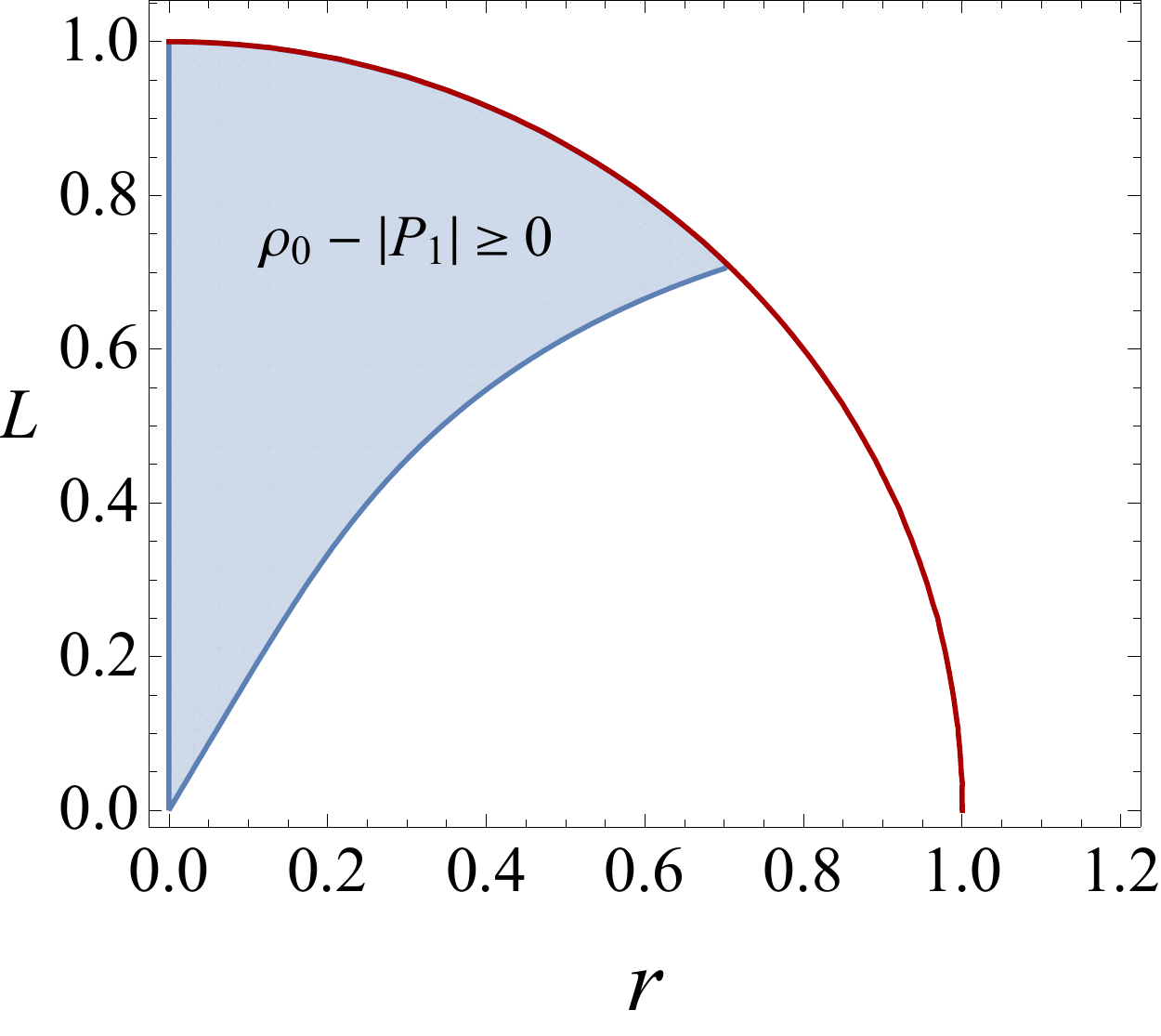}
		\caption{$\rho_0-|{P_1}|\ge0$}
		\label{fig:ec-s0-d1}
	\end{subfigure}
	\begin{subfigure}[b]{0.3\textwidth}
		\centering
		\includegraphics[width=\textwidth]{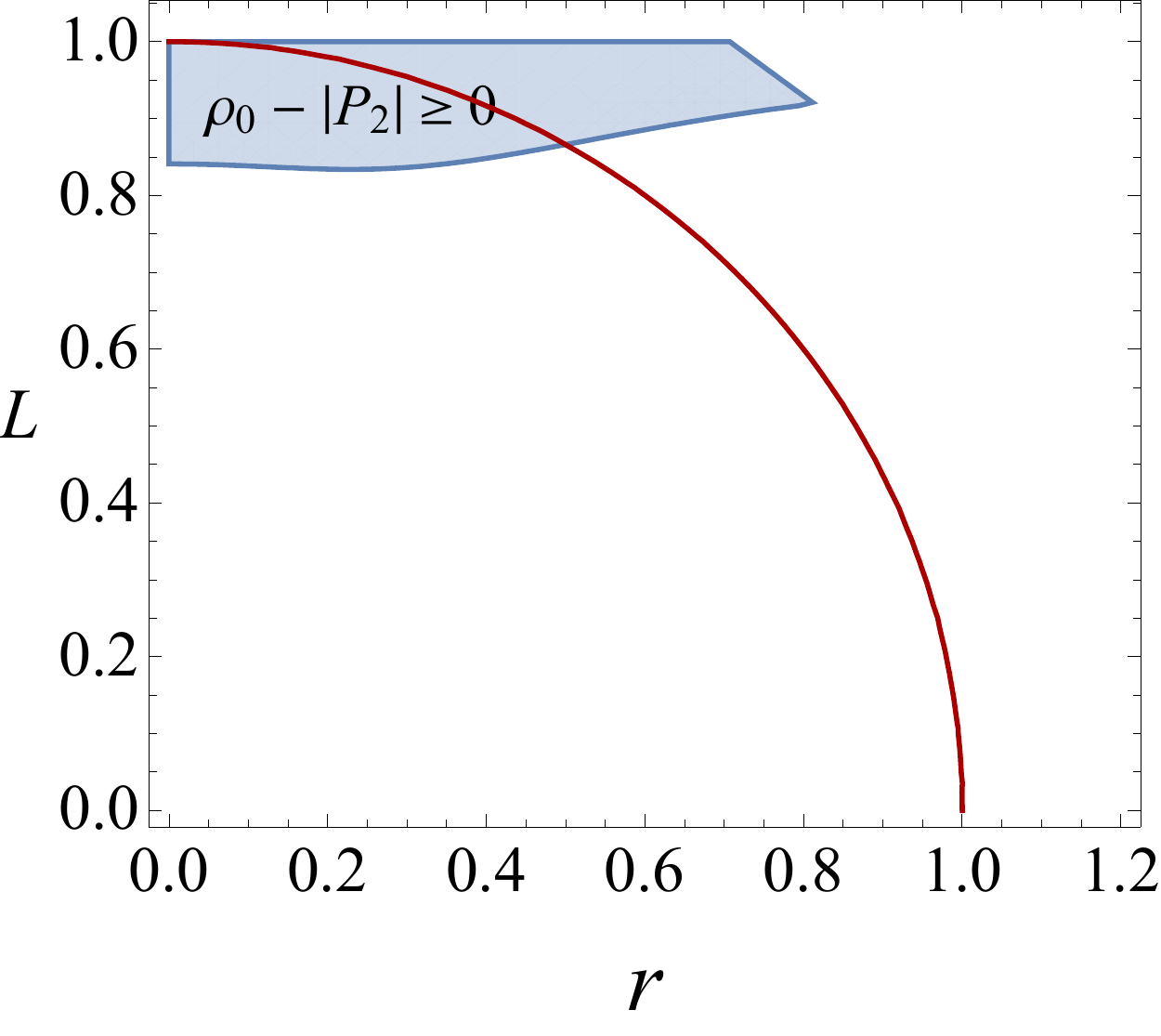}
		\caption{$\rho_0-|P_2|\ge0$}
		\label{fig:ec-s0-d2}
	\end{subfigure}
	\captionsetup{width=.9\linewidth}
	\caption{The blue shadows show the physical regions in which the corresponding inequalities are satisfied
	for the case of $N=1/2$, where the red curves are horizons. Note that the existence of horizons gives the
	constraint, $L^2<1$, in this case.}
	\label{fig:ec-s0}
\end{figure}

For the case of $N=1/2$, we compute the six independent quantities,
\begin{equation}
	\rho_0 = \frac{L^8 - 3 (L^4 + 1) r^4 - 2 L^2 r^6 + 4 L^2 r^2}{8\pi(L^2 +r^2)^5},
\end{equation}
\begin{equation}
	\rho_0 + P_1 = - \frac{6 L^2 r^2  \left[(L^2 + r^2)^2 - 1\right]}{ 8\pi(L^2 +r^2)^5},
\end{equation}
\begin{equation}
	\rho_0 + P_2 = \frac{2 L^8 + 6 L^6 r^2 + L^4  (6 r^4 - 1) + 2 L^2 r^2  (r^4- 1) - 9 r^4}{ 8\pi(L^2 + r^2)^5},
\end{equation}
\begin{equation}
	\rho_0 + \sum_{i = 1}^3 P_i = \frac{2L^8 + 6 L^6 r^2 + 2L^4  (3 r^4 - 1) + 2L^2r^2  (r^4 - 3) - 12 r^4}{8\pi (L^2 + r^2)^5},
\end{equation}
\begin{eqnarray}
	\rho_0 - |P_1 | &=& \frac{L^8 - 3 (L^4 + 1) r^4 - 2 L^2 r^6 + 4 L^2 r^2}{8\pi(L^2 + r^2)^5} \nonumber \\
& &- \left| \frac{L^8 + 3 (3 L^4 - 1) r^4 + 4 L^2 r^6 + 2 (3L^4 - 1) L^2 r^2}{ 8\pi(L^2 + r^2)^5} \right|,
\end{eqnarray}
\begin{eqnarray}
	\rho_0 - |P_2 | &= & \frac{L^8 - 3 (L^4 + 1) r^4 - 2 L^2 r^6 + 4 L^2 r^2}{ 8\pi(L^2 + r^2)^5} \nonumber \\
& &- \left| \frac{L^8 + 6 L^6 r^2 + L^4 (9 r^4 - 1) + L^2(4 r^6 - 6 r^2) - 6 r^4}{8\pi (L^2 + r^2)^5} \right|.
\end{eqnarray}
The energy conditions require that these quantities should be nonnegative. We plot the allowed  regions on the $r-L$
plane in Fig.~\ref{fig:ec-s0}.

\begin{figure}[!ht]
	\centering
	\begin{subfigure}[b]{0.33\textwidth}
		\centering
		\includegraphics[width=\textwidth]{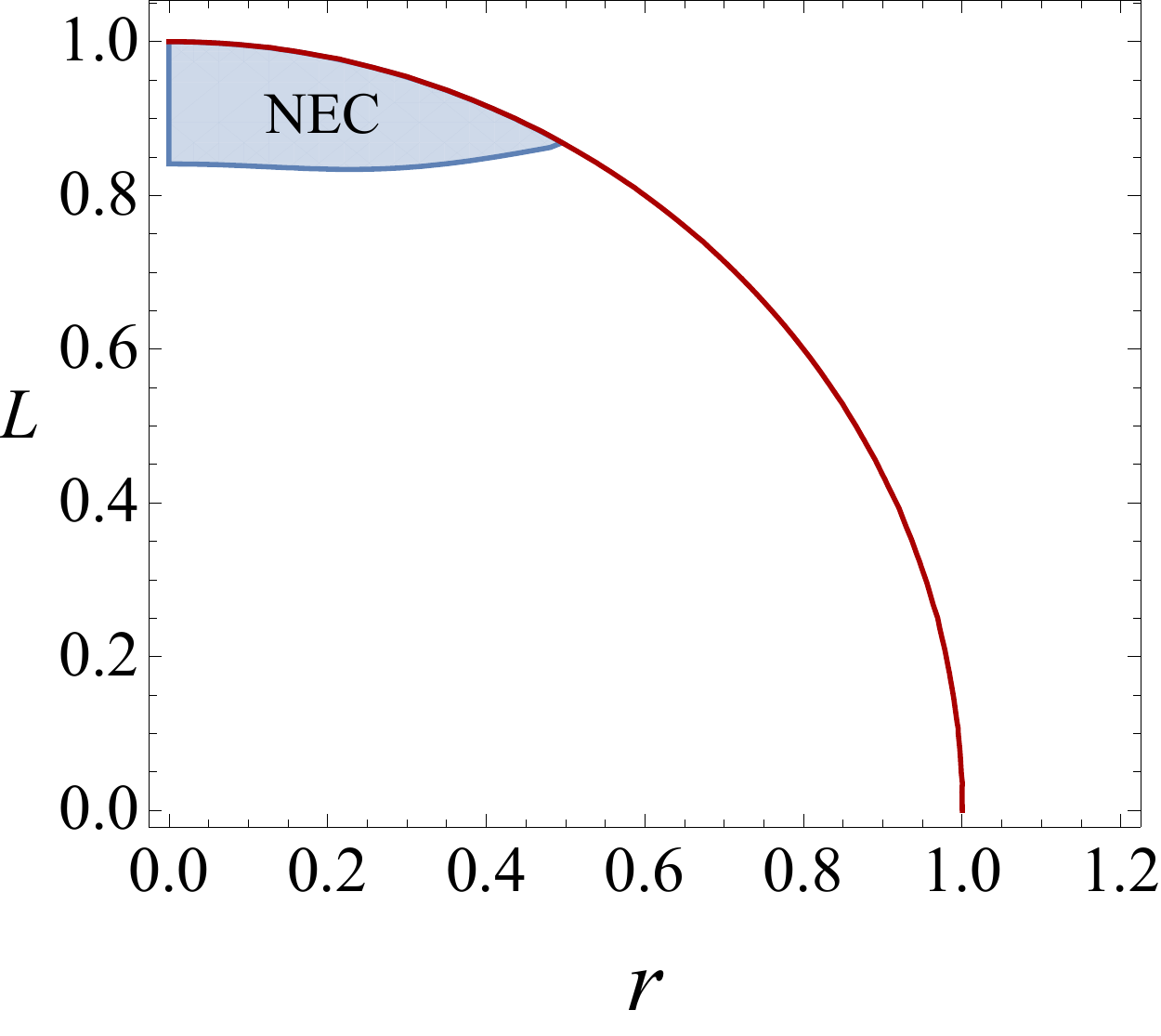}
		\caption{Null energy condition}
		\label{fig:null}
	\end{subfigure}
	\begin{subfigure}[b]{0.33\textwidth}
		\centering
		\includegraphics[width=\textwidth]{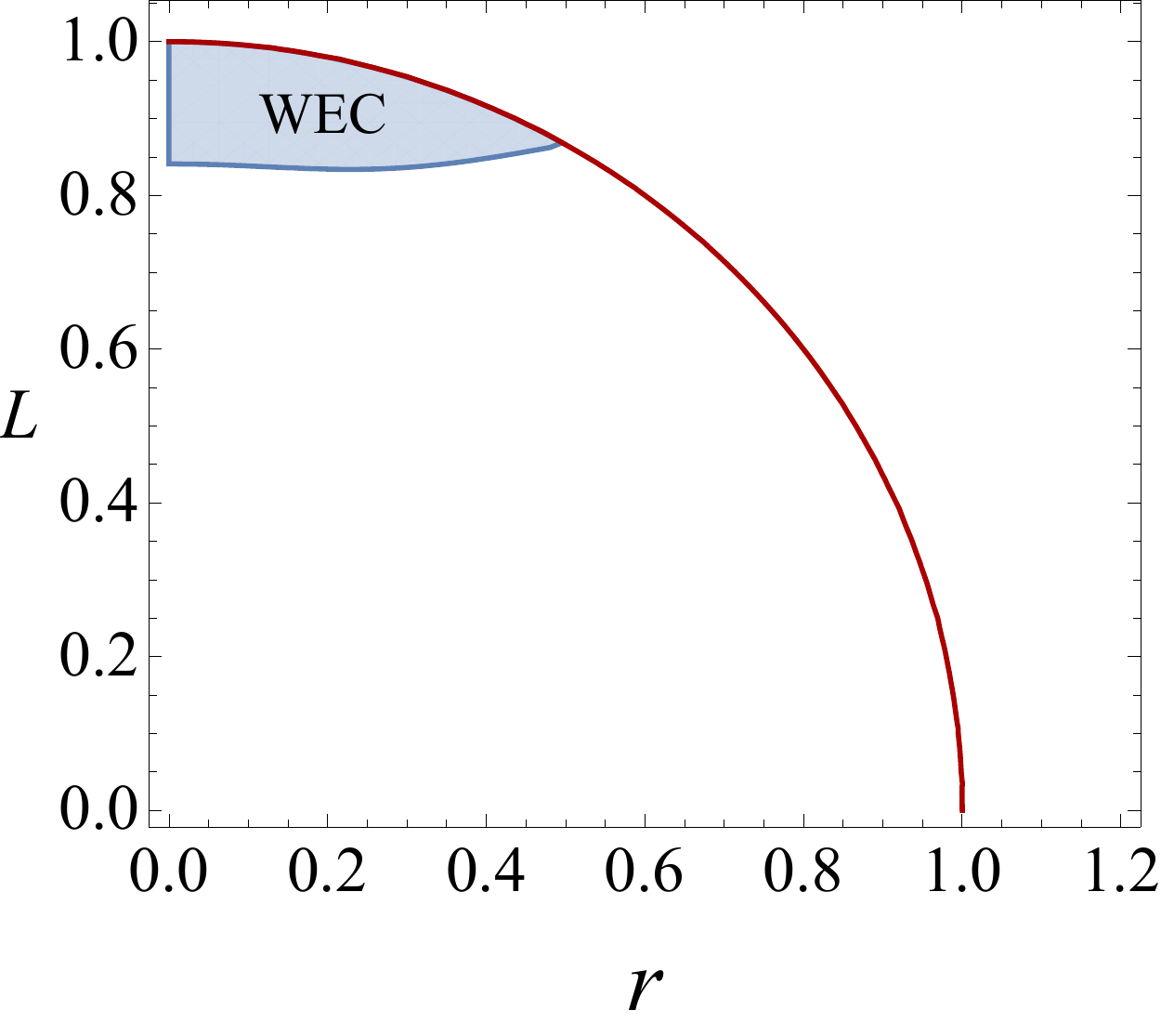}
		\caption{Weak energy condition}
		\label{fig:weak-null}
	\end{subfigure}
	\begin{subfigure}[b]{0.33\textwidth}
		\centering
		\includegraphics[width=\textwidth]{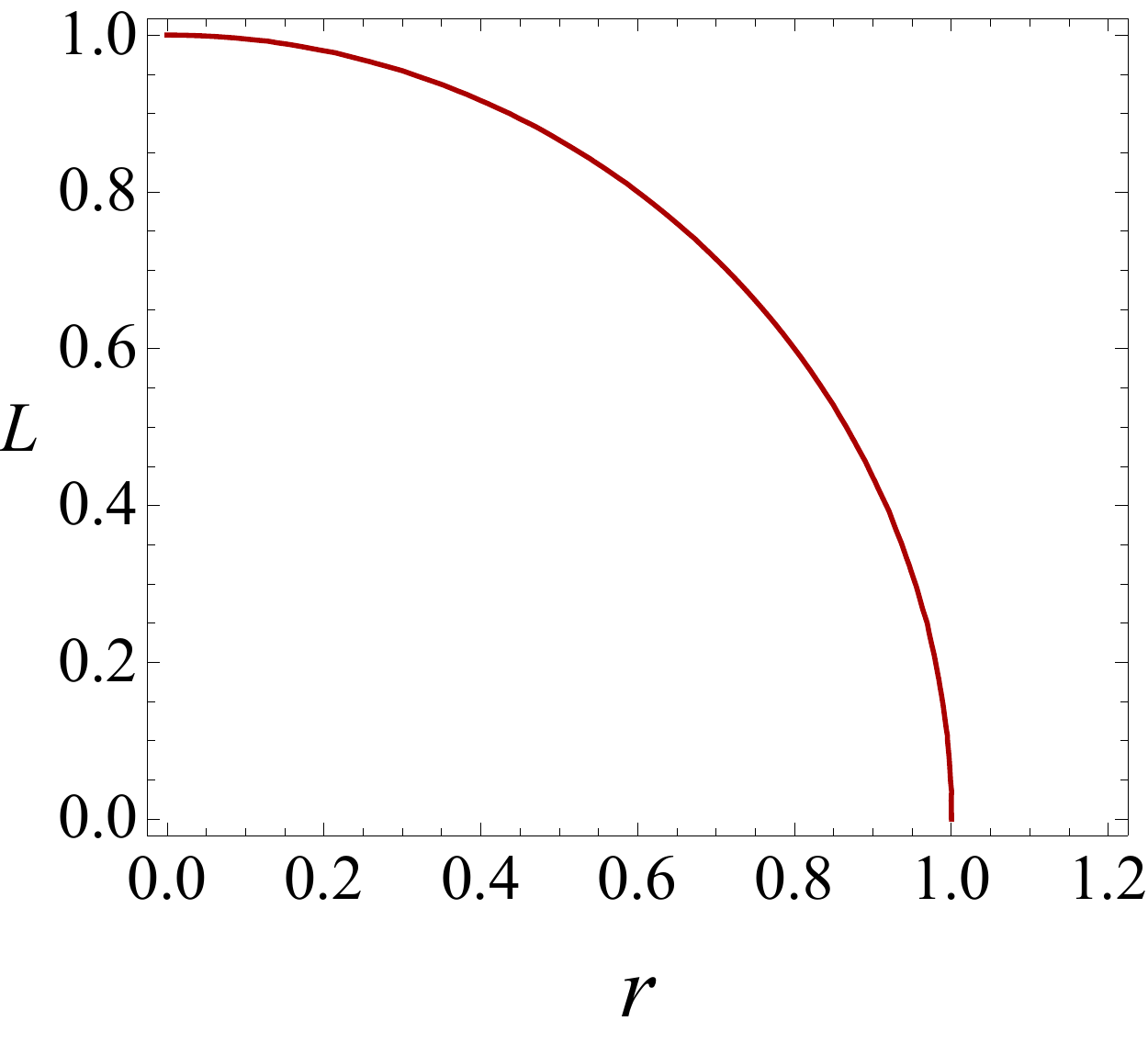}
		\caption{Strong energy condition}
		\label{fig:strong}
	\end{subfigure}
	\begin{subfigure}[b]{0.33\textwidth}
		\centering
		\includegraphics[width=\textwidth]{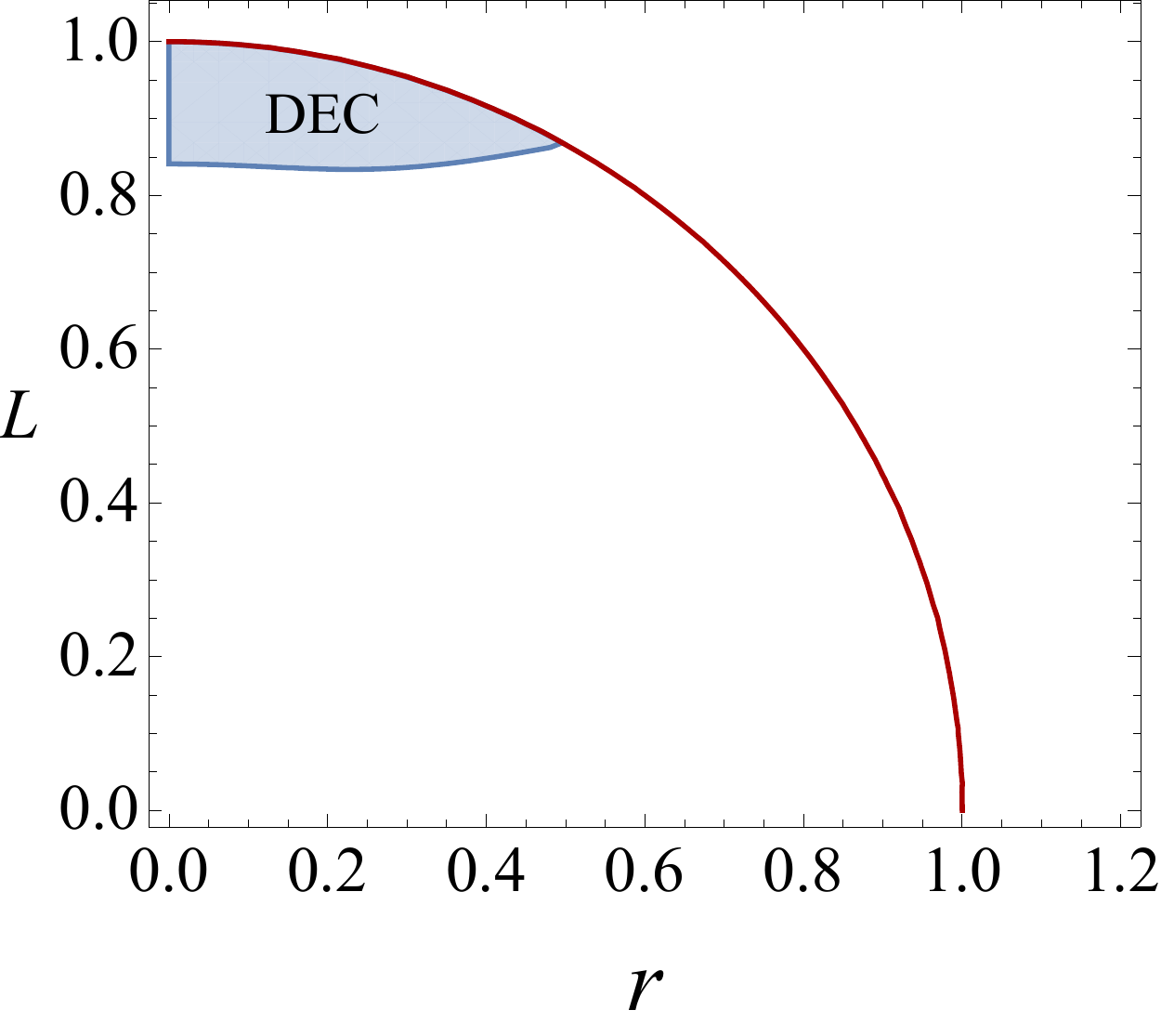}
		\caption{Dominant energy condition}
		\label{fig:dominant}
	\end{subfigure}
	\captionsetup{width=.9\linewidth}
	\caption{The blue shadows show the valid domains of  NEC, WEC, and DEC for the case of $N=1/2$, where no
		valid domains exist for the SEC. The red curves are horizons in this case.}
	\label{fig:energy-condition-full}
\end{figure}

Combining the six subfigures in Fig.~\ref{fig:ec-s0} with the four energy conditions in
Eq.~(\ref{encon}), we can determine the domains that the energy conditions are satisfied for the case of
$N=1/2$, which is plotted in Fig.~\ref{fig:energy-condition-full}.
We can see from Fig.~\ref{fig:energy-condition-full} that the SEC is completely violated in the entire parameter
range and spacetime,  $L^2<1$ and $ r\in [0,\infty)$. This is actually what we expected because 
the spacetime with $f=1-1/r^4$ is asymptotic to the metric of our ARBH model, see 
Eqs.~(\ref{eq14}) and (\ref{eq:metric s=1/2}) in the limit of $r \gg L$, and such a spacetime is of repulsive
interaction which breaks the SEC, see App.~\ref{appendix:B} for a detailed explanation. However, the situation of our ARBH model is more complicated than usual. We see in Fig.~\ref{fig:ec-s0-s1-Inc} that $\rho_0 + \sum_{i = 1}^3 P_i\geqslant 0$ is
satisfied in one region outside the horizon, i.e., the ARBH produces an attractive interaction outside the horizon
although the SEC is violated based on Ref.~\cite{Carroll:2004st}. The reason that makes the SEC invalid is that
$\rho_0+{P_1}\ge 0$ is violated outside the horizon, which is different from the situation in the usual BH models with $f=1-1/r^4$. In
addition, the NEC, WEC, and DEC are satisfied in  a piece of domains inside the horizon (also including the
horizon as boundary) for the parameter range $0.8<L\le 1.0$. 

For the case of $N=1$, we compute the six independent quantities,
\begin{equation}
	\rho_0 = - \frac{r^4 \left[4 L^{10} + (16 L^4 + 3) r^6 + (16 L^4 + 3) L^4 r^2 + 4
		L^2 r^8 + 2 (12 L^4 - 7) L^2 r^4\right]}{ 8\pi(L^2 + r^2)^8},
\end{equation}
\begin{equation}
	\rho_0 + P_1 = \frac{4L^2 r^2  (L^2 + r^2 - r)  (L^2 - 3 r^2)  (L^2 + r^2 +
		r)  [(L^2 + r^2)^2 + r^2]}{8\pi(L^2 + r^2)^8},
\end{equation}
\begin{equation}
	\rho_0 + P_2 = \frac{r^2  \left[4 L^{12} + 20 L^{10} r^2 + (20 L^4 - 9) r^8 + (40
		L^4 - 1) L^4 r^4 + 4 L^2 r^{10} + 2 (20 L^4 + 7) L^2 r^6\right]}{ 8\pi(L^2 +r^2)^8},
\end{equation}
\begin{equation}
	\rho_0 + \sum_{i = 1}^3 P_i = \frac{4r^2  \left[3 L^{12} + 13 L^{10} r^2 + 22 L^8
		r^4 + 18 L^6 r^6 + 7 L^4 r^8 + L^2 r^6  (r^4 + 3) - 3 r^8\right]}{8\pi (L^2 +r^2)^8},
\end{equation}
\begin{equation}
\begin{split}
	\rho_0 - | P_1 | = &- \frac{r^4 \left[4 L^{10} + (16 L^4 + 3) r^6 + (16 L^4 + 3)
	L^4 r^2 + 4 L^2 r^8 + 2 (12 L^4 - 7) L^2 r^4\right]}{8\pi (L^2 + r^2)^8} \\
 & - \left|\frac{r^2 \left[- 4 L^{12} - 8 L^{10} r^2 + 8 L^8 r^4 + 32 L^6 r^6 + L^4  (28
	r^8 + r^4) + 2 L^2  (4 r^{10} + r^6) - 3 r^8\right]}{8\pi(L^2 + r^2)^8} \right|,
\end{split}
\end{equation}

\begin{equation}
    \begin{split}
        \rho_0 - | P_2 | =& - \frac{r^4 \left[4 L^{10} + (16 L^4 + 3) r^6 + (16 L^4 + 3)
	L^4 r^2 + 4 L^2 r^8 + 2 (12 L^4 - 7) L^2 r^4\right]}{ 8\pi(L^2 + r^2)^8} \\
& - \left|\frac{2r^2 \left[2 L^{12} + 12 L^{10} r^2 + 32 L^6 r^6 + 3 (6 L^4 - 1) r^8 + (28
	L^4 + 1) L^4 r^4 + 4 L^2 r^{10}\right]}{ 8\pi(L^2 + r^2)^8} \right|.
    \end{split}
\end{equation}

\begin{figure}[!ht]
	\centering
	\begin{subfigure}[b]{0.3\textwidth}
		\centering
		\includegraphics[width=\textwidth]{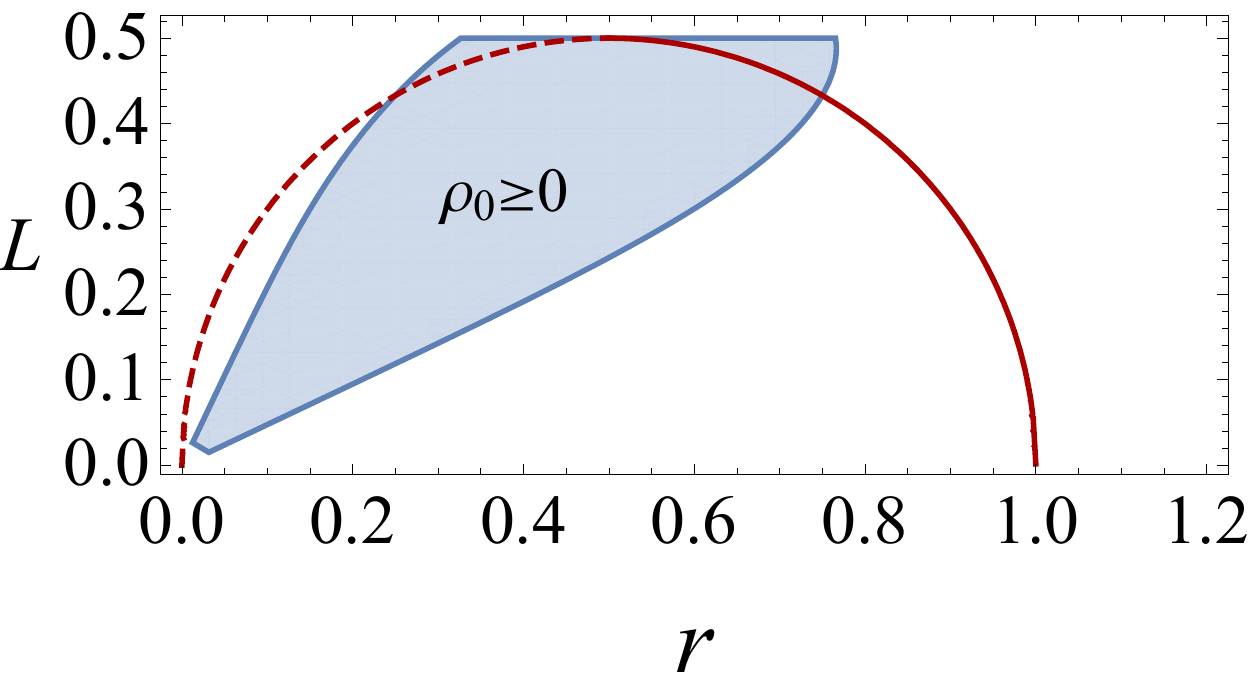}
		\caption{$\rho_0\ge0$}
		\label{fig:rho0,S=1}
	\end{subfigure}
	\begin{subfigure}[b]{0.3\textwidth}
		\centering
		\includegraphics[width=\textwidth]{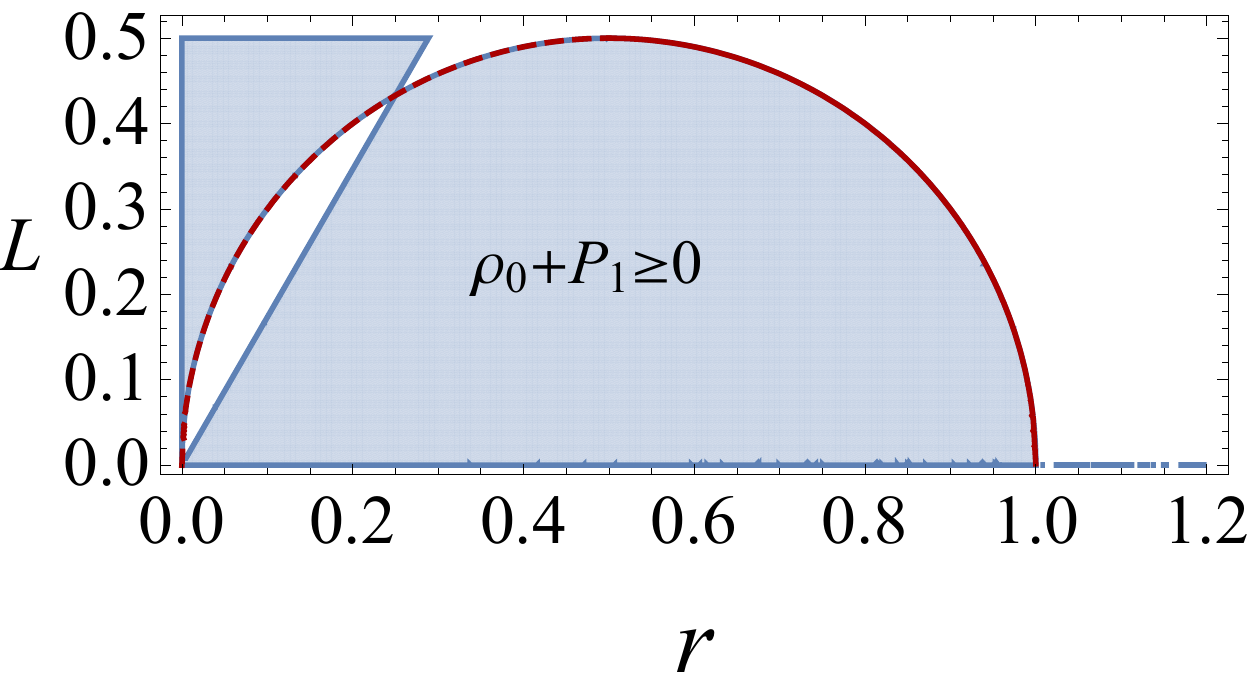}
		\caption{$\rho_0+P_1\ge0$}
		\label{fig:rhoP10,S=1}
	\end{subfigure}
	\begin{subfigure}[b]{0.3\textwidth}
		\centering
		\includegraphics[width=\textwidth]{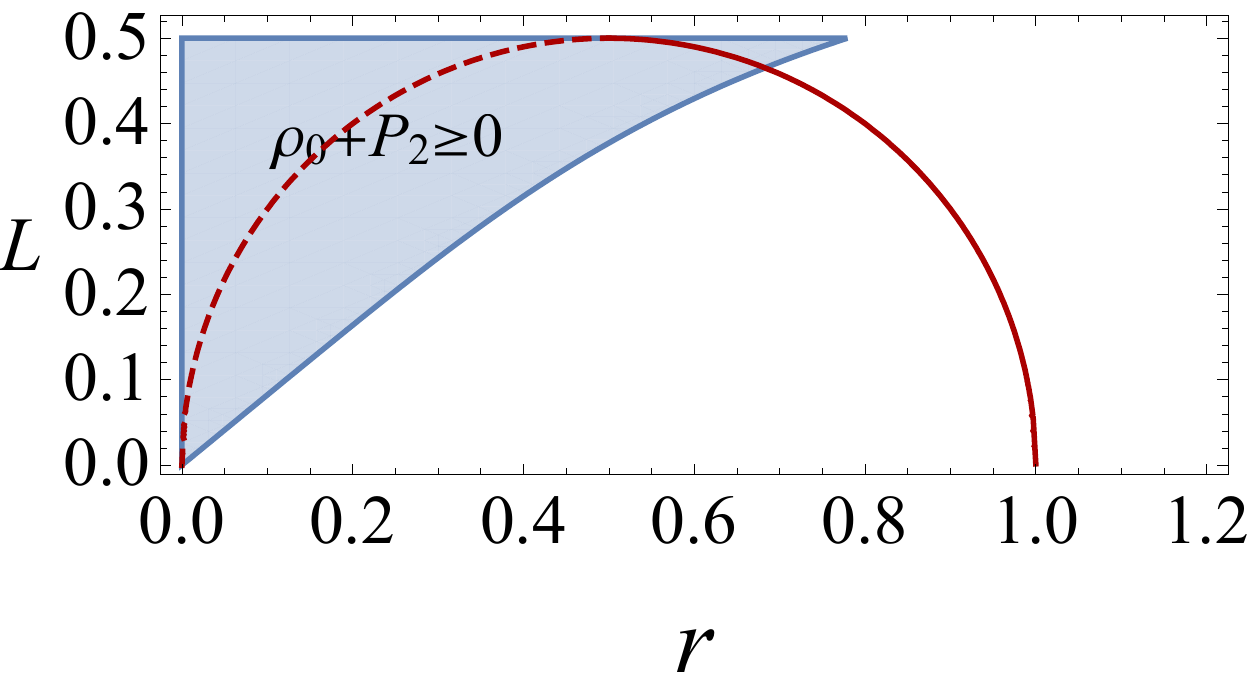}
		\caption{$\rho_0+P_2\ge0$}
		\label{fig:rhoP20,S=1}
	\end{subfigure}
\begin{subfigure}[b]{0.37\textwidth}
	\centering
	\includegraphics[width=\textwidth]{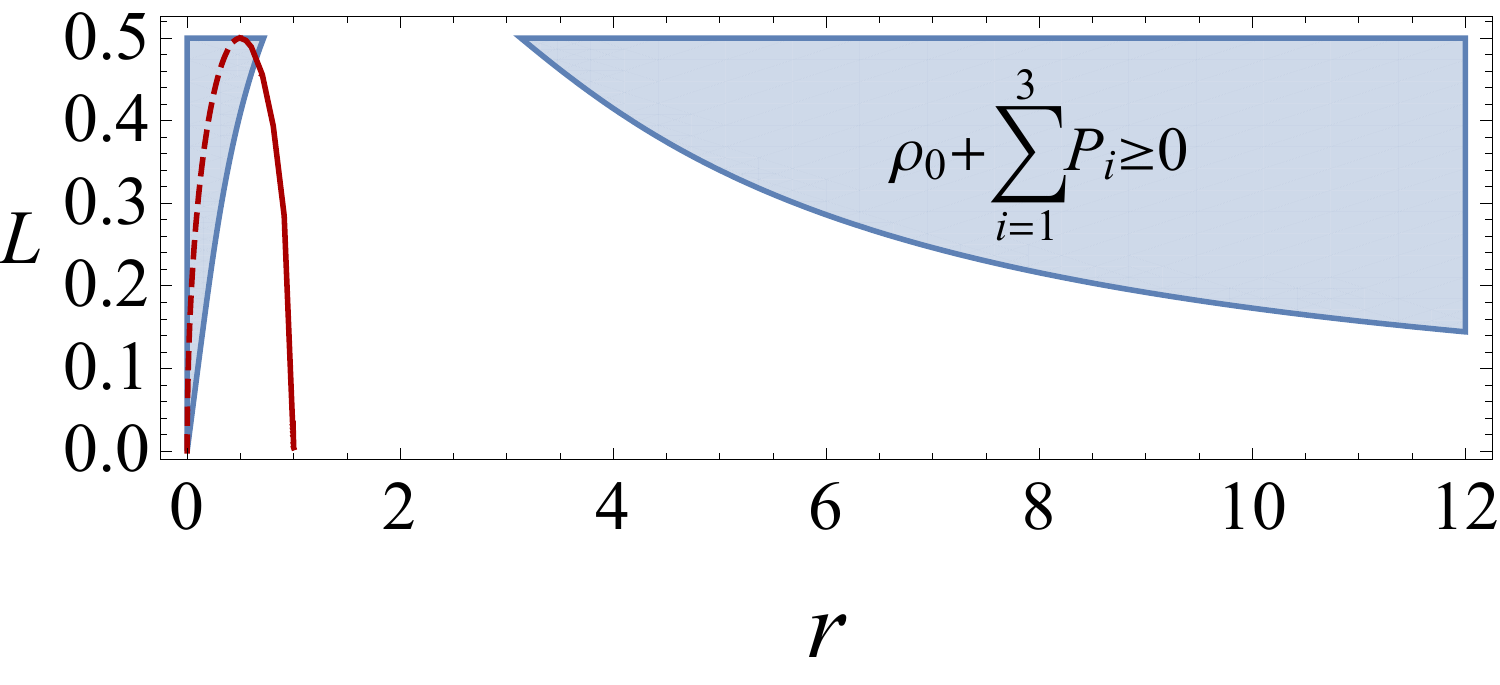}
	\caption*{(d1) $\rho_0+\sum_{i=1}^3 P_i\ge0$}
	\label{fig:rhoP1P2P20,S=1-Inc}
\end{subfigure}
	\begin{subfigure}[b]{0.3\textwidth}
		\centering
		\includegraphics[width=\textwidth]{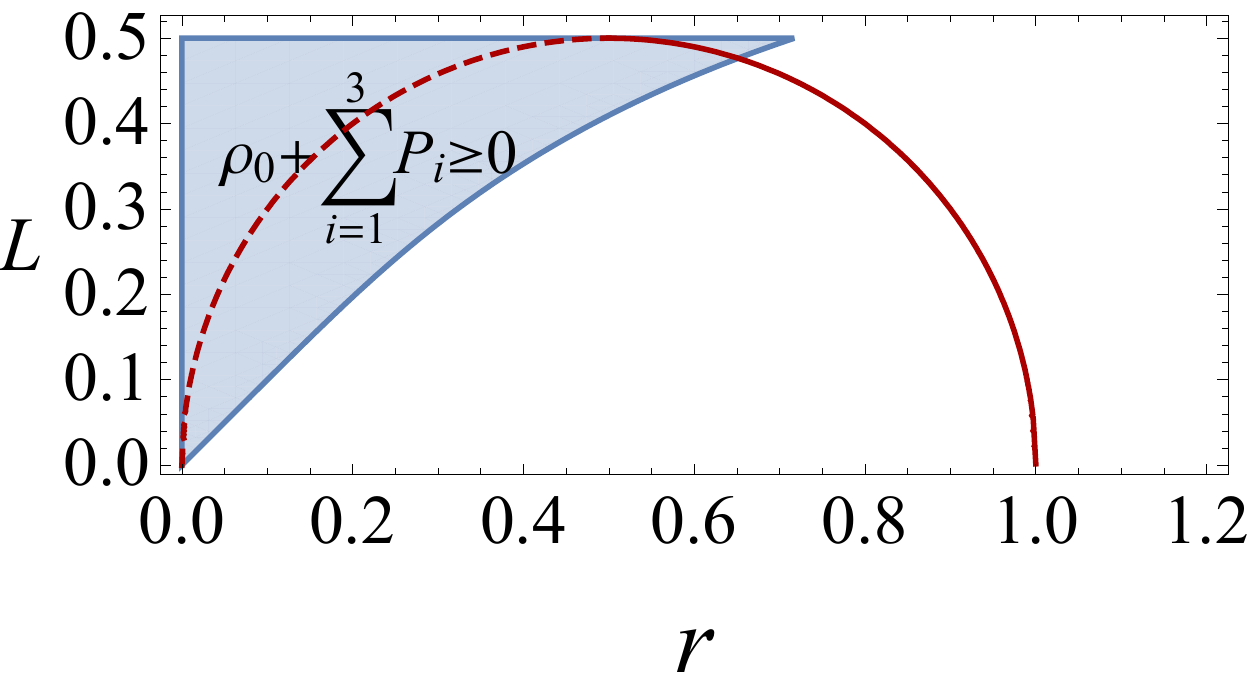}
		\caption*{(d2) $\rho_0+\sum_{i=1}^3 P_i\ge0$}
		\label{fig:rhoP1P2P20,S=1}
	\end{subfigure}
	
\begin{subfigure}[b]{0.3\textwidth}
		\centering
		\includegraphics[width=\textwidth]{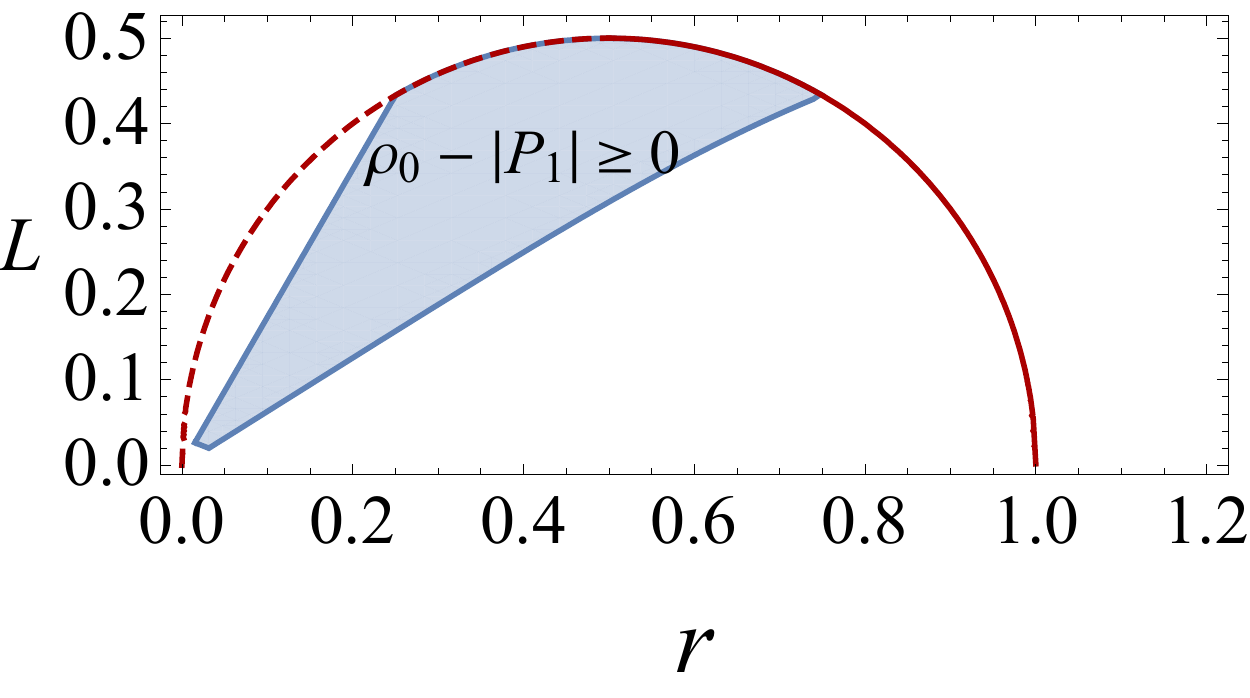}
		\caption*{(e) $\rho_0-|{P_1}|\ge0$}
		\label{fig:rhoABSP10,S=1}
	\end{subfigure}
	\begin{subfigure}[b]{0.3\textwidth}
		\centering
		\includegraphics[width=\textwidth]{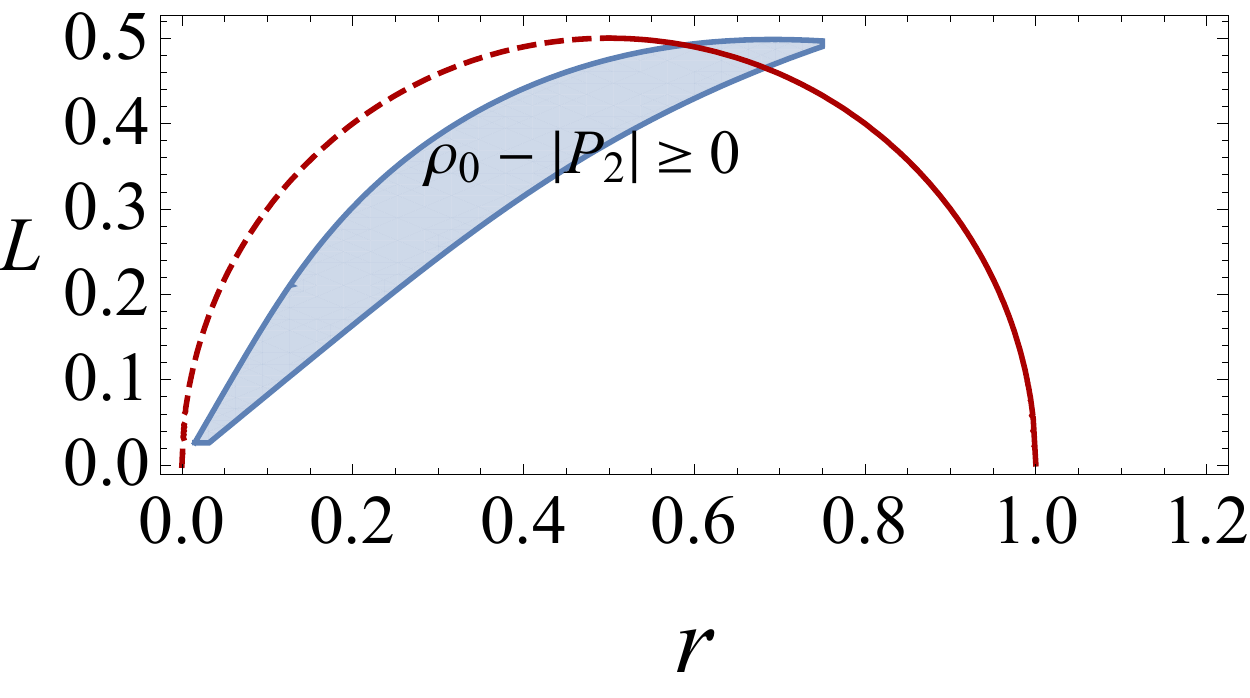}
		\caption*{(f) $\rho_0-|P_2|\ge0$}
		\label{fig:rhoABSP20,S=1}
	\end{subfigure}
	\captionsetup{width=.9\linewidth}
	\caption{The blue shadows show the physical regions in which the corresponding inequalities are satisfied
	for the case of $N=1$, where the dotted red curves are inner horizons while the solid ones outer horizons. Note
	that the existence of horizons gives the constraint, $L^2\le1/4$, in this case. Subfigure (d2) shows the detail features of the left shadow in subfigure (d1).}
	\label{fig:ec-s1}
\end{figure}

\begin{figure}[!ht]
	\centering
	\begin{subfigure}[b]{0.33\textwidth}
		\centering
		\includegraphics[width=\textwidth]{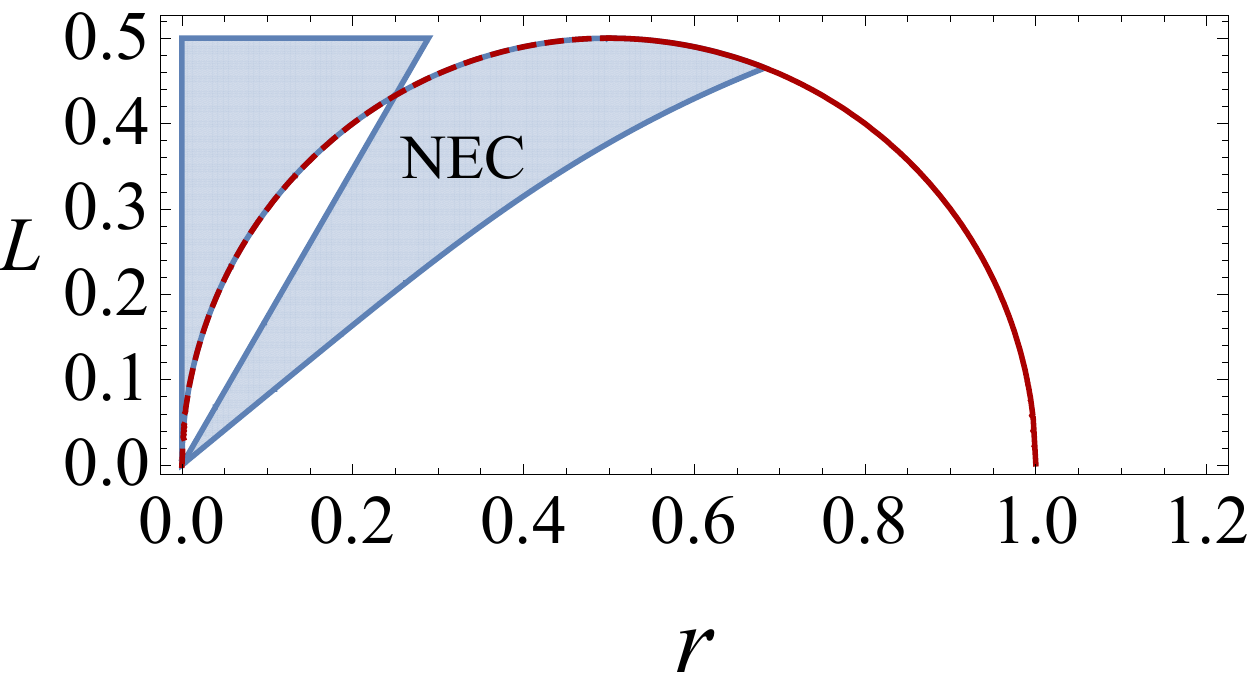}
		\caption{Null energy condition}
		\label{fig:nullS=1}
	\end{subfigure}
	\begin{subfigure}[b]{0.33\textwidth}
		\centering
		\includegraphics[width=\textwidth]{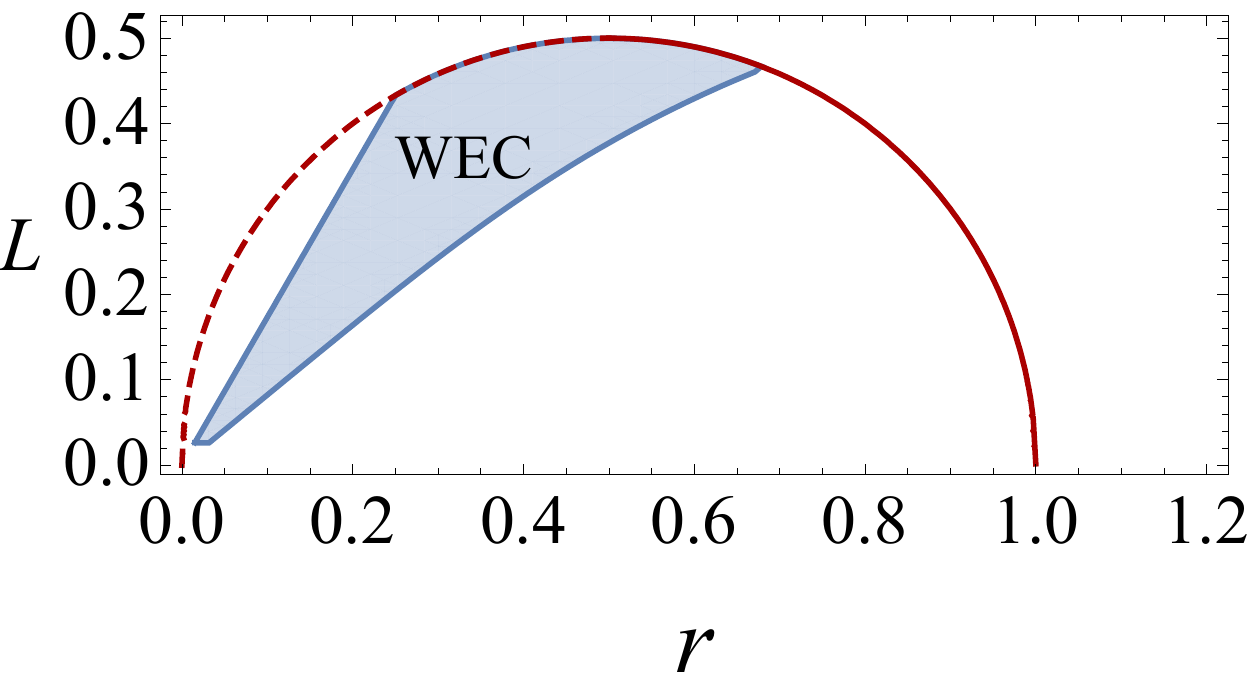}
		\caption{Weak energy condition}
		\label{fig:weak-nullS=1}
	\end{subfigure}
	\begin{subfigure}[b]{0.33\textwidth}
		\centering
		\includegraphics[width=\textwidth]{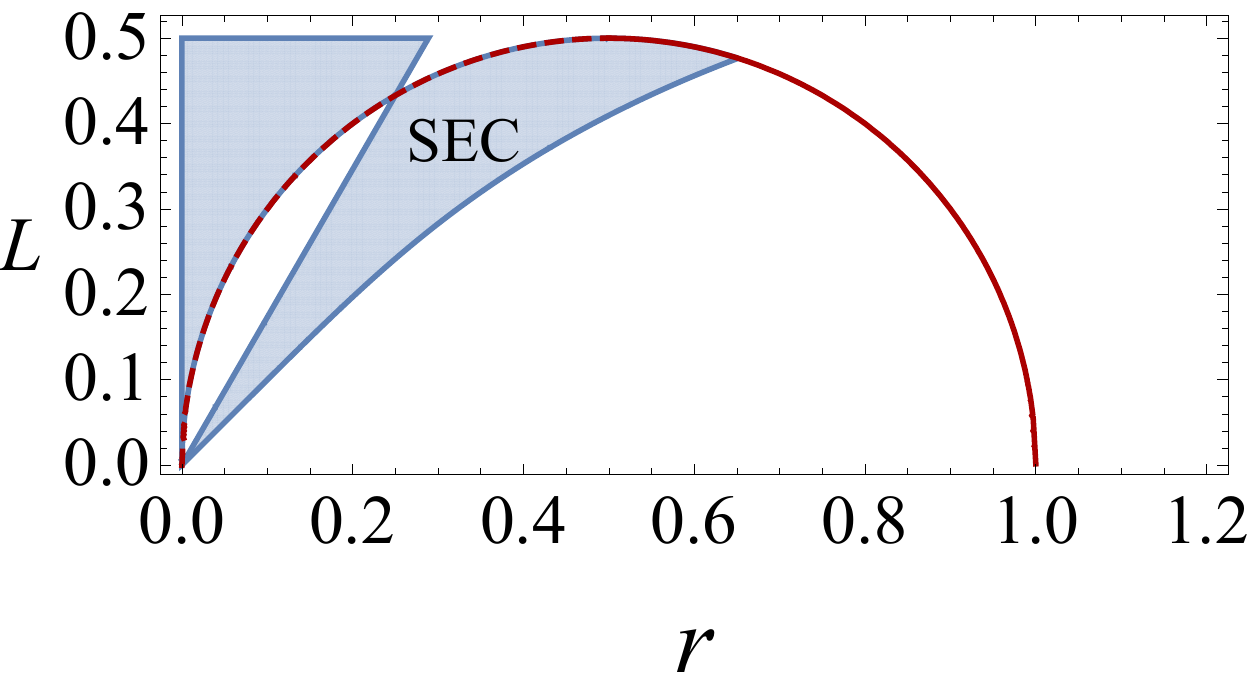}
		\caption{Strong energy condition}
		\label{fig:strongS=1}
	\end{subfigure}
	\begin{subfigure}[b]{0.33\textwidth}
		\centering
		\includegraphics[width=\textwidth]{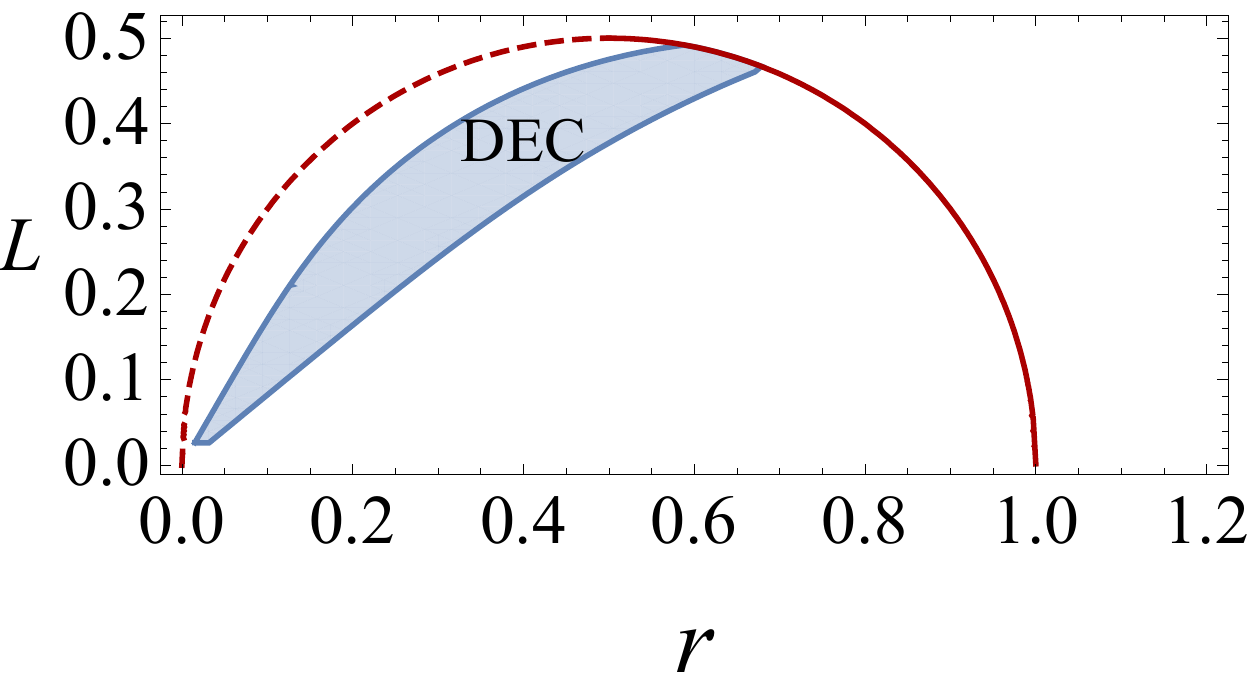}
		\caption{Dominant energy condition}
		\label{fig:dominantS=1}
	\end{subfigure}
	\captionsetup{width=.9\linewidth}
	\caption{The blue shadows show the valid domains of  various energy conditions for the case of $N=1$. The
		dotted red curves are inner horizons while the solid ones outer horizons in this case.}
	\label{fig:energy-condition-fullS=1}
\end{figure}

The energy conditions require that these quantities should be nonnegative. We plot the allowed  regions on the $r-L$
plane in Fig.~\ref{fig:ec-s1}. Similarly, the corresponding valid domains of energy conditions are shown in
Fig.~\ref{fig:energy-condition-fullS=1} for the case of $N=1$ when Fig.~\ref{fig:ec-s1} is combined with
Eq.~(\ref{encon}).
We can see from Fig.~\ref{fig:energy-condition-fullS=1} that the NEC and SEC are  satisfied in two pieces of
domains for the parameter range $0<L\le 1/2$, where one is located inside the inner horizon and the other between the inner and
outer horizons (also including the horizons as boundaries). It is worthy to emphasize that the
situation of SEC in our model is a counterexample of the work \cite{Zaslavskii:2010qz} in which the breaking domain of SEC for a regular black hole with metric $g_{\mu\nu}={\rm diag}\left\{-f,
f^{-1}, r^2, r^2 \sin^{2 } \theta\right\}$ must be located  inside horizon. The reason is that our ARBH model does not satisfy the simple relation, $-g_{tt}g_{rr}=1$.
Therefore, our situation of SEC becomes complicated. Moreover, the WEC and DEC are satisfied in only one piece of
domains between the inner and outer horizons (also including the two horizons as boundaries for WEC and only
the outer horizon as boundary for DEC)  for the parameter ranges $0<L\le 1/2$ (WEC) and $0<L< 1/2$ (DEC), respectively.

\begin{figure}[!ht]
	\centering
	\begin{subfigure}[b]{0.33\textwidth}
		\centering
		\includegraphics[width=\textwidth]{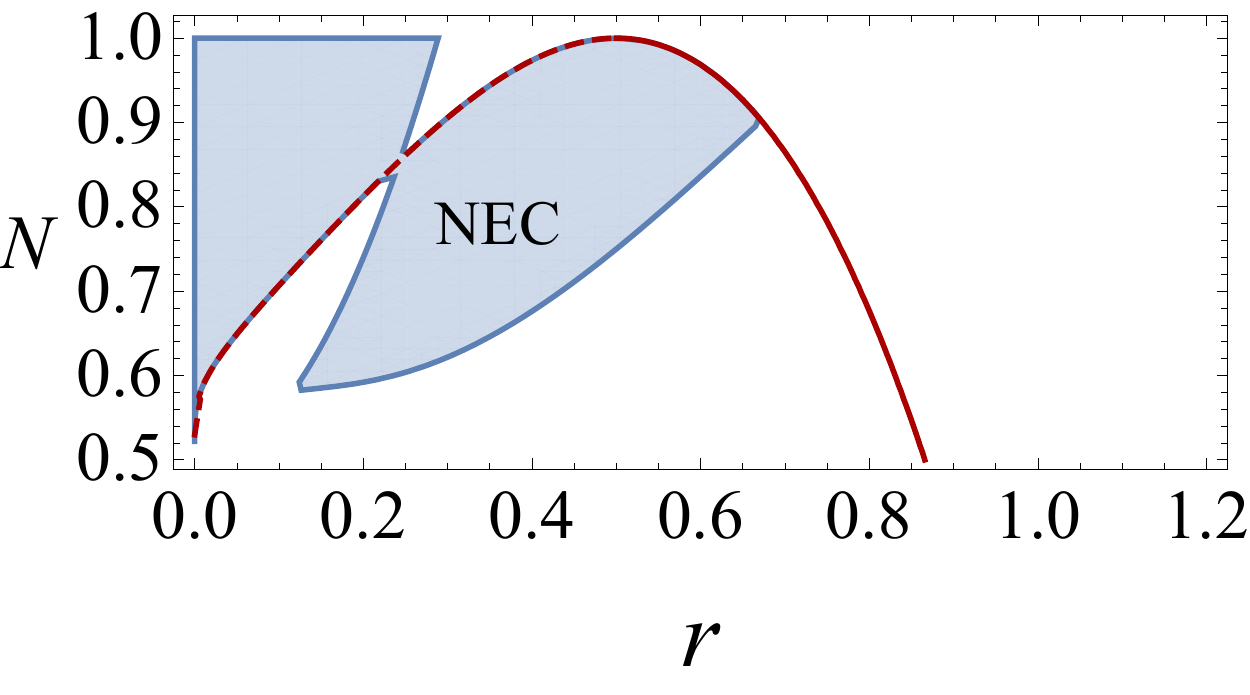}
		\caption{Null energy condition}
		\label{fig:nullS=N}
	\end{subfigure}
	\begin{subfigure}[b]{0.33\textwidth}
		\centering
		\includegraphics[width=\textwidth]{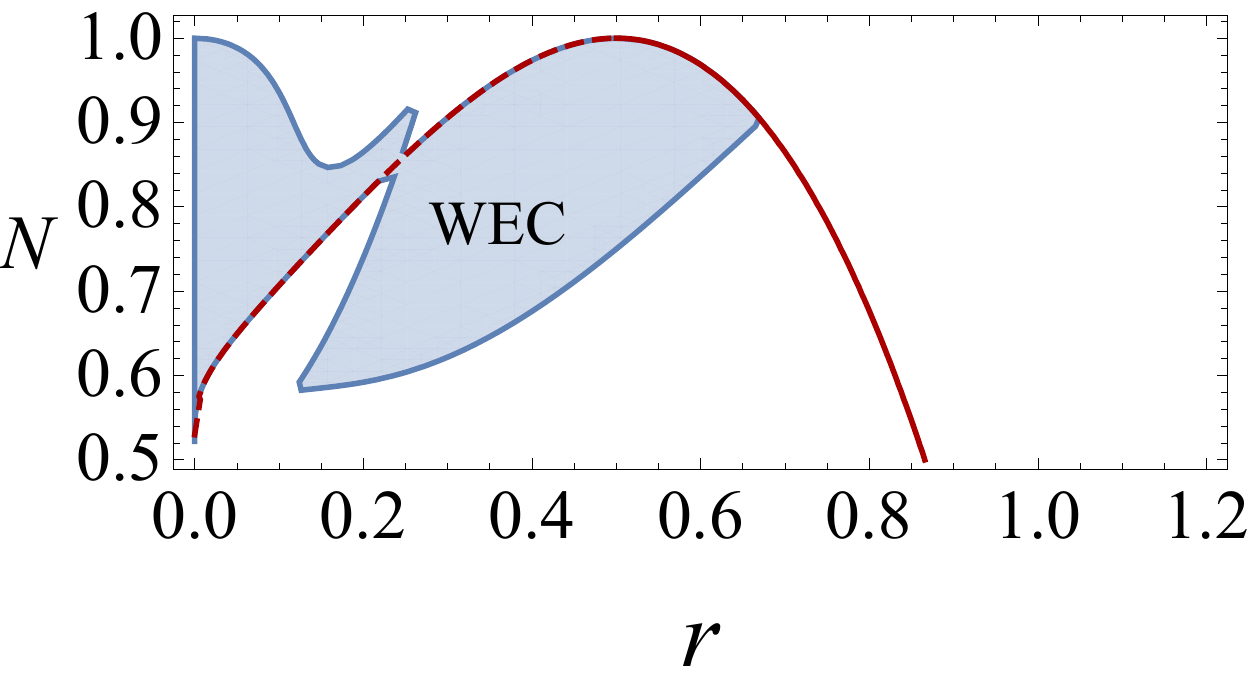}
		\caption{Weak energy condition}
		\label{fig:weak-nullS=N}
	\end{subfigure}
	\begin{subfigure}[b]{0.33\textwidth}
		\centering
		\includegraphics[width=\textwidth]{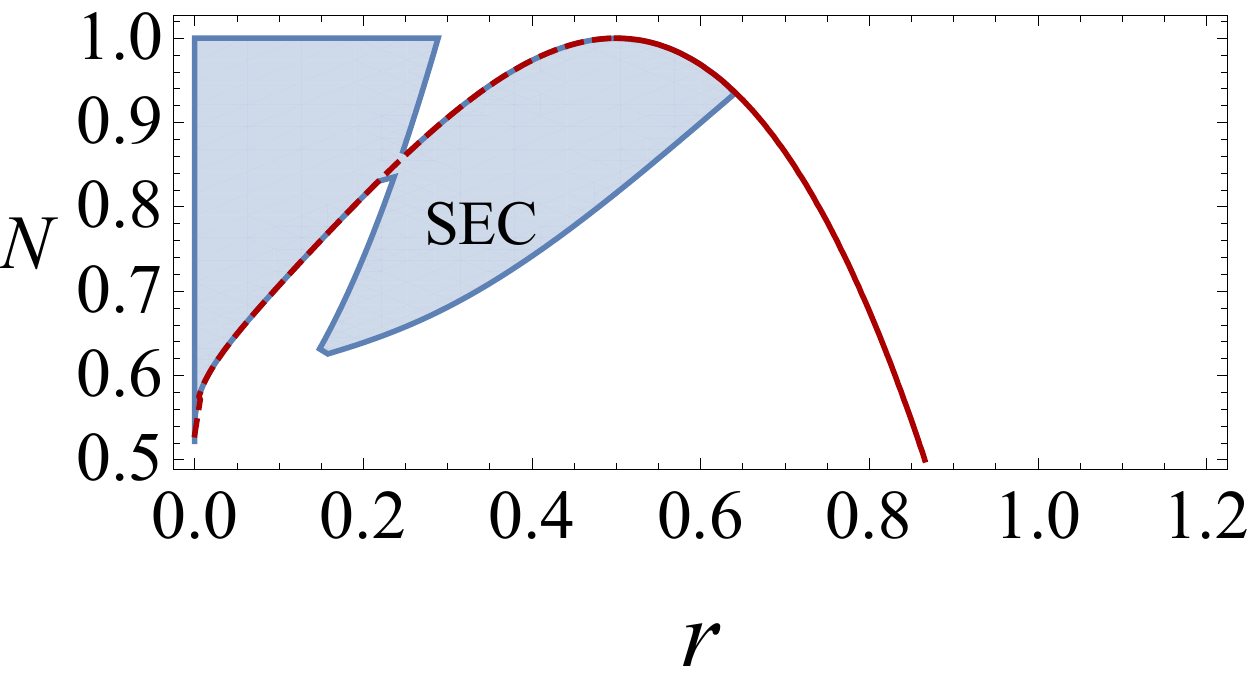}
		\caption{Strong energy condition}
		\label{fig:strongS=N}
	\end{subfigure}
	\begin{subfigure}[b]{0.33\textwidth}
		\centering
		\includegraphics[width=\textwidth]{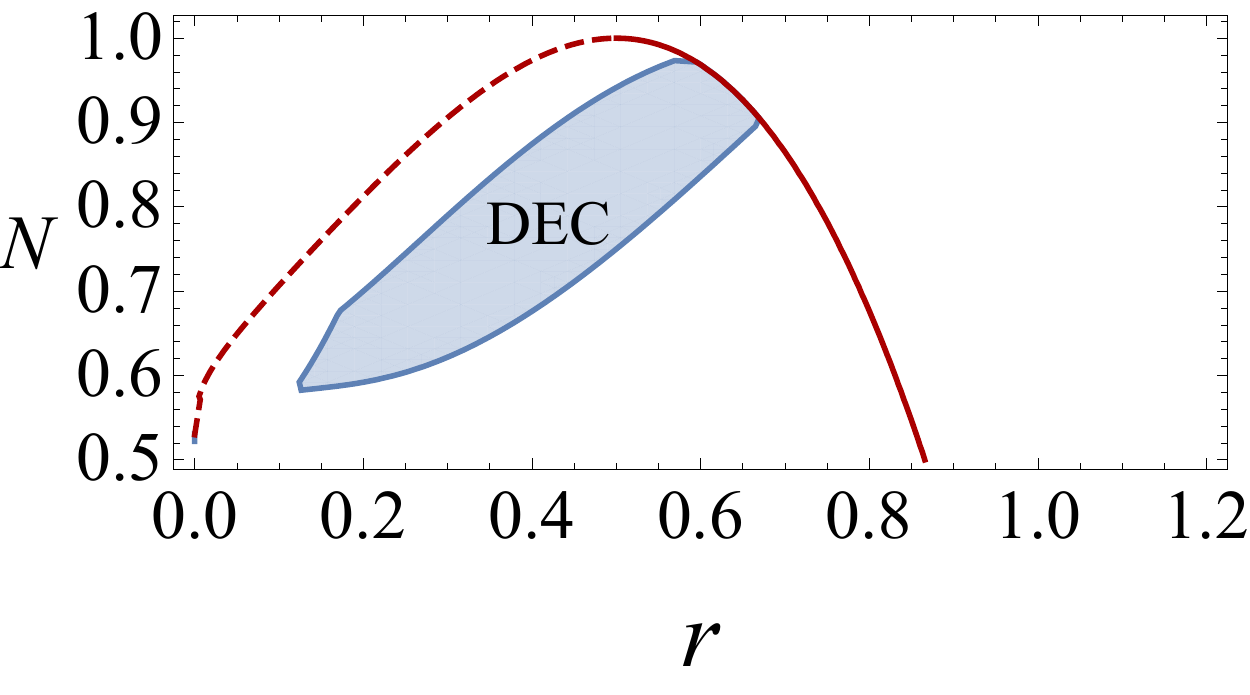}
		\caption{Dominant energy condition}
		\label{fig:dominantS=N}
	\end{subfigure}
	\captionsetup{width=.9\linewidth}
	\caption{The blue shadows show the valid domains of  various energy conditions for the case of $L=1/2$. The
		dotted red curves are inner horizons while the solid ones outer horizons in this case.  The existence of
		horizons gives the	constraint, $1/2 \leqslant N \leqslant 1$, in this case.}
	\label{fig:energy-condition-fullS=N}
\end{figure}

Besides the above discussions of energy conditions on the $r-L$ plane, for our ARBH model depicted by Eq.~(\ref{eq:vre}), we further investigate its energy conditions by plotting the valid domains on the $r-N$ plane in Fig.~\ref{fig:energy-condition-fullS=N}. The
NEC, WEC and SEC are satisfied in two pieces of domains for the parameter range $1/2< N\le 1$ and  $L=1/2$, where one piece is located
inside the inner horizon and the other between	the inner and outer horizons (also including the horizons as
boundaries). However, the DEC is satisfied in only one piece of domains between the inner and outer horizons (also
including the outer horizon as boundary) for the parameter range $1/2< N< 1$ and  $L=1/2$. In particular,  Fig.~\ref{fig:strongS=N} shows that the SEC is completely violated in the entire spacetime $ r\in [0,\infty)$ in the vicinity of $N=1/2$.

\newpage
\subsection{Energy conditions of conformally related Schwarzschild black holes}
In Sec.~\ref{sec2} and Sec.~\ref{sec:bardeen}, we have seen that our ARBH model can be regarded as a conformally related  BH in the perspective of  finiteness of curvature invariants and completeness of geodesics, where the density of fluid acts as the scale factor.
It is just a seeming reason that the line element of our ARBH model, Eq.\ \eqref{lineele} and Eq.\ \eqref{eq:vre}, looks like that of a conformally related BH.  The virtual reason is that the acoustic analog leads to a non-vanishing partition function
 if it is interpreted in the context of conformally invariant theory. Let us extend this discussion. If the Euclidean action of our ARBH model were constructed  \cite{Gibbons:1976ue} by
\begin{equation}
 \tilde I=\int\dif x^4\sqrt{-g}\, R+\cdots,   
\end{equation}
where the ellipsis represents the surface term and matter sectors, it would be divergent since $\sqrt{-g}$ is divergent at $r=0$. 
As a result,  all the thermodynamic variables computed
by the path-integral method would be trivial because the partition function $Z=\me^{-\tilde I}$ vanishes.
Nevertheless, if we construct our ARBH model in the conformal theory~\cite{DGB2009},  i.e.,
\begin{equation}
    I=\int\dif x^4\sqrt{-g}\left[ \frac{1}{12}\varphi^2
R-\frac{1}{2}\varphi g^{\mu\nu}\nabla_{\mu}\nabla_{\nu}\varphi\right]+\cdots,
\end{equation} where $\varphi$ is a massless scalar field and $\nabla_{\mu}$  covariant derivative,
the situation will be improved because the scalar field $\varphi$ can absorb the divergence of the measure $\sqrt{-g}$ based on the conformal symmetry.

Here we intend to emphasize that this analogue BH has its own specific
properties in the energy conditions that are distinct from those of a conformally related  BH. We shall take
CRSBHs as an example, analyze its energy conditions and compare them with our ARBH model's.

The scale factor of CRSBHs takes~\cite{Chen:2019iuo} the form,
\begin{equation}
	S (r) = \left( 1 + \frac{\bar{L}^2}{r^2} \right)^{2 \bar{N}},
\end{equation}
where $\bar{N}$ and $\bar{L}$ have the same meanings as those of $N$ and $L$ in Eq.~(\ref{eq:vre}), and $\bar L$ and $\bar N$ are independent of each other but $L$ and $N$ are related to each other due to the existence of horizons. We can verify that the regularity of CRSBHs requires $\bar{N}\geqslant3/4$.

\begin{figure}[!ht]
	\centering
	\begin{subfigure}[b]{0.30\textwidth}
		\centering
		\includegraphics[width=\textwidth]{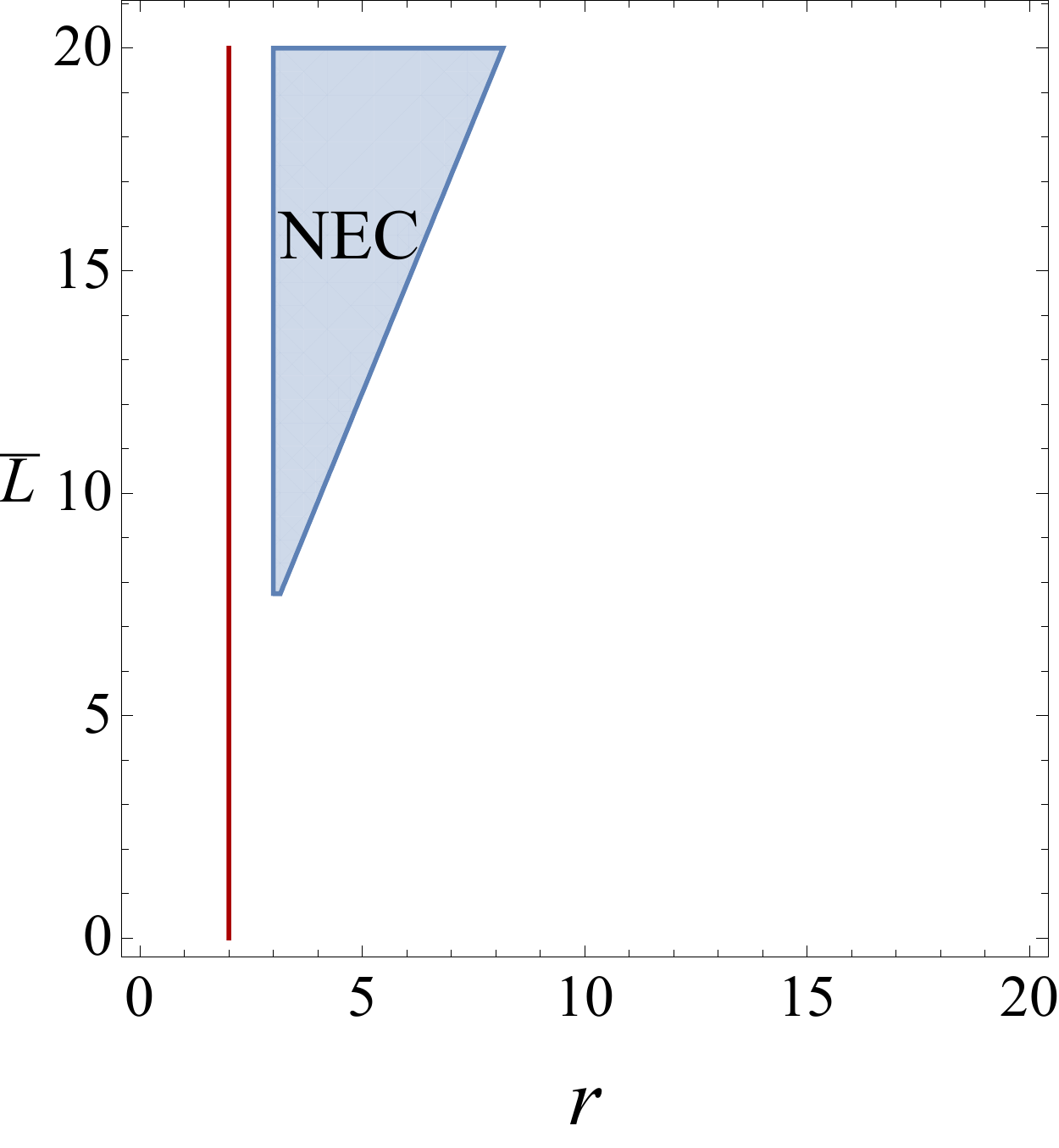}
		\caption{Null energy condition}
		\label{fig:null-cs}
	\end{subfigure}
	\begin{subfigure}[b]{0.30\textwidth}
		\centering
		\includegraphics[width=\textwidth]{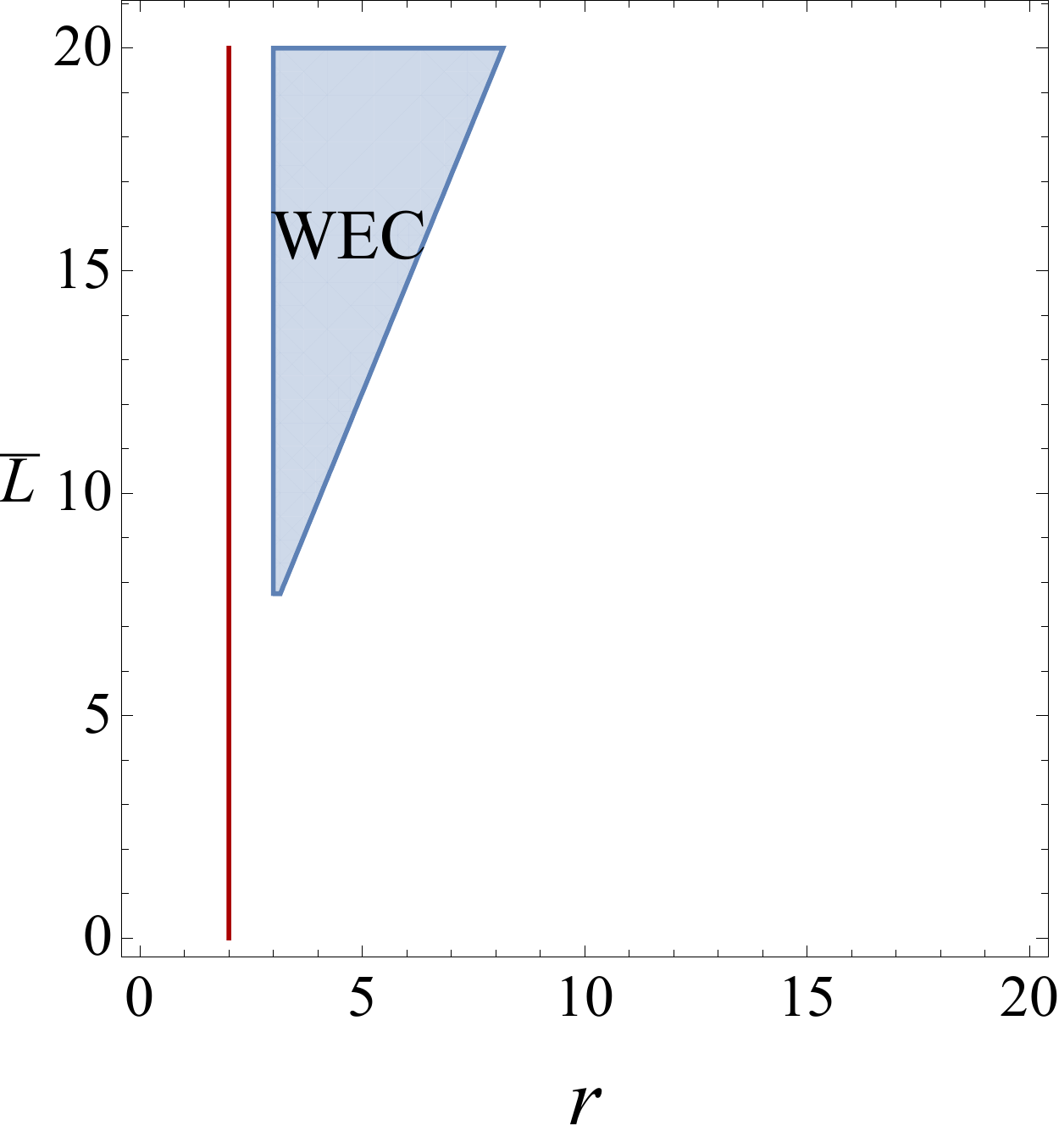}
		\caption{Weak energy condition}
		\label{fig:weak-null-cs}
	\end{subfigure}
	
\begin{subfigure}[b]{0.30\textwidth}
		\centering
		\includegraphics[width=\textwidth]{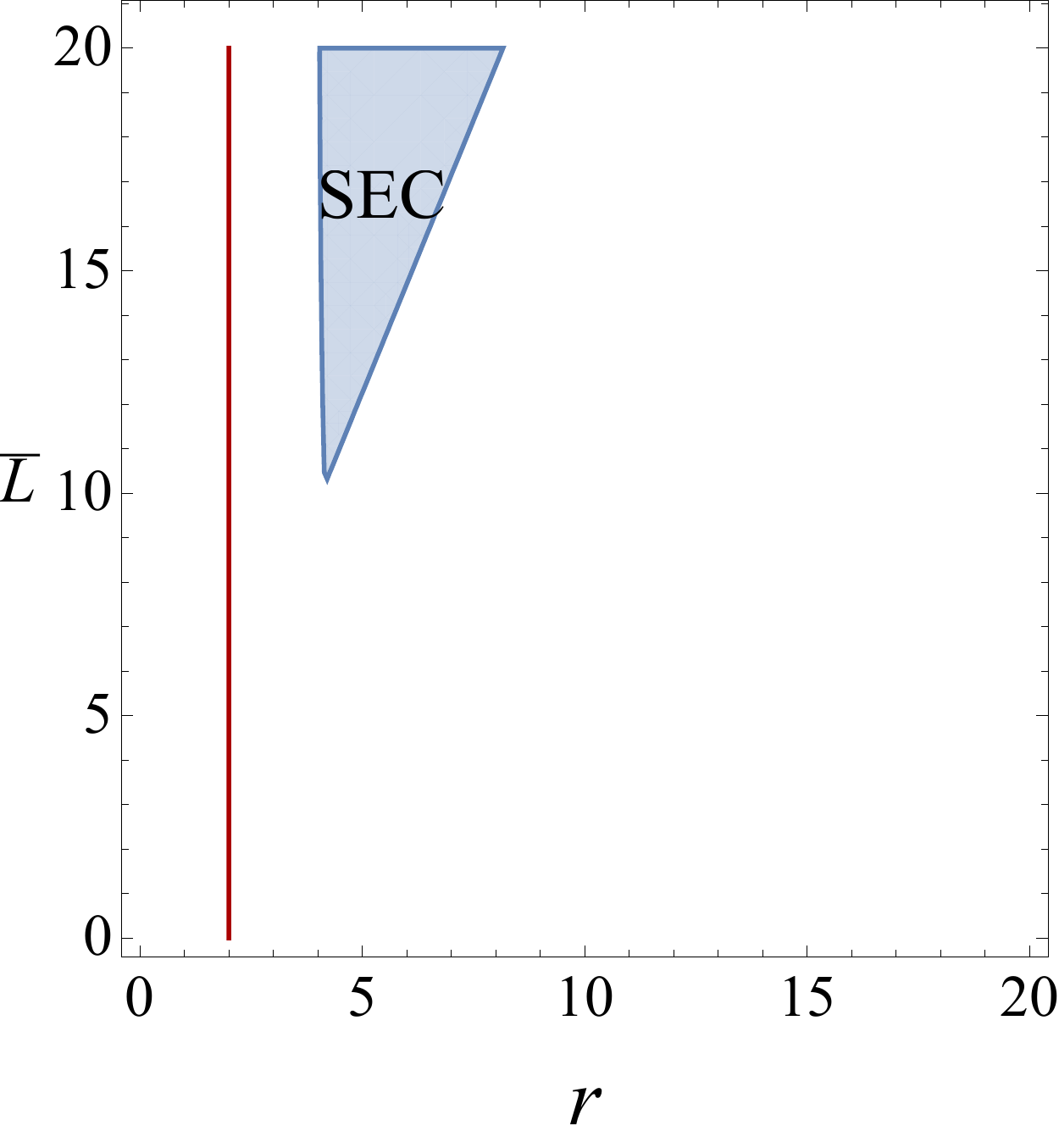}
		\caption{Strong energy condition}
		\label{fig:strong-cs}
	\end{subfigure}
	\begin{subfigure}[b]{0.30\textwidth}
		\centering
		\includegraphics[width=\textwidth]{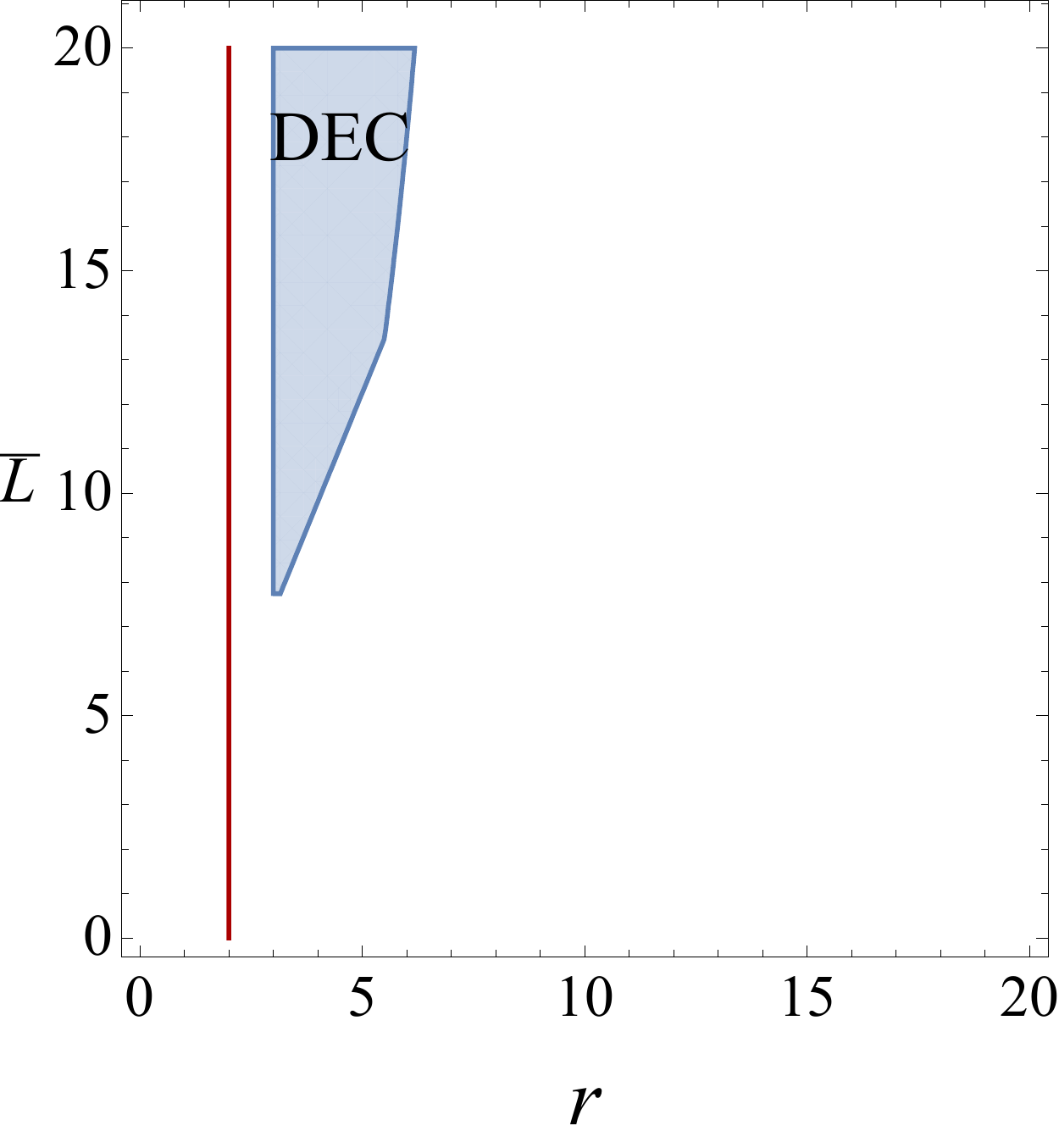}
		\caption{Dominant energy condition}
		\label{fig:dominant-cs}
	\end{subfigure}
	\captionsetup{width=.9\linewidth}
	\caption{The blue shadows show the valid domains of  various energy conditions for the case of
		$\bar{N}=3/4$. The red lines are horizons.}
	\label{fig:energy-condition-cs}
\end{figure}

\begin{figure}[!ht]
	\centering
	\begin{subfigure}[b]{0.30\textwidth}
		\centering
		\includegraphics[width=\textwidth]{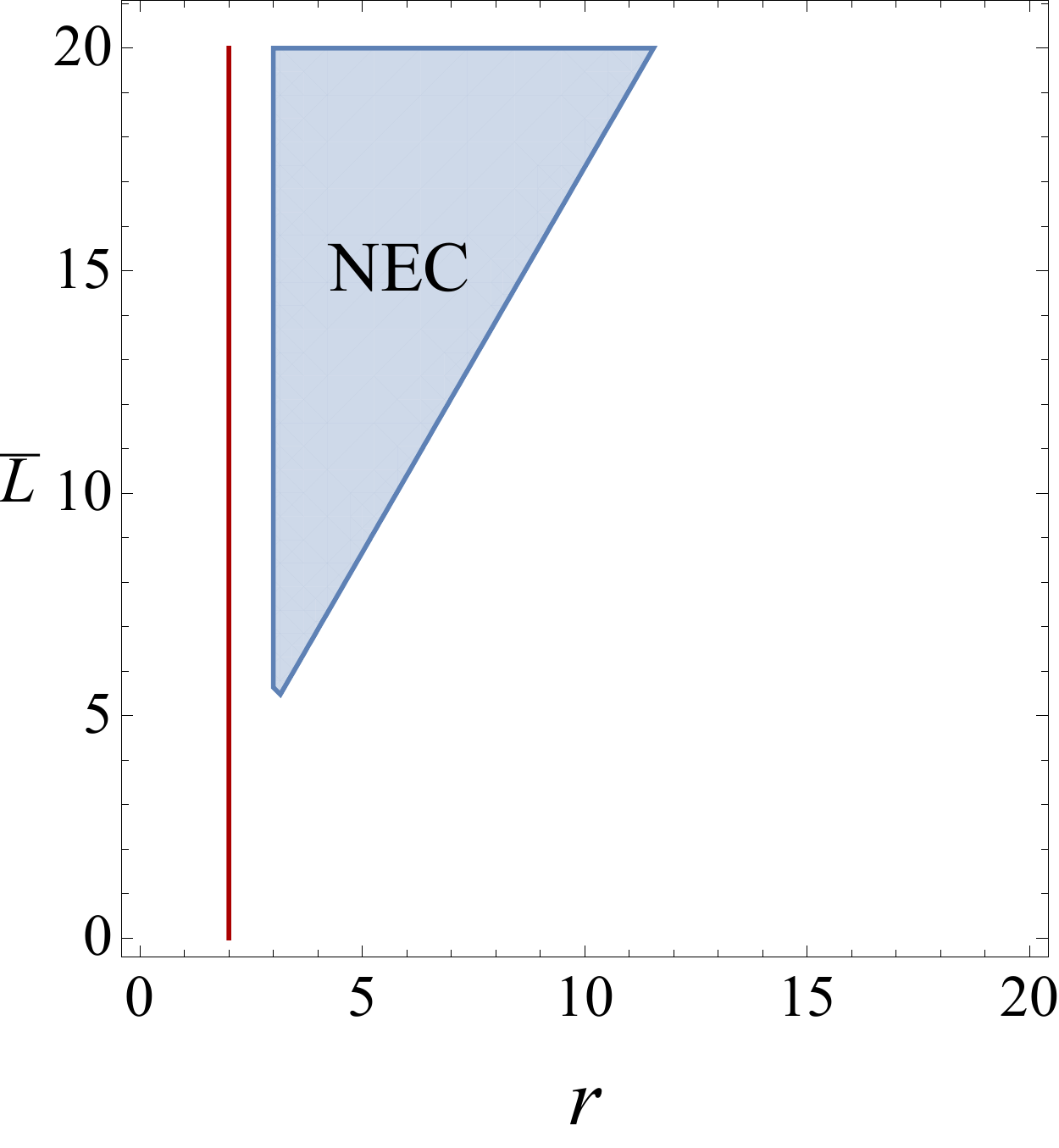}
		\caption{Null energy condition}
		\label{fig:null-cs-1}
	\end{subfigure}
	\begin{subfigure}[b]{0.30\textwidth}
		\centering
		\includegraphics[width=\textwidth]{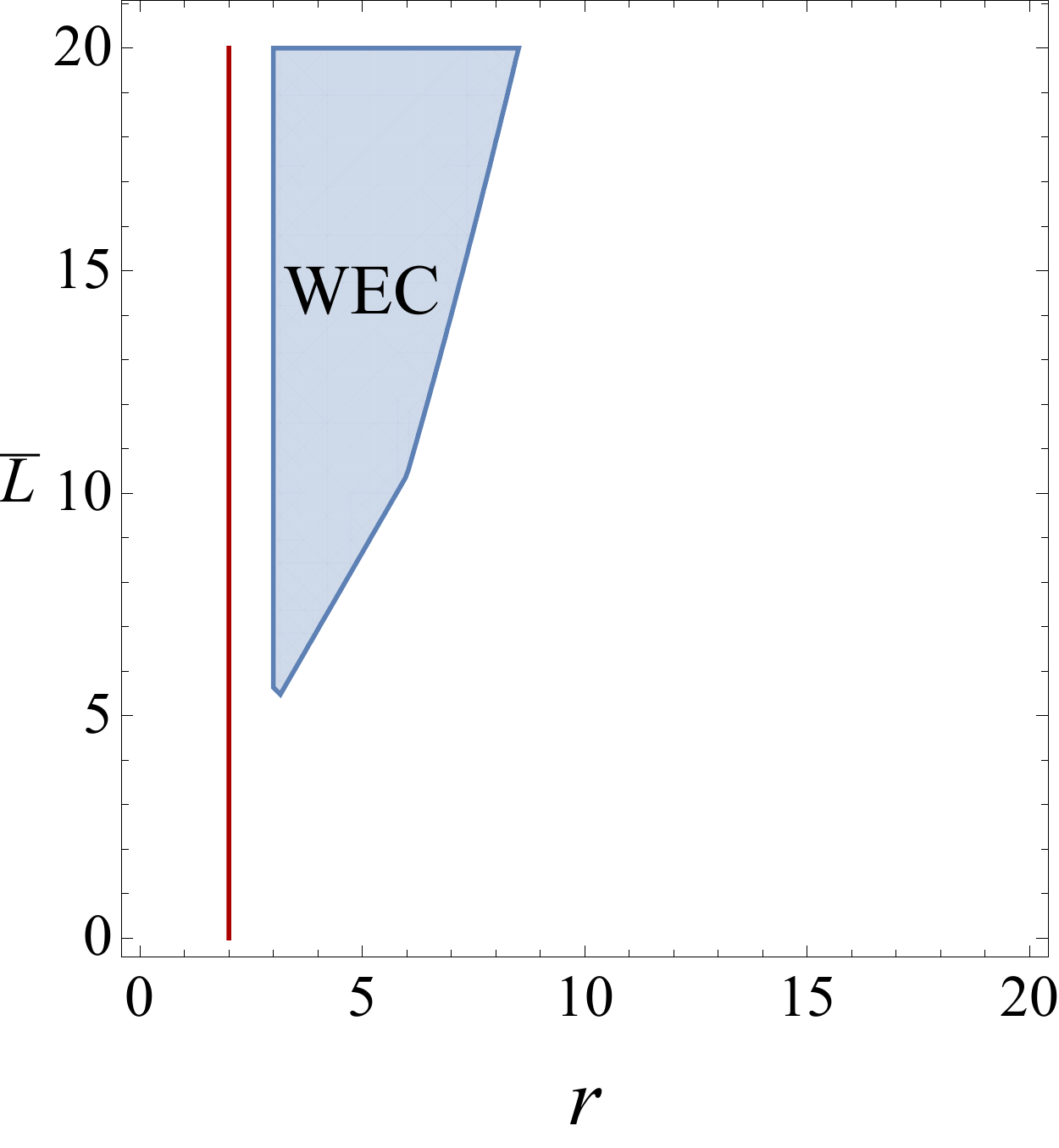}
		\caption{Weak energy condition}
		\label{fig:weak-null-cs-1}
	\end{subfigure}
	
\begin{subfigure}[b]{0.30\textwidth}
		\centering
		\includegraphics[width=\textwidth]{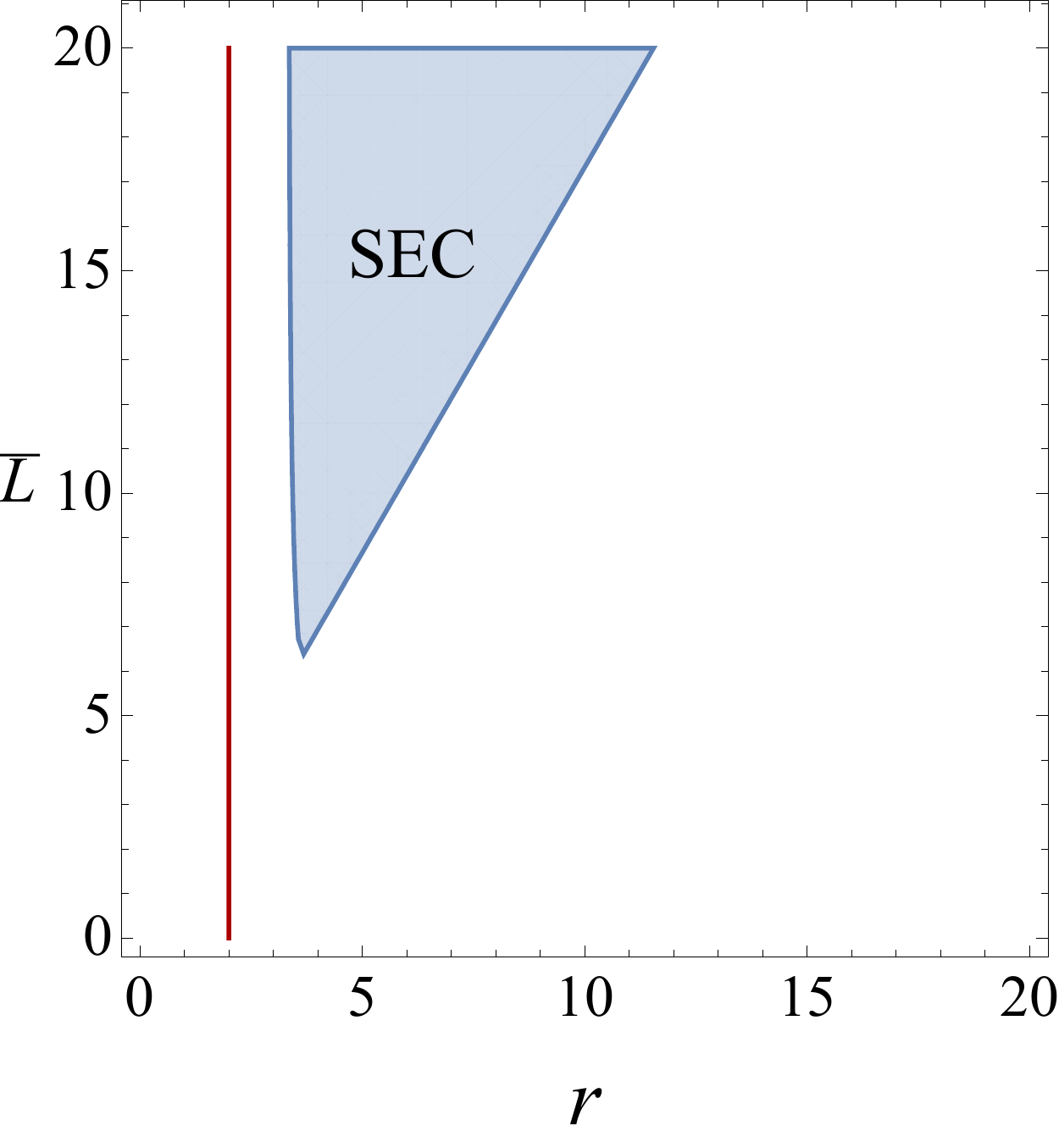}
		\caption{Strong energy condition}
		\label{fig:strong-cs-1}
	\end{subfigure}
	\begin{subfigure}[b]{0.30\textwidth}
		\centering
		\includegraphics[width=\textwidth]{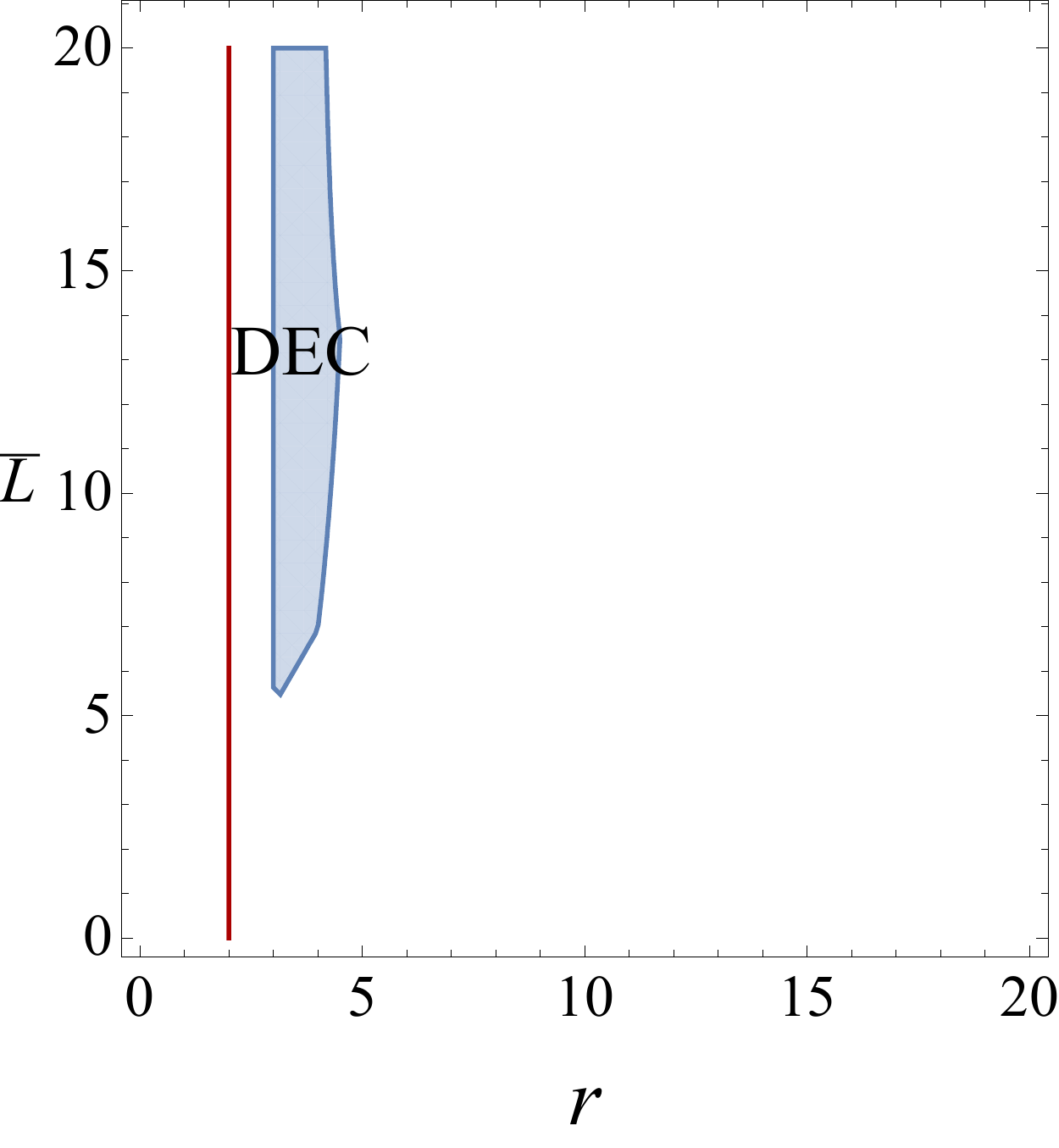}
		\caption{Dominant energy condition}
		\label{fig:dominant-cs-1}
	\end{subfigure}
	\captionsetup{width=.9\linewidth}
	\caption{The blue shadows show the valid domains of  various energy conditions for the case of
		$\bar{N}=1$. The  red lines are horizons.}
	\label{fig:energy-condition-cs-1}
\end{figure}

Following the same procedure as that in the above subsection, we plot the valid domains of energy conditions of CRSBHs\footnote{The energy conditions of CRSBHs were analyzed in Ref.~\cite{Toshmatov:2017kmw} in which the sign of energy density is wrong. See App.~\ref{appendix:A} for our explanations.} on the $r-\bar{L}$ plane in Figs.~\ref{fig:energy-condition-cs} and \ref{fig:energy-condition-cs-1} for the two cases of $\bar{N}= 3/4$ and $\bar{N}= 1$, respectively. We can see that the energy conditions are satisfied only outside the horizon of CRSBHs, which is completely different from the situation of our ARBH model in Figs.~\ref{fig:energy-condition-full} and \ref{fig:energy-condition-fullS=1}.
We also notice that the valid domains in Figs.~\ref{fig:energy-condition-cs} and \ref{fig:energy-condition-cs-1} are located in the areas with a minimum value of $\bar{L}$, and that they expand when $\bar{L}$ increases. However, it is obvious that the expansion of domains does not happen in our ARBH model, see Figs.~\ref{fig:energy-condition-full} and \ref{fig:energy-condition-fullS=1}, because $L$ is constrained by the value of $N$. Especially, the NEC and SEC are satisfied at $r = 0$ for our ARBH model, see Figs.~\ref{fig:energy-condition-fullS=1} and ~\ref{fig:energy-condition-fullS=N}, which does not appear in the CRSBHs. This feature (the SEC is not violated at  $r = 0$) implies that the interaction is attractive in the vicinity of $r = 0$ in our ARBH model, which presents the characteristic of this acoustic analog.

\begin{figure}[!ht]
	\centering
	\begin{subfigure}[b]{0.30\textwidth}
		\centering
		\includegraphics[width=\textwidth]{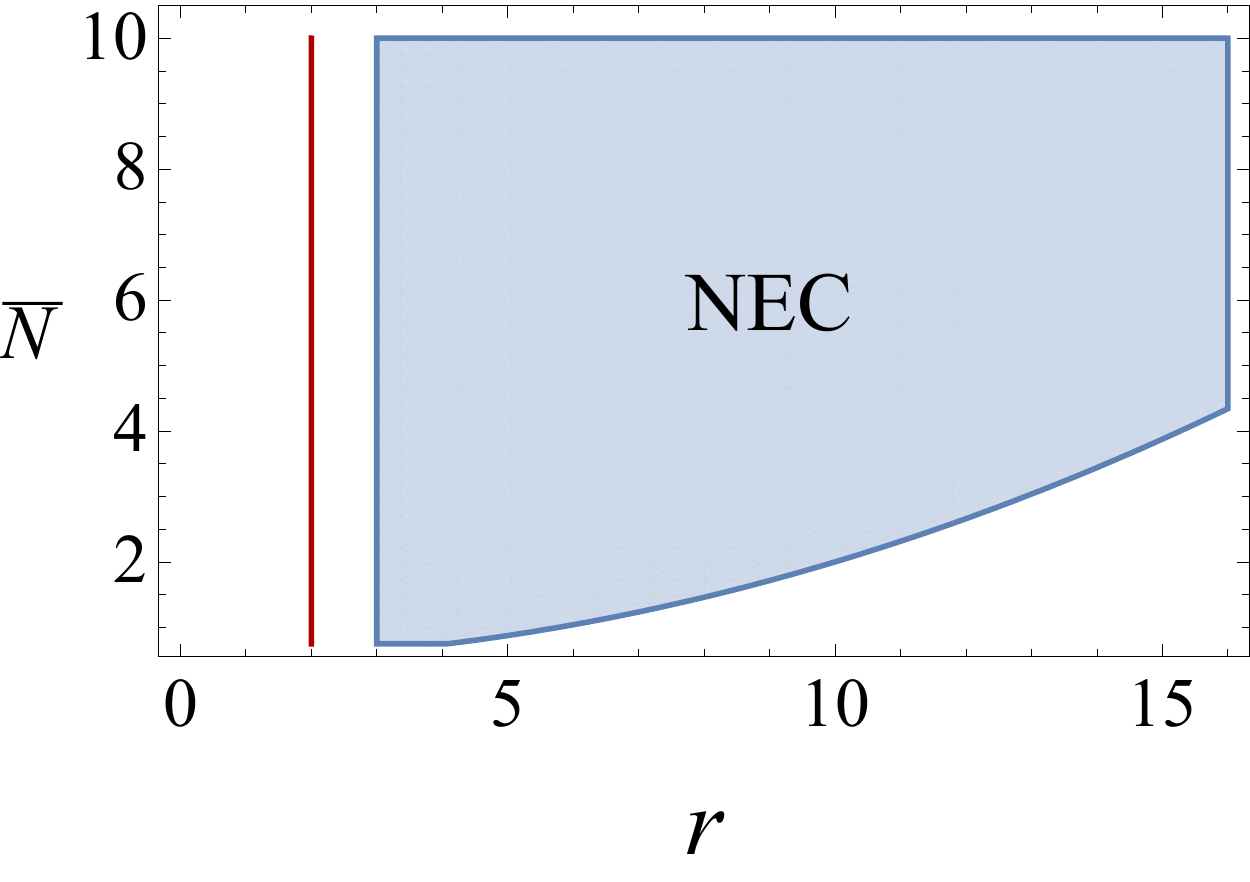}
		\caption{Null energy condition}
		\label{fig:null-cs-N}
	\end{subfigure}
	\begin{subfigure}[b]{0.30\textwidth}
		\centering
		\includegraphics[width=\textwidth]{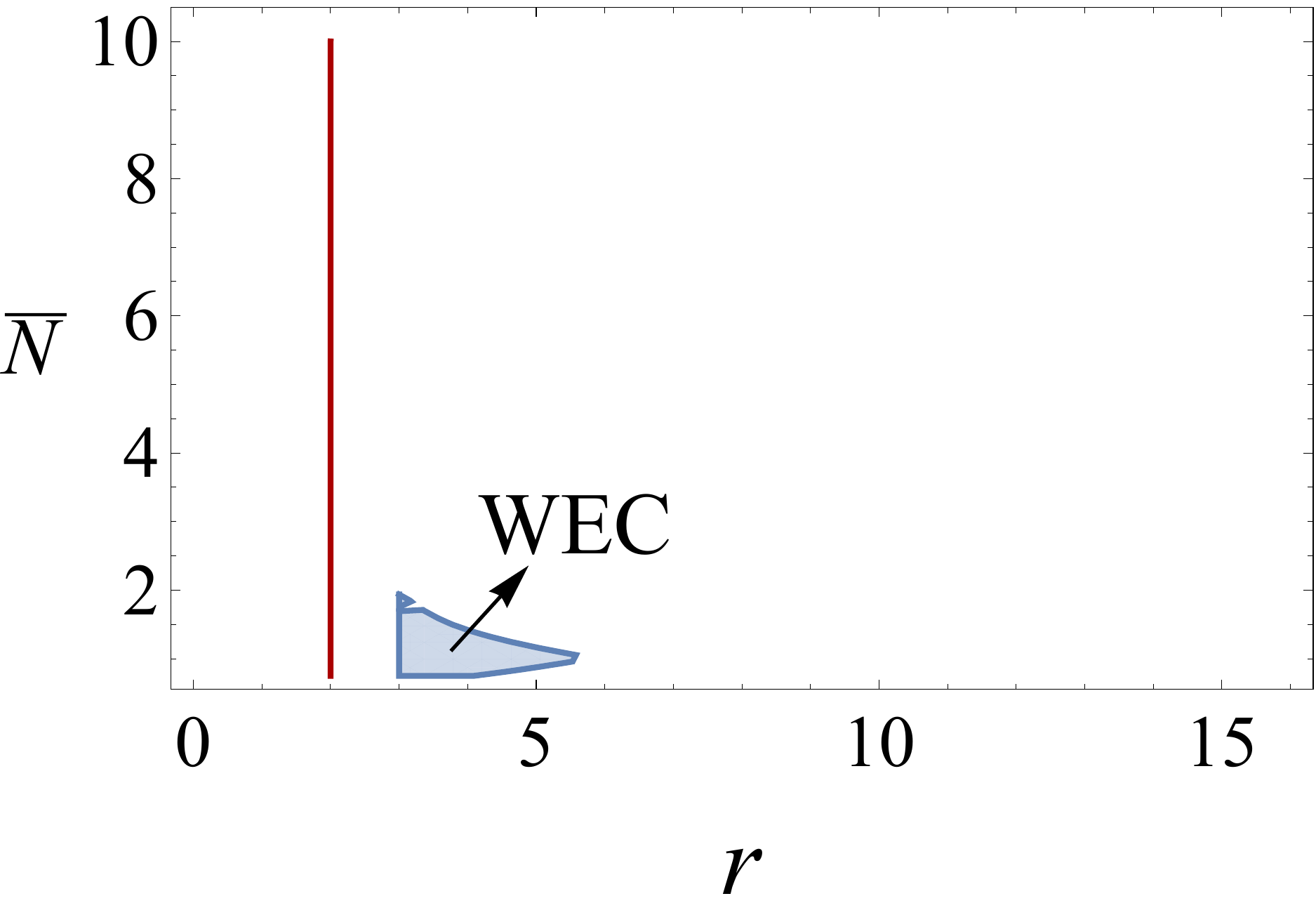}
		\caption{Weak energy condition}
		\label{fig:weak-null-cs-N}
	\end{subfigure}
	
\begin{subfigure}[b]{0.30\textwidth}
		\centering
		\includegraphics[width=\textwidth]{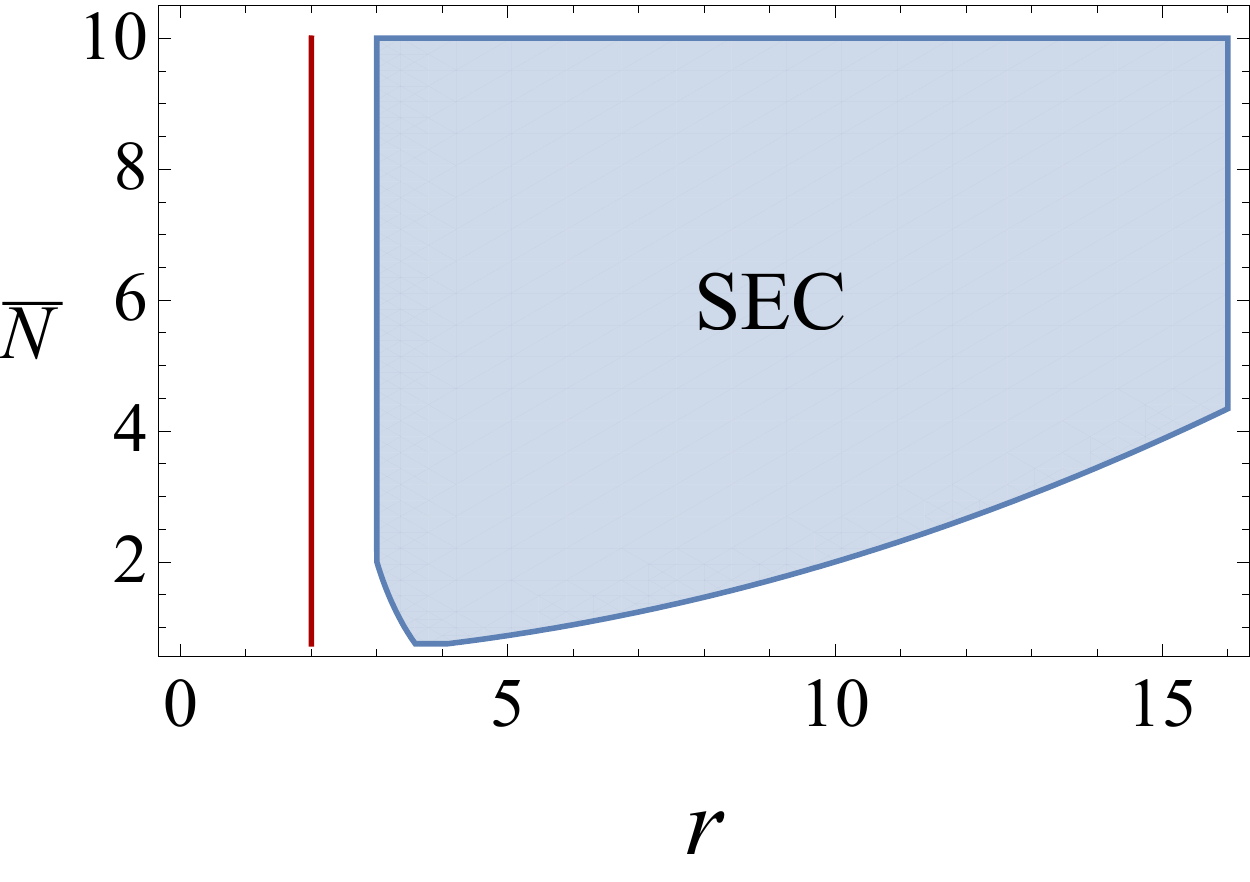}
		\caption{Strong energy condition}
		\label{fig:strong-cs-N}
	\end{subfigure}
	\begin{subfigure}[b]{0.30\textwidth}
		\centering
		\includegraphics[width=\textwidth]{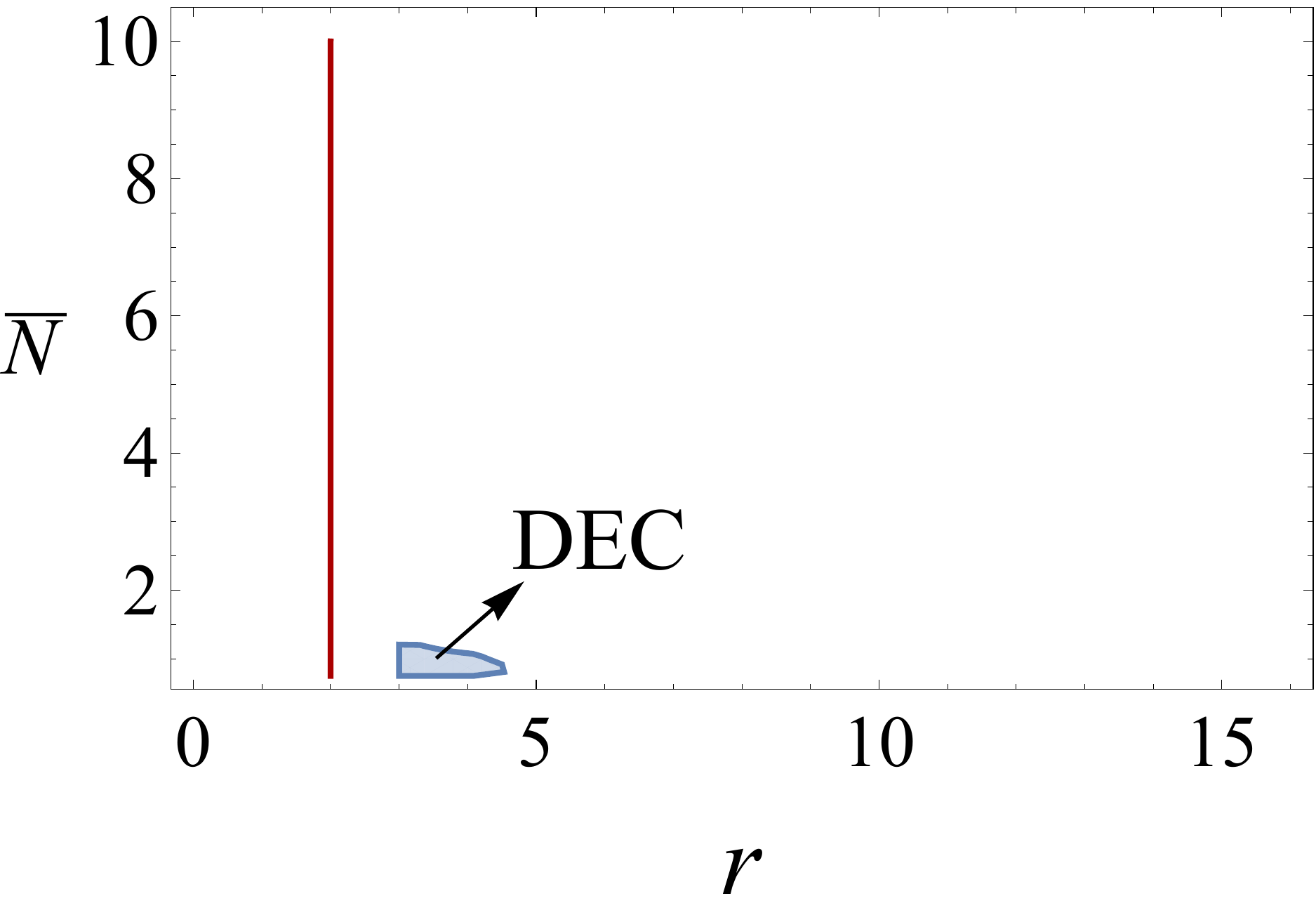}
		\caption{Dominant energy condition}
		\label{fig:dominant-cs-N}
	\end{subfigure}
	\captionsetup{width=.9\linewidth}
	\caption{The blue shadows show the valid domains of  various energy conditions for the case	of $\bar{L}=10$.
	The red lines are horizons.}
	\label{fig:energy-condition-cs-N}
\end{figure}

In addition, we  plot the valid domains of energy conditions of CRSBHs on the $r-\bar{N}$ plane in Fig.~\ref{fig:energy-condition-cs-N}. When comparing it with  Fig.~\ref{fig:energy-condition-fullS=N}, we find that the domains of energy conditions of CRSBHs are located outside the horizon while those of our ARBH model inside the outer horizon. This is the main difference between the ARBHs and CRSBHs in the energy conditions, and the other differences are similar to those mentioned above between $r-{L}$ and $r-\bar{L}$ graphs.

\section{Quasinormal modes of acoustic regular black holes}
\label{sec3}
\begin{figure}[!ht]
	\centering
	\begin{subfigure}[b]{0.45\textwidth}
		\centering
		\includegraphics[width=\textwidth]{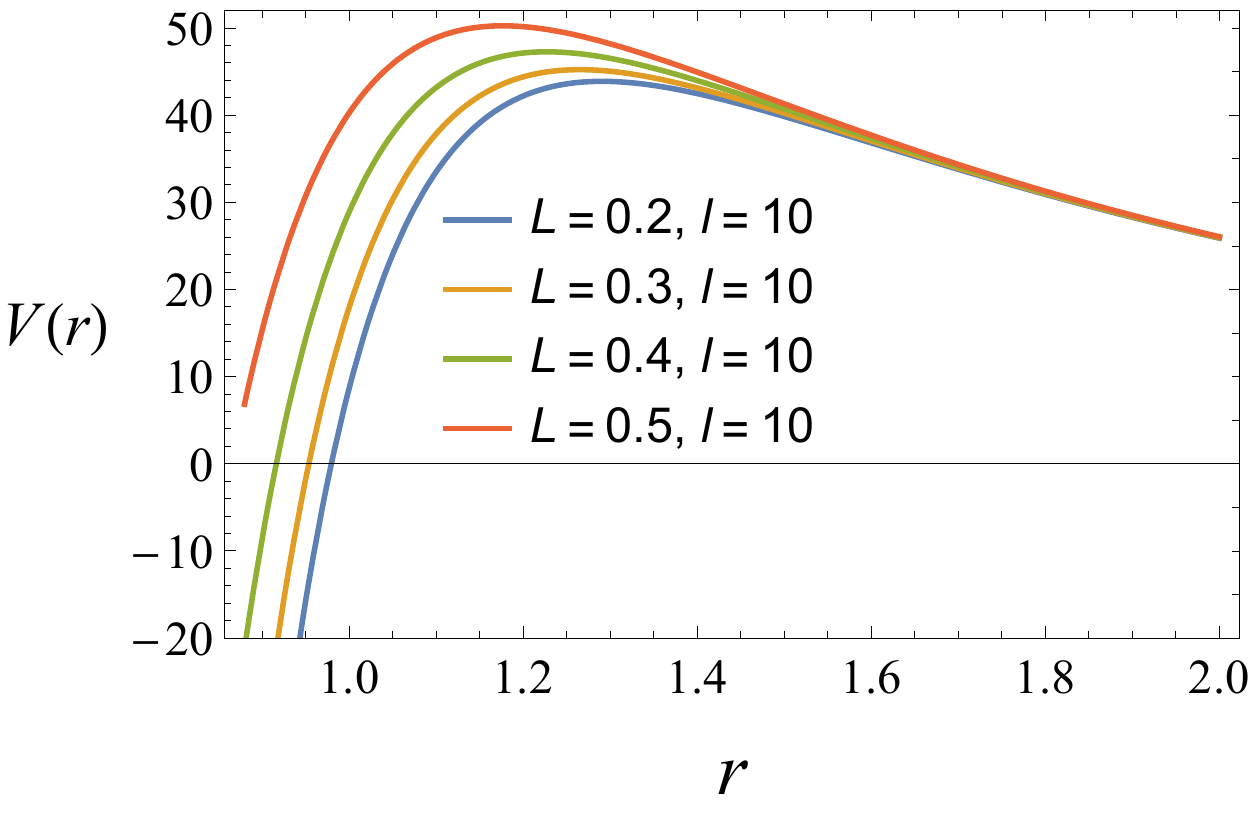}
	\end{subfigure}
	\begin{subfigure}[b]{0.45\textwidth}
		\centering
		\includegraphics[width=\textwidth]{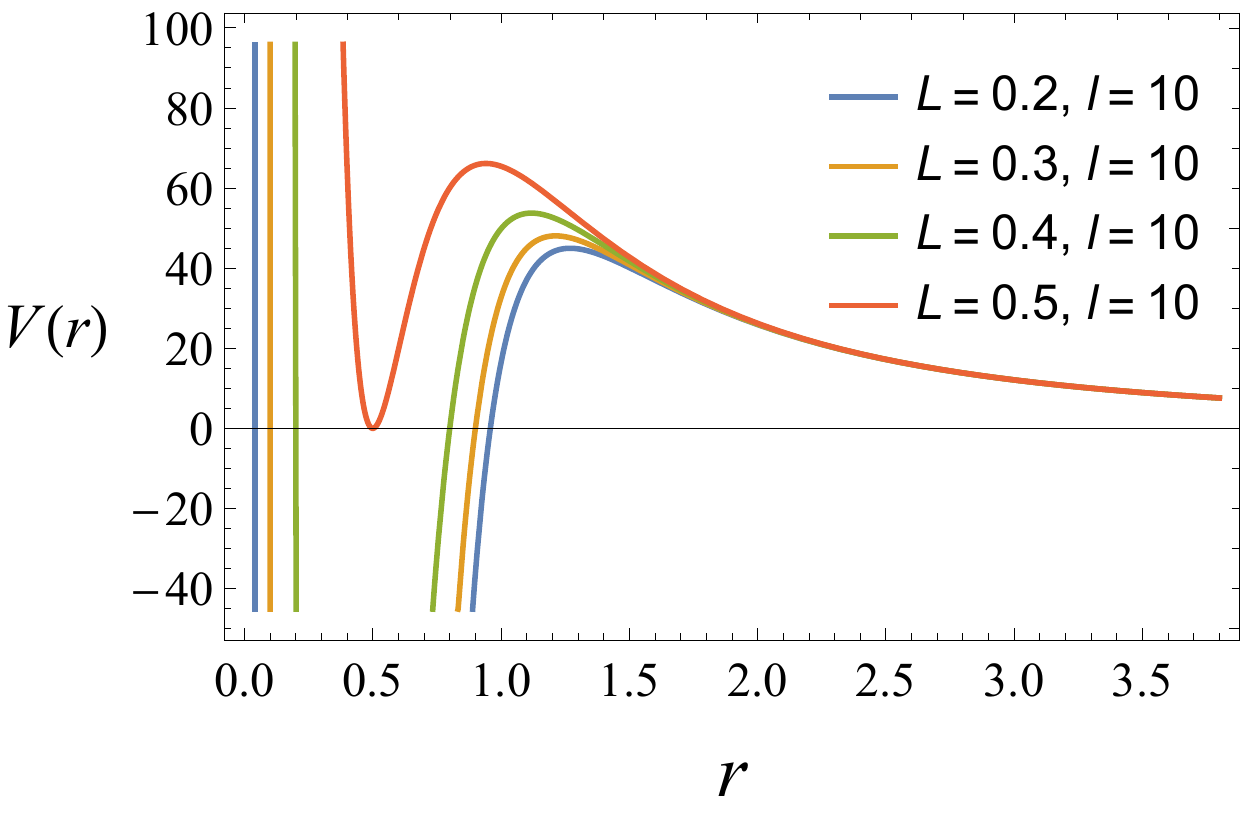}
	\end{subfigure}
	\captionsetup{width=.9\linewidth}
	\caption{$V(r)$ with respect to $r$ for the cases of $N=1/2$ (left) and $N=1$ (right).}
	\label{fig4}
\end{figure}

In this section, we discuss the sound propagation in the spacetime of our ARBH model. As mentioned in Introduction, the equation of motion for an acoustic disturbance is identical~\cite{unruh1981experimental} to the d'Alembertian equation of a massless scalar field propagating in a curved spacetime. That is, the sound propagation in our ARBH spacetime
manifests as the propagation of a massless scalar field in an effective curved spacetime, which is described by
the Klein-Gordon equation. As a result, we can analyze the stability of our ARBH model by computing its QNMs in
terms of the WKB
method~\cite{schutz1985black,iyer1987blackhole,iyer1987blackholea,berti2009quasinormal,konoplya2003quasinormal},
 where the 6th-order WKB method is adopted in order to have the balance between precision and complexity of
numerical calculations.
The Klein-Gordon equation for a massless scalar field $\Phi$ in a curved spacetime can be written as
\begin{equation}
	\frac{1}{\sqrt{- g}} \partial_{\mu}  \left( \sqrt{- g} g^{\mu \nu}
	\partial_{\nu} \Phi \right) = 0, \label{eq26}
\end{equation}
where $\Phi$ represents the disturbance to the background fluid, i.e., the potential function of acoustic waves
\cite{visser1998acoustic}. In order to separate the variables in Eq.~\eqref{eq26}, the function $\Phi$ can be
chosen as
\begin{equation}
	\Phi = \frac{1}{r \sqrt{\rho}} \Psi (r) Y_{\ell}^m (\theta, \phi) \me^{- i
		\omega t},\label{eq27}
\end{equation}
where $Y_{\ell}^m (\theta, \phi)$ is spherical harmonic function of degree $l$ and order $m$, and $l$ is also called the multipole number. Substituting Eq.~\eqref{eq27} into Eq.~\eqref{eq26}, we get the Schr\"odinger-like equation~\cite{Chen:2019iuo},
\begin{equation}
	\frac{d^2 \Psi}{dr_{\ast}^2} + \omega^2 \Psi = V (r) \Psi, \label{eq28}
\end{equation}
with the effective potential,
\begin{equation}
	V (r) = f (r)  \left[ \frac{l (l + 1)}{r^2} + \frac{1}{Z}  \frac{d}{dr}
	\left( f (r) \frac{dZ}{dr} \right) \right], \label{eq29}
\end{equation}
where $Z \equiv r\sqrt{\rho} $ and $r_{\ast} $ is the tortoise coordinate defined by $ dr_{\ast} =dr / f (r)$. For our ARBH model, substituting
Eq.~\eqref{eq:metric s=1/2} for the case of $N=1/2$ and Eq.~\eqref{eq:metric s=1} for the case of $N=1$ into Eq.~(\ref{eq29}), we write down explicitly the effective potentials,
\begin{equation}
V_{N=1/2} (r) = \left[1 - \frac{1}{(L^2 + r^2)^2} \right]  \left[ \frac{l (l +
	1)}{r^2} + \frac{L^2 (L^2 + r^2)^2 - L^2 + 4 r^2}{(L^2 + r^2)^4} \right],\label{pot12}
\end{equation}
and
\begin{equation}
	\begin{aligned}
	V_{N=1}(r) =& \left[ 1 - \frac{1}{r^4  \left(1+ \frac{L^2}{r^2} \right)^4}
\right]  \\
& \times \left[ \frac{(l + 1) l}{r^2} + \frac{2 (L^{12} + 5 L^{10} r^2 + 10 L^8 r^4 +
	10 L^6 r^6 + L^4  (5 r^8 + r^4) + L^2 r^6  (r^4 - 5) + 2 r^8)}{r^2  (L^2 +
	r^2)^6} \right].\label{pot1}
	\end{aligned}
\end{equation}
Now we plot the effective potential $V (r)$ with respect to radial coordinate $r$ for different values of
parameter $L$ but a fixed $l=10$ in Fig.~\ref{fig4}. It can be seen that the potential has only one maximum
value for the case of $N=1/2$, while it has one minimum value and one maximum value for the case of $N=1$.
When $l$ is fixed, both minimum and maximum values increase as $L$ increases.

The QNMs solved from Eq.~\eqref{eq28} together with the effective potential Eq.~\eqref{eq29} can be cast in the complex form, $\omega ={\rm Re}\,\omega + i\, {\rm Im}\,\omega$, where the real part, ${\rm Re}\,\omega$, represents the oscillation of perturbation, while the imaginary part, ${\rm Im}\,\omega$, characterizes the dissipation of perturbation. We use the 6th-order WKB method to provide numerical solutions. 
It should be noted that the WKB method requires that the effective potential $V (r)$ has one single maximum outside the horizon and that the multipole number $l$ is larger than the overtone number which is taken to be zero for the fundamental mode of scalar field perturbation~\cite{berti2009quasinormal}.  We can see  from Fig.~\ref{fig4} that our ARBH model meets the requirement.

The QNMs satisfy~\cite{konoplya2003quasinormal} the following formula in the 6th-order WKB method,
\begin{equation}
	i \frac{\left(\omega^{2}-V_{0}\right)}{\sqrt{-2 V_{0}^{\prime \prime}}}-\sum_{i=2}^{6} \Lambda_{i}=n+\frac{1}{2},\label{wkb6}
\end{equation}
where $V_0$ is the maximum of the effective potential $V(r)$, $V_{0}^{\prime \prime}=\left.\frac{\mathrm{d}^{2}
	V\left(r\right)}{\mathrm{d} r_{*}^{2}}\right|_{r_{*}=(r_{*})_{0}}$, $r_0$ is the position of the peak value of the
effective potential, $n$ is overtone number, and $\Lambda_i$  $(i=2, 3, \dots, 6)$ are constant coefficients related to the  corrections from the 2nd- to 6th-orders. By substituting Eq.~(\ref{pot12}) or Eq.~(\ref{pot1}) into Eq.~(\ref{wkb6}), we can obtain the QNMs numerically for our ARBH model in the case of $N=1/2$ or $N=1$.

\begin{figure}[!ht]
	\centering
	\begin{subfigure}[b]{0.45\textwidth}
		\centering
		\includegraphics[width=\textwidth]{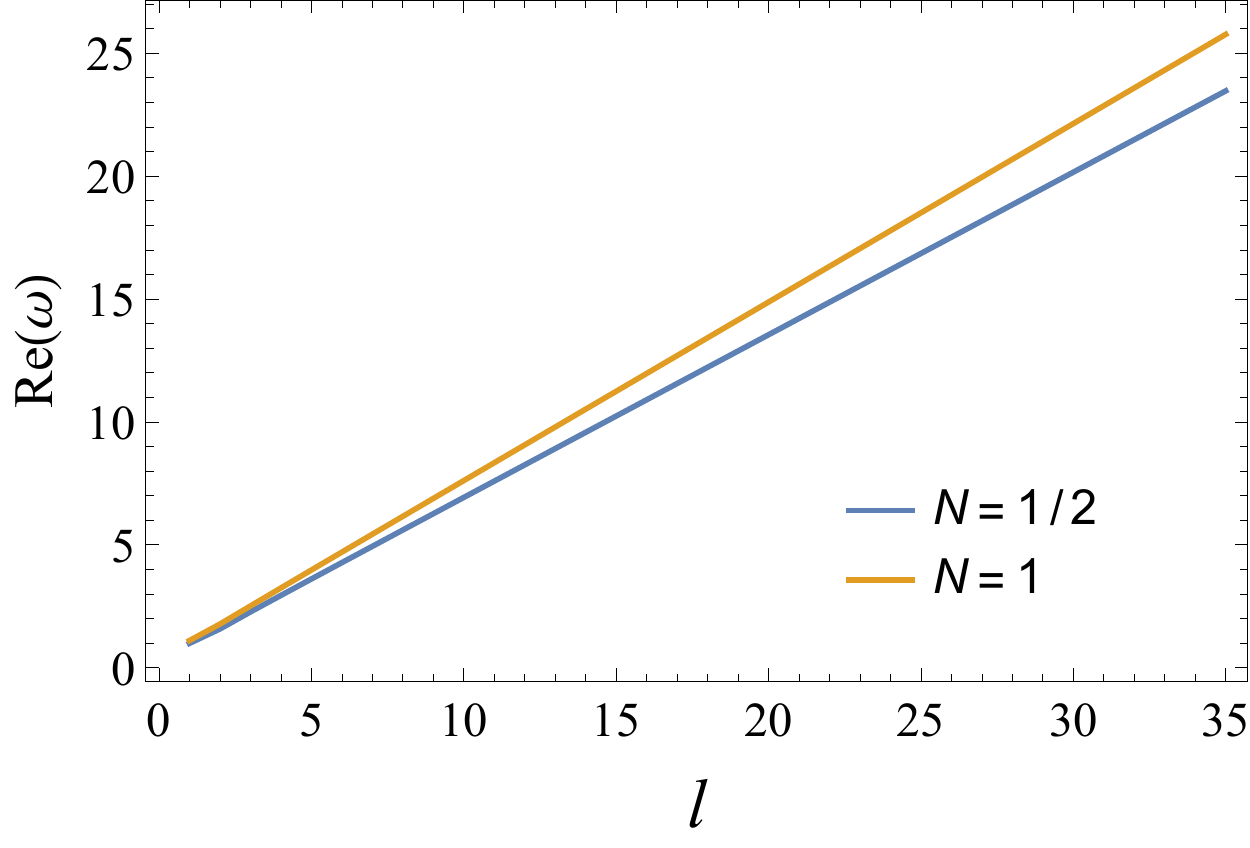}
	\end{subfigure}
	\begin{subfigure}[b]{0.45\textwidth}
		\centering
		\includegraphics[width=\textwidth]{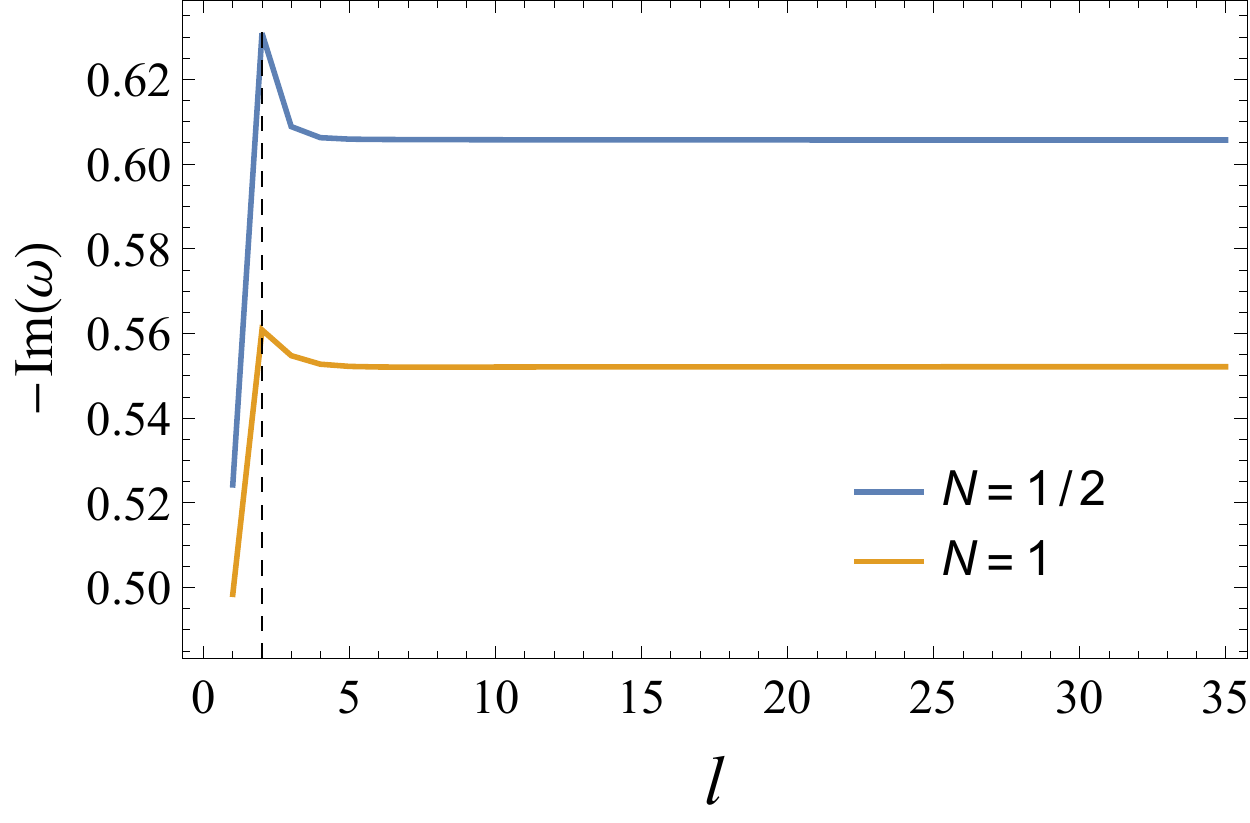}
	\end{subfigure}
	\captionsetup{width=.9\linewidth}
	\caption{QNMs with respect to $l$, where $L=0.45$ and $n=0$ are set. The left diagram represents the real parts of $\omega$ with respect to $l$ for the cases of $N=1/2$ and $N=1$, respectively; the right diagram represents the negative imaginary parts of $\omega$ with respect to $l$ for the cases of $N=1/2$ and $N=1$, respectively.}
	\label{fig7}
\end{figure}

In Fig.~\ref{fig7}, we
show the results of the QNMs depending on the characteristic parameter $l$, where $L=0.45$ and $n=0$ are set.
The left diagram of Fig.~\ref{fig7} correspond to the change of ${\rm Re}\,\omega$ with respect to $l$ for the cases of $N=1/2$ and $N=1$,  respectively, and the right diagram of Fig.~\ref{fig7} correspond to the change of $-{\rm Im}\,\omega$ with respect to $l$ for the cases of $N=1/2$ and $N=1$, respectively. We note that the real parts of two cases have similar behaviors, so do the negative imaginary parts. In the left
diagram ${\rm Re}\,\omega$ depends on $l$ linearly, and the slope is approximately $0.66$ and $0.73$  for the
cases of $N=1/2$ and $N=1$, respectively. We deduce that the oscillating frequency of case $N=1/2$ is smaller
than that of case $N=1$ for a fixed $l$, and that the difference of oscillating frequency between the two cases
becomes large when $l$ increases. In the right diagram $-{\rm Im}(\omega)$ has a peak at $l=2$, where the
peak is approximately $0.63$ for the case of $N=1/2$ and $0.56$  for the case of $N=1$; in particular, $-{\rm
Im}(\omega)$ goes to constant  when $l\ge 5$, which equals $0.61$ and $0.55$ for the cases of $N=1/2$ and
$N=1$, respectively. We deduce that the damping time  (inversely proportional to $-{\rm Im}(\omega)$) of the
former case is smaller than that of the latter, and that there exists a minimum damping time at $l=2$ for the two
cases. We further know that our ARBH model is more stable in the case of $N=1$ than in the case of $N=1/2$ for
a fixed $l$, where the minimum damping time at the peak corresponds to the state with the least stability, and
that the stability decreases quickly when $l$ takes values from one to two, and increases slowly when $l$ takes
values from two to five, and finally maintains unchanged when $l\ge 5$  for the two cases.

\begin{figure}[!ht]
	\centering
	\begin{subfigure}[b]{0.455\textwidth}
		\centering
		\includegraphics[width=\textwidth]{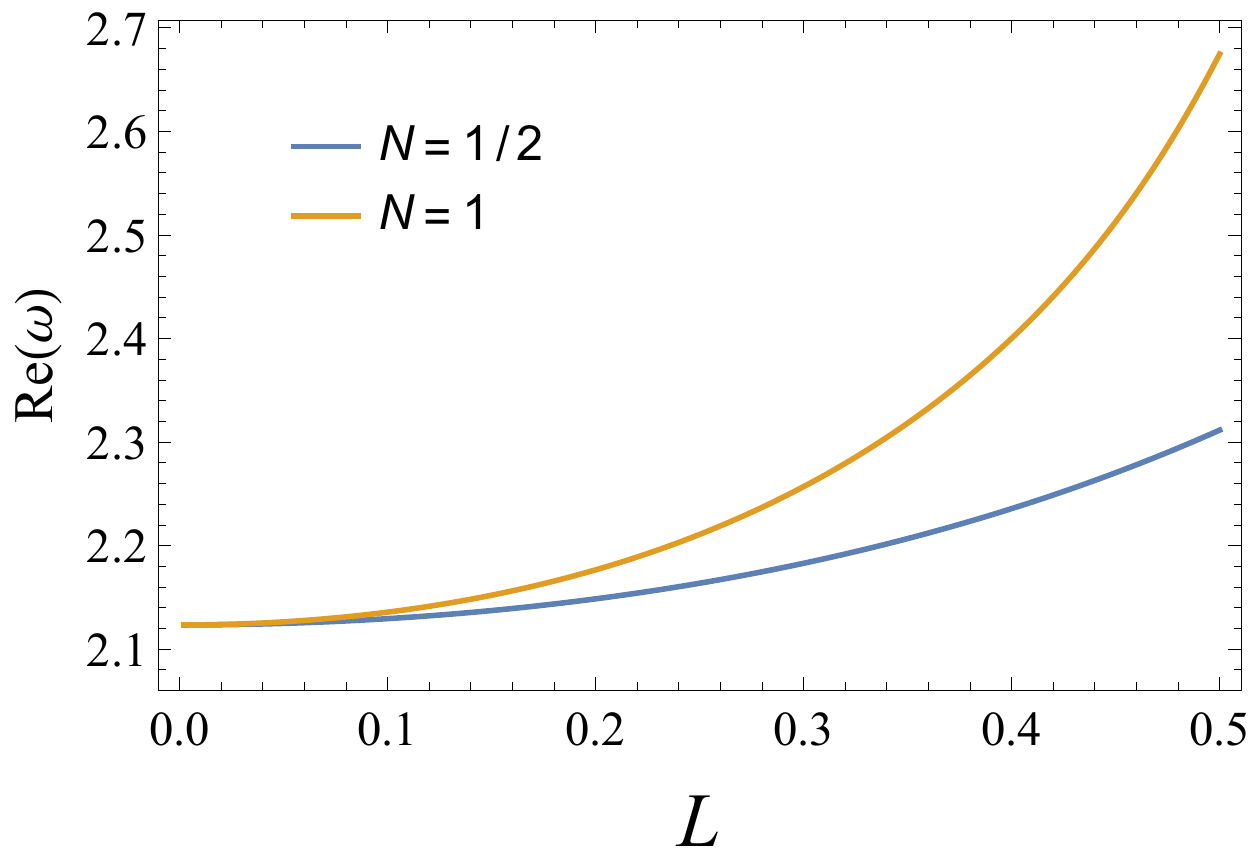}
	\end{subfigure}
	\begin{subfigure}[b]{0.46\textwidth}
		\centering
		\includegraphics[width=\textwidth]{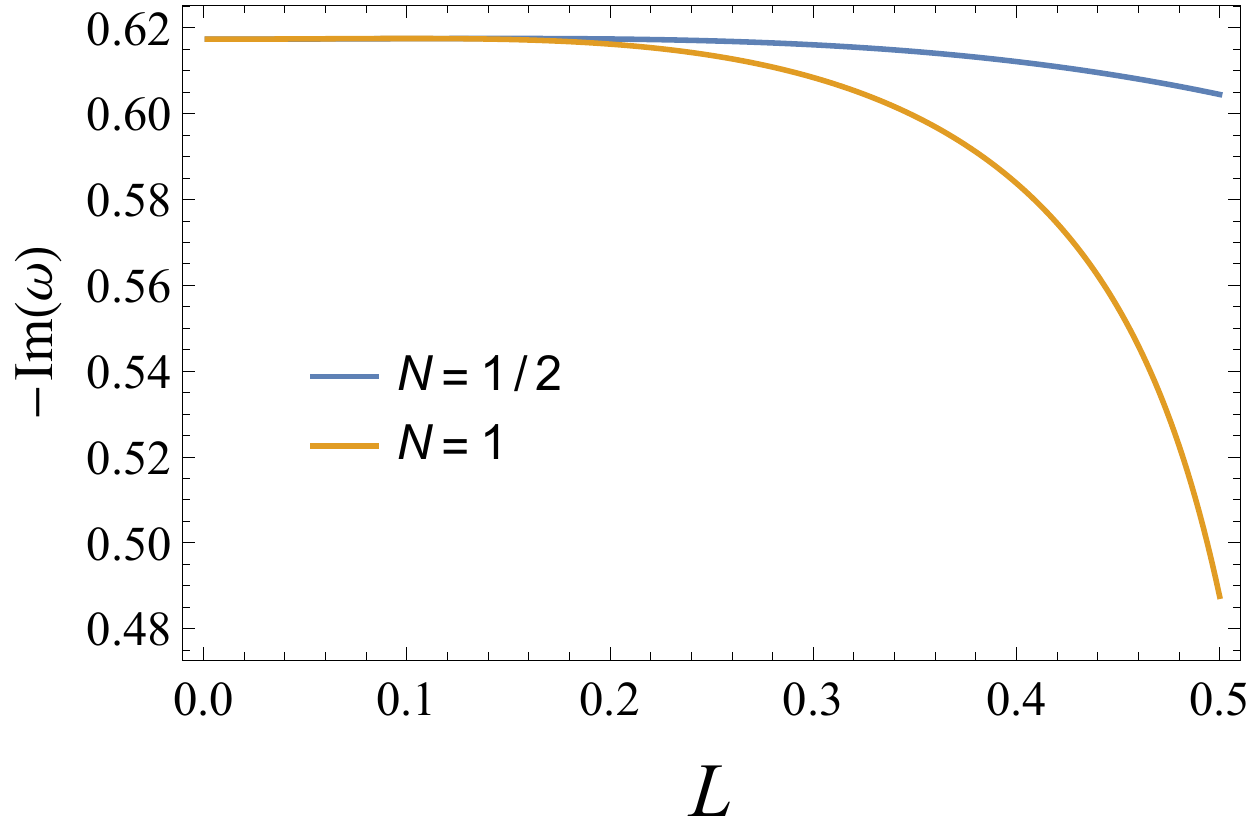}
	\end{subfigure}
	\captionsetup{width=.9\linewidth}
	\caption{QNMs with respect to $L$, where $l=3$ and $n=0$ are set. The left diagram represents the real parts
	of $\omega$ with respect to $L$ for the cases of $N=1/2$ and $N=1$, respectively; the right diagram
	represents the negative imaginary parts of $\omega$ with respect to $L$ for the cases of $N=1/2$ and $N=1$,
	respectively.}
	\label{fig: with rspect B}
\end{figure}

In Fig.~\ref{fig: with rspect B}, we
draw the results of the QNMs depending on the characteristic parameter $L$, where $l=3$ and $n=0$ are set. For
the two cases of $N=1/2$ and $N=1$, the real parts increase while the negative imaginary
parts decrease when $L$ increases.  For a fixed $L$, the real part of  case $N=1/2$ is smaller than that of case
$N=1$, which shows that the oscillating frequency for the former is smaller than that for the latter after our
ARBH model is perturbed; when $L$ becomes large, the difference of oscillating frequency between the two
cases becomes large. However, the negative imaginary part of  case $N=1/2$ is larger than that of case $N=1$ for a
fixed $L$, which shows that the damping time for the former is smaller than that for the latter;  when $L$
becomes large, the difference of damping time between the two cases also becomes large. In addition, our ARBH model
is stable after it is perturbed because the imaginary part is negative, and it is more stable in the case of $N=1$
than in the case of $N=1/2$ for a fixed $L$, and on the other hand  it is more stable for a larger $L$ in the both
cases.

\begin{figure}[!ht]
	\centering
	\begin{subfigure}[b]{0.45\textwidth}
		\centering
		\includegraphics[width=\textwidth]{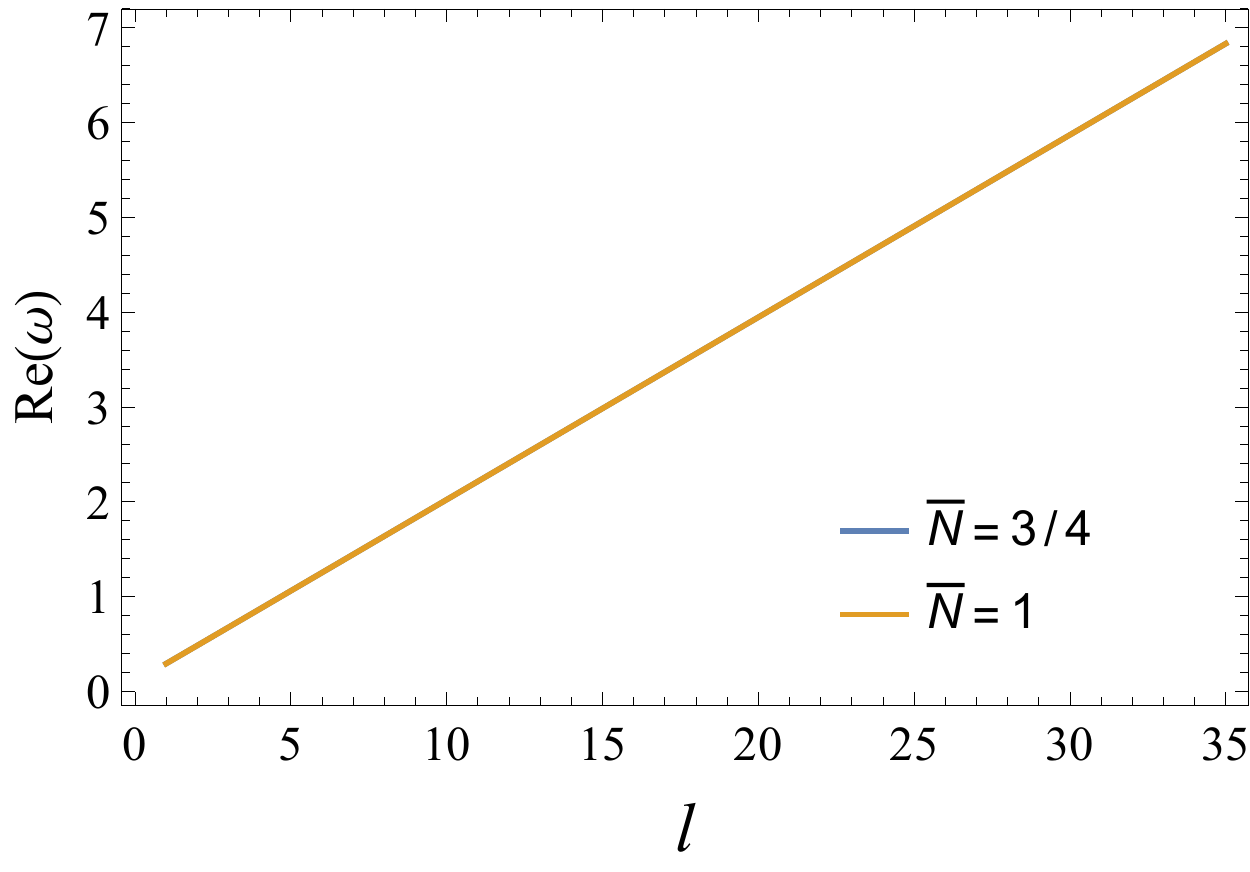}
	\end{subfigure}
	\begin{subfigure}[b]{0.485\textwidth}
		\centering
		\includegraphics[width=\textwidth]{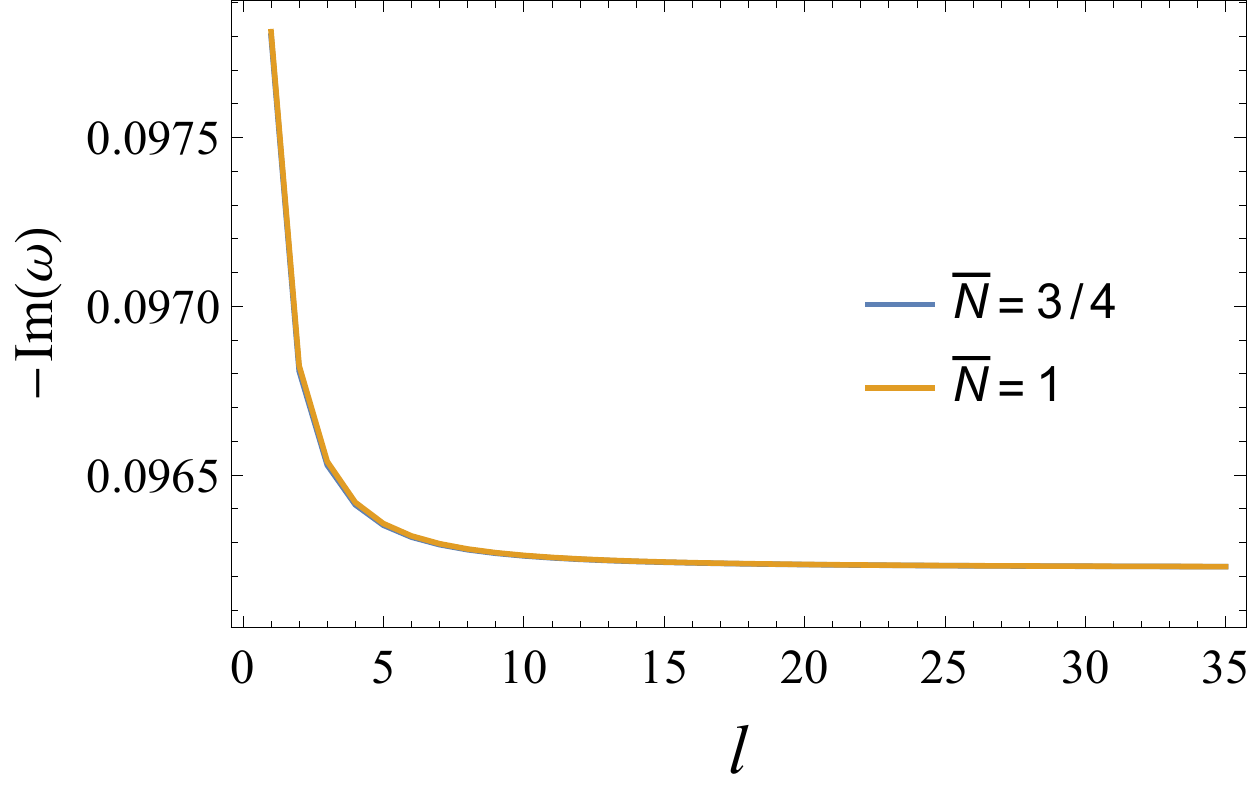}
	\end{subfigure}
	\captionsetup{width=.9\linewidth}
	\caption{QNMs of CRSBHs with respect to $l$, where $\bar{L}=0.45$ and $n=0$ are set. The left diagram
	represents
	the real parts
	of $\omega$ with respect to $l$ for the cases of $\bar{N}=3/4$ and $\bar{N}=1$, respectively; the right
	diagram
	represents the negative imaginary parts of $\omega$ with respect to $l$ for the cases of $\bar{N}=3/4$ and
	$\bar{N}=1$,
	respectively. Note that the curves of the two cases are almost overlapped.}
	\label{fig: cs-l}
\end{figure}

We also calculate the QNMs of CRSBHs and compare them with those of the ARBH. In
Fig.~$\ref{fig: cs-l}$, we draw the results with respect to the multiple number $l$ in the cases of $\bar{N} = 3/4$ and $\bar{N} = 1$, respectively.
In the left diagram of Fig.~\ref{fig: cs-l},  we find that the real parts of QNMs increase when
$l$ increases, which is similar to that of the ARBH. The right diagram of Fig.~\ref{fig: cs-l} shows that the negative
imaginary parts decrease monotonically when $l$ increases. However, we note that the negative imaginary parts of
the ARBH oscillate when $l$ increases and reach the maximum at $l=2$, see Fig.~\ref{fig7} for the details.

\begin{figure}[!ht]
	\centering
	\begin{subfigure}[b]{0.45\textwidth}
		\centering
		\includegraphics[width=\textwidth]{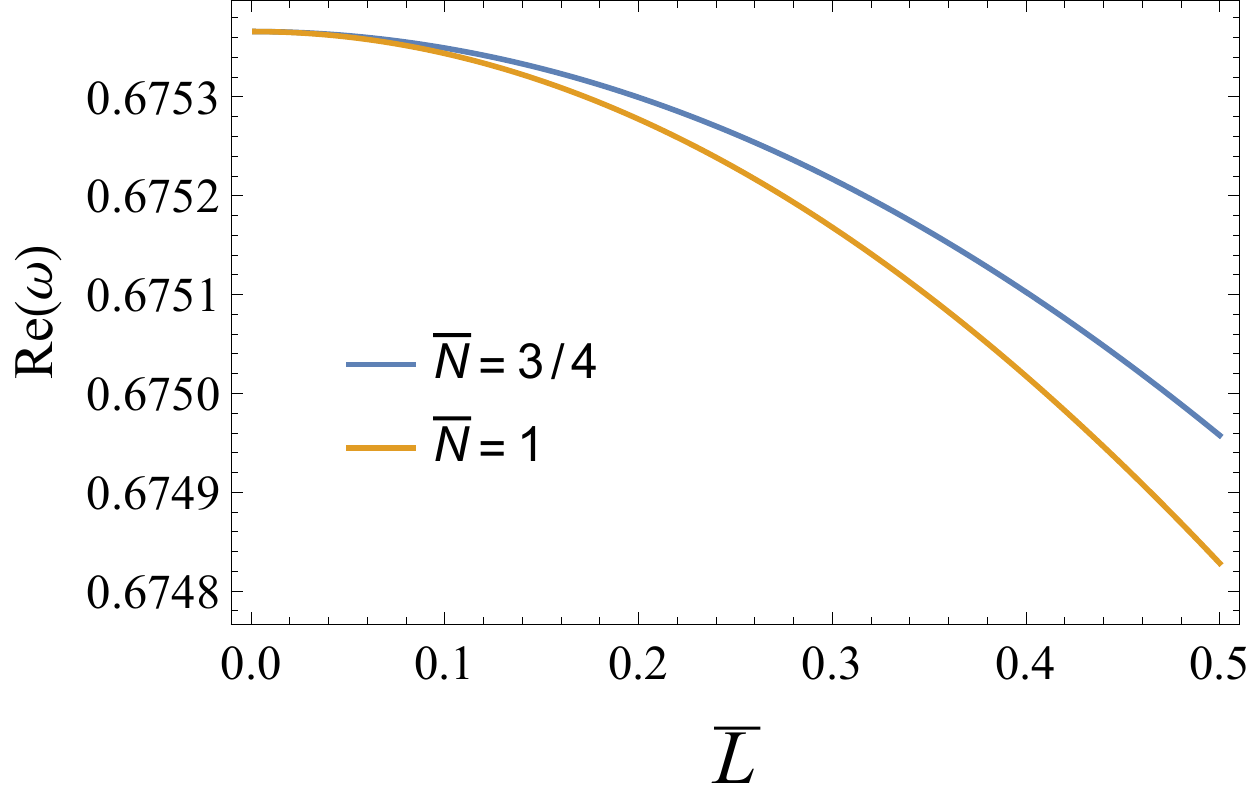}
	\end{subfigure}
	\begin{subfigure}[b]{0.46\textwidth}
		\centering
		\includegraphics[width=\textwidth]{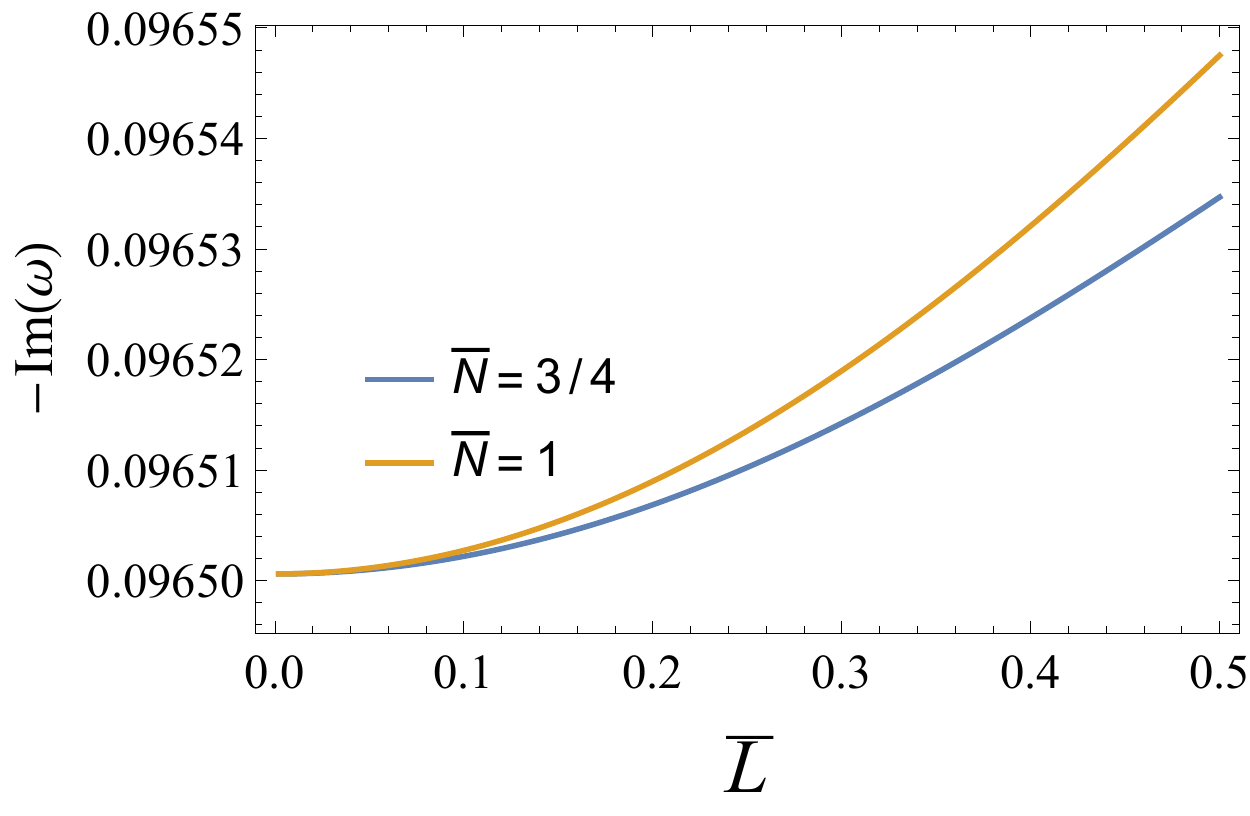}
	\end{subfigure}
	\captionsetup{width=.9\linewidth}
\caption{QNMs of CRSBHs with respect to $\bar{L}$, where $l=3$ and $n=0$ are set. The left diagram
represents the real parts of $\omega$ with respect to $\bar{L}$ for the cases of $\bar{N}=3/4$ and
$\bar{N}=1$, respectively; the right diagram represents the negative imaginary parts of $\omega$ with respect
to $\bar{L}$ for the cases of $\bar{N}=3/4$ and $\bar{N}=1$, respectively.}
\label{fig: cs-LL}
\end{figure}

At last, we investigate the QNMs of CRSBHs with respect to $\bar{L}$ and compare them with those of the ARBH. We plot Fig.~\ref{fig: cs-LL} for the two cases of $\bar{N} = 3/4$ and $\bar{N}= 1$, where the real parts
decrease while the negative imaginary parts increase when $\bar{L}$ increases. For a fixed
$\bar{L}$, the real part of case $\bar{N}=3/4$ is larger than that of case $\bar{N}= 1$, which shows that the
oscillating
frequency for the former is larger than that for the latter after a CRSBH is  perturbed; when $\bar{L}$ becomes
large, the difference of oscillating frequency between the two cases also becomes large. On the other hand, the negative
imaginary part of case $\bar{N}=3/4$ is smaller than that of case $\bar{N}= 1$ for a fixed $\bar{L}$, which shows
that the damping time for the former is larger than that for the latter; when $\bar{L}$ becomes large, the difference of
damping time between the two cases also becomes large. Comparing Fig.~\ref{fig: with rspect B} with Fig.~\ref{fig: cs-LL}, we find that the relative positions of the blue and orange curves are just opposite and the changes of them with respect to $L$ and $\bar L$ are opposite, too.

\section{Summary}
\label{sec4}
In the present work, we construct a general ARBH model in the spherically symmetric fluid. Unlike the current ABH model~\cite{berti2004quasinormala} whose velocity of fluid diverges at $r = 0$, our model has a finite
 velocity but divergent density, where the density plays the role of the scale factor of a conformally related BH. The fluid flow is realized with the aid of a certain
external field, which may offer a
 possibility to produce ARBHs in laboratory. Moreover, we give the valid domains of various energy conditions. 
  As we have shown in Fig.~\ref{fig:energy-condition-fullS=N}, the violated domains of the 
strong energy condition are located outside the horizon rather than inside the horizon, which may change our 
current knowledge on the relation  between the regularity and strong energy condition.
In addition, we compare our ARBH model with conformally related BHs in the aspect of energy conditions, and find the similarity and diversity between the two types of BHs.

In order to study ARBHs experimentally, it is necessary to analyze the
QNMs of ARBHs. Using the WKB method, we calculate the QNMs of our ARBH model characterized by  Eq.~\eqref{eq:vre} in the cases of $N=1/2$ and $N=1$. The results show that the imaginary parts of QNMs are
negative, which implies that our ARBH model is stable after it is perturbed. Moreover, the detail features of
oscillating frequency and damping time are also given. In particular, we reveal the dependence of stability on the
characteristic length of the scale factor (the density of fluid), $L$, i.e., our ARBH model is more stable for a larger
$L$.  When $N$ is larger, the oscillation is faster. 
In summary, we have shown that the acoustic gravity is able to be employed as a means 
of studying the scalar perturbation of RBHs.

The simulation method we proposed is suitable for a large class of RBHs and provides a basis for further researches of the Hawking radiation and superradiance. Meanwhile, there is plenty of room for improvement in our method if we strictly follow certain physical principles, such as maintaining the energy conditions, which will be reported soon in our next work.

In addition, our further considerations also focus on the divergence of the classical action in the ARBH model we construct.
This issue may lead to a vanishing partition function. Since the metric of our ARBH model has a conformal structure,
we try to deal with the issue in the framework of conformal gravity, where the divergence will be improved when
 a scalar field is introduced. This will be reported elsewhere.
\section*{Acknowledgments}
The authors are grateful to Y. Li for valuable remarks on conformal gravity. In particular,  the authors would like to thank the anonymous referee for the helpful comments that improve this work greatly.
This work was supported in part by the National Natural Science Foundation of China under Grant Nos. 11675081 and 12175108.
\appendix
\section{Energy conditions of CRSBHs}
\label{appendix:A}

In this appendix, we reanalyze the energy conditions of CRSBHs which have been considered in Ref.~\
\cite{Toshmatov:2017kmw}. Because the sign of the energy density is wrong  in  Ref.~\	
\cite{Toshmatov:2017kmw}, all the energy conditions related to it have to be reconsidered.
	
\subsection{The difference between ${T^{\mu}}_{\nu}$ and	$e_{\mu}^{(a)}T^{\mu\nu}e_{\nu}^{(b)}$}
Let us start with the perfect fluid whose energy-momentum tensor takes the form,
\begin{equation}
	T^{\mu\nu} = (\rho_0 +p)U^\mu U^\nu+p g^{\mu\nu},
\end{equation}
where $g_{\mu\nu} U^\mu U^\nu=-1$. In the rest frame, one can set $U^\mu=(1/\sqrt{-g_{00}}, 0, 0, 0)$,
thus the diagonalized form can be written as
\begin{equation}
	{T^{\mu}}_{\nu} = (\rho_0 +p)U^\mu U_\nu+p
	\delta^{\mu}_{\nu},
\end{equation}
namely,
\begin{equation}
	{T^0}_0=\frac{{G^0}_0}{8\pi}=-\rho_0,\qquad
	{T^i}_j=\frac{{G^i}_j}{8\pi}=p \delta^i_j.
\end{equation}
Therefore, one has
\begin{equation}
	{T^{\mu}}_{\nu}=\diag\{-\rho_0, p, p, p\},
\end{equation}
where the $00$ component of ${T^{\mu}}_{\nu}$ is negative energy density and the trace of ${T^{\mu}}_{\nu}$ equals
\begin{equation}
\Tr\, {T^\mu}_\nu=-\rho_0+3p.
\end{equation}

Alternatively, one can diagonalize $T^{\mu\nu}$ by using orthonormal tetrads. If the metric is diagonal,
$g_{\mu\nu}=\diag\{g_{tt}, g_{rr}, g_{\theta\theta}, g_{\phi\phi}\}$, the tetrads $e_{\mu}^{(a)}$ are of the
following form,
\begin{equation}
	e_{\mu}^{(a)}=\diag
	\left\{\sqrt{-g_{tt}}, \sqrt{g_{rr}}, \sqrt{g_{\theta\theta}}, \sqrt{g_{\phi\phi}}\right\}.\label{eq:ee}
\end{equation}
Using Eq.~\eqref{eq:ee}, one can get
\begin{equation}
	e_{\mu}^{(0)}T^{\mu\nu}e_{\nu}^{(0)}=\rho_0,\qquad
	e_{\mu}^{(i)}T^{\mu\nu}e_{\nu}^{(j)}=p \delta^i_j.
\end{equation}
Therefore, an alternative diagonalized form is
\begin{equation}
	e_{\mu}^{(a)}T^{\mu\nu}e_{\nu}^{(b)}=\diag\{\rho_0, p, p, p\},
\end{equation}
and the corresponding trace is
\begin{equation}
	\Tr\, e_{\mu}^{(a)}T^{\mu\nu}e_{\nu}^{(b)}
	=\rho_0+3p.
\end{equation}

\subsection{The correct sign of the energy density of CRSBHs}	
For the CRSBHs, the metric is
	\begin{equation}
		g_{\mu\nu}= S(r)\diag\{-f, f^{-1}, r^2, r^2 \sin^2\theta\},
	\end{equation}
where
	\begin{equation}
		S(r)=\left(1+\frac{\bar{L}^2}{r^2}\right)^{2\bar{N}},\qquad
		f(r)=1-\frac{2M}{r}.
	\end{equation}
The $00$ component of ${T^{\mu}}_{\nu}$ can be computed,
	\begin{equation}
		{T^0}_0=-\rho_0=-e_{\mu}^{(0)}T^{\mu\nu}e_{\nu}^{(0)}=\frac{4 \bar{L}^2 \bar{N} r^{4 \bar{N}-3}
		}{8\pi\left(\bar{L}^2+r^2\right)^{2
		(\bar{N}+1)}}
		\left[\bar{L}^2 \left(-2 M \bar{N}+M+\bar{N}r- r\right)+r^2 (r-3 M)\right],
	\end{equation}
namely, the energy density is
	\begin{equation}
		\rho_0 = -\frac{4 \bar{L}^2 \bar{N} r^{4 \bar{N}-3}
		}{8\pi\left(\bar{L}^2+r^2\right)^{2
				(\bar{N}+1)}}
		\left[\bar{L}^2 (-2 M \bar{N}+M+\bar{N}r-r)+ r^2(r-3 M)\right],
	\end{equation}
which is different from Eq.~(A.1) of Ref.~\cite{Toshmatov:2017kmw} up to a minus sign. Thus, all the
inequalities will be different, e.g.,
	\begin{equation}
 \rho_0+\sum_{i=1}^{3} p_i= -\frac{4 \bar{L}^2 \bar{N} r^{4 \bar{N}-3}}{ 8\pi\left(\bar{L}^2+r^2\right)^{2
 (\bar{N}+1)}}
		\left[\bar{L}^2 (M (8 \bar{N}+2)-(4 \bar{N} -1)r)- (r-6 M)r^2\right]
	\end{equation}
is different from Eq.~(33) of Ref.~\cite{Toshmatov:2017kmw}.
\section{Energy conditions of the toy model $f(r)=1-r^n$}
\label{appendix:B}
In order to analyze the SEC and DEC outside the event horizon, we analyze the model with the metric,
	\begin{equation}
		g_{\mu\nu}={\rm diag}\left\{-(1-r^n), (1-r^n)^{-1}, r^2, r^2
		\sin^{2 } \theta\right\},
	\end{equation}
where $n\in \mathbb{R}$. The SEC is then represented via $\rho_0$ and $P_i$  $(i=1,2,3)$,
	\begin{equation}
		\rho_0 + P_1 = 0,\label{eq:n1}
	\end{equation}
	\begin{equation}
		\rho_0 + P_2 = \rho_0 + P_3 = - \frac{(n - 2)  (n + 1)}{16 \pi} r^{n - 2},\label{eq:n2}
	\end{equation}
	\begin{equation}
		\rho_0 + \sum_{i = 1}^3 P_i = - \frac{n (n + 1)}{8 \pi} r^{n - 2}.\label{eq:n3}
	\end{equation}
From Eqs.~\eqref{eq:n1}-\eqref{eq:n3}, we note that $n$ must satisfy $- 1 \leq n \leq 0$ in order to ensure that
the SEC is satisfied. For the DEC, one has the following form,
	\begin{equation}
		\rho_0 - | P_1 | = \frac{(n + 1 - | n + 1 |) }{8 \pi} r^{n - 2},\label{eq:4}
	\end{equation}
	\begin{equation}
		\rho_0 - | P_2 | = \frac{(2 n + 2 - | n |  | n + 1 |) }{16 \pi} r^{n - 2},\label{eq:5}
	\end{equation}
which leads to $- 1 \leq n \leq 2$ if the DEC is satisfied. For our ARBH model, the metric function is asymptotic to $1-1/r^{4}$ at
infinity, i.e., $n=-4$, which means that both SEC and DEC are violated outside the event horizon.
	
From the point of view of Raychaudhuri's equation \cite{Carroll:2004st}, when the expansion, rotation, and shear can be
neglected, one obtains
	\begin{equation}
		\frac{\dif \xi}{\dif \tau} =-4\pi \left(\rho_0+\sum_{i=1}^{3} P_i\right),
	\end{equation}
where $\xi$ denotes expansion of geodesics and $\tau$ affine parameter. The violation of Eq.~\eqref{eq:n3} implies  $\dif\xi/\dif \tau>0$, i.e.,	the gravity is repulsive outside the horizon for $n=-4$.


\end{document}